\UseRawInputEncoding
\documentclass[journal]{IEEEtran}
\usepackage[linesnumbered,ruled]{algorithm2e}

\renewcommand{\IEEEQED}{\IEEEQEDclosed}
\usepackage{color}
\usepackage[table]{xcolor}
\usepackage{graphicx}
\usepackage{epstopdf}
\usepackage{booktabs}
\usepackage{subfigure}
\usepackage{graphicx,cite,epsfig,amssymb,amsmath,multirow,lettrine,flushend,extarrows,makecell}
\usepackage{graphics}
\usepackage{array}

\usepackage{threeparttable}

\begin{document}



\title{A Review of Hydrogen-Enabled Resilience Enhancement for Multi-Energy Systems}

\author{{Liang~Yu,~\IEEEmembership{Senior Member,~IEEE}, Haoyu Fang, Goran Strbac,~\IEEEmembership{Fellow,~IEEE}, Dawei Qiu,~\IEEEmembership{Senior Member,~IEEE}, Dong Yue,~\IEEEmembership{Fellow,~IEEE}, Xiaohong Guan,~\IEEEmembership{Life Fellow,~IEEE}, and Gerhard P. Hancke,~\IEEEmembership{Life Fellow,~IEEE}}
\thanks{
\newline L. Yu is with the College of Automation, Nanjing University of Posts and Telecommunications, Nanjing 210023, China. He is also with the Department of Electrical and Electronic Engineering, Imperial College London, London SW7 2AZ, U.K. (e-mail: liang.yu@njupt.edu.cn)
\newline H. Fang and G. Strbac are with the Department of Electrical and Electronic Engineering, Imperial College London, London SW7 2AZ, U.K. (e-mail: haoyu.fang22@imperial.ac.uk; g.strbac@imperial.ac.uk)
\newline D. Qiu is with the Department of Engineering, University of Exeter, Exeter, EX4 4PY, U.K. (e-mail: d.qiu@exeter.ac.uk)
\newline D. Yue is with the Advanced Technology Institute for Carbon Neutrality, Nanjing University of Posts and Telecommunications, Nanjing 210023, China (e-mail: medongy@vip.163.com)
\newline X. Guan is with the Systems Engineering Institute, Ministry of Education Key Laboratory for Intelligent Networks and Network Security, Xi’an Jiaotong University, Xi'an 710049, China (xhguan@sei.xjtu.edu.cn)
\newline G. P. Hancke is with the College of Automation, Nanjing University of Posts and Telecommunications, Nanjing 210023, China (e-mail: g.hancke@ieee.org)
}}


\maketitle

\begin{abstract}
Ensuring resilience in multi-energy systems (MESs) has become increasingly urgent and challenging due to the growing frequency and severity of extreme events, such as natural disasters, extreme weather, and cyber-physical attacks. Among the various approaches to enhancing MES resilience, hydrogen integration offers significant potential in cross-temporal, cross-spatial, and cross-sector flexibility, as well as black-start capability. Although considerable efforts have been devoted to this area, a systematic review of resilience enhancement in hydrogen-enabled MESs is still lacking. To address this gap, this paper presents a comprehensive review of hydrogen-enabled MES resilience enhancement. First, advantages, vulnerabilities, and challenges related to hydrogen-enabled MES resilience enhancement are summarized. Next, a resilience enhancement framework for hydrogen-enabled MESs is proposed, based on which existing resilience metrics and event-oriented contingency models are reviewed and discussed. Planning measures are then classified according to the types of hydrogen-related facilities, together with uncertainty handling methods, scenario generation methods, and planning problem formulation frameworks. In addition, operational enhancement measures are categorized into three response stages: prevention, emergency response, and restoration. Finally, research gaps are identified and future directions are discussed, including comprehensive resilience metric design, advanced extreme-event scenario generation, spatiotemporal cyber-physical contingency modeling under compound extreme events, coordinated planning and operation across multiple networks and timescales, low-carbon resilient planning and operation, and large language model-assisted whole-process resilience enhancement.
\end{abstract}

\begin{IEEEkeywords}
Multi-energy systems, hydrogen-enabled resilience enhancement, resilience metrics, contingency modeling, planning, operation, extreme events, low-carbon, large language models, deep reinforcement learning
\end{IEEEkeywords}

\section{Introduction}\label{s1}
It is well recognized that power and energy systems are indispensable to modern society\cite{YuY2024}. Compared with single-energy systems, multi-energy systems (MESs), which integrate electricity, heat, gas, and other energy carriers, can offer more advantages in terms of efficiency, flexibility, economic performance, decarbonization, and energy security. Thus, MESs play an important role in shaping our energy future\cite{Malley2020}. However, the operation of MESs is facing growing challenges due to the rising frequency and intensity of extreme events, such as natural disasters (e.g., earthquakes, floods, wildfires), climate-induced extreme weather (e.g., heatwaves, hurricanes, cold spells), and cyber-physical attacks targeting critical infrastructure\cite{Perera2020,Nik2021,Stankovi2023,Do2023}. These events can significantly affect interconnected energy supply networks, and their correlated occurrence across time and space (e.g., concurrent heatwaves and cyber threats) often leads to compound risks with amplified socio-economic impacts\cite{XuLuo2024}. For instance, blackout-related economic losses from extreme weather events are estimated at 20–55 billion dollars per year in the United States\cite{Nik2021}. To address these challenges, the concept of resilience has been introduced, emphasizing an energy system’s ability to withstand, adapt to, and recover from extreme events while maintaining essential functions\cite{Stankovi2023}, \cite{Jasi2021,Afgan2012,Yang2022}. As MESs couple multiple energy carriers and play a critical role in supporting both critical loads and economic activities, ensuring their resilience is not only a technical necessity but also a societal imperative.

\begin{figure}[!htb]
\centering
\includegraphics[scale=0.3]{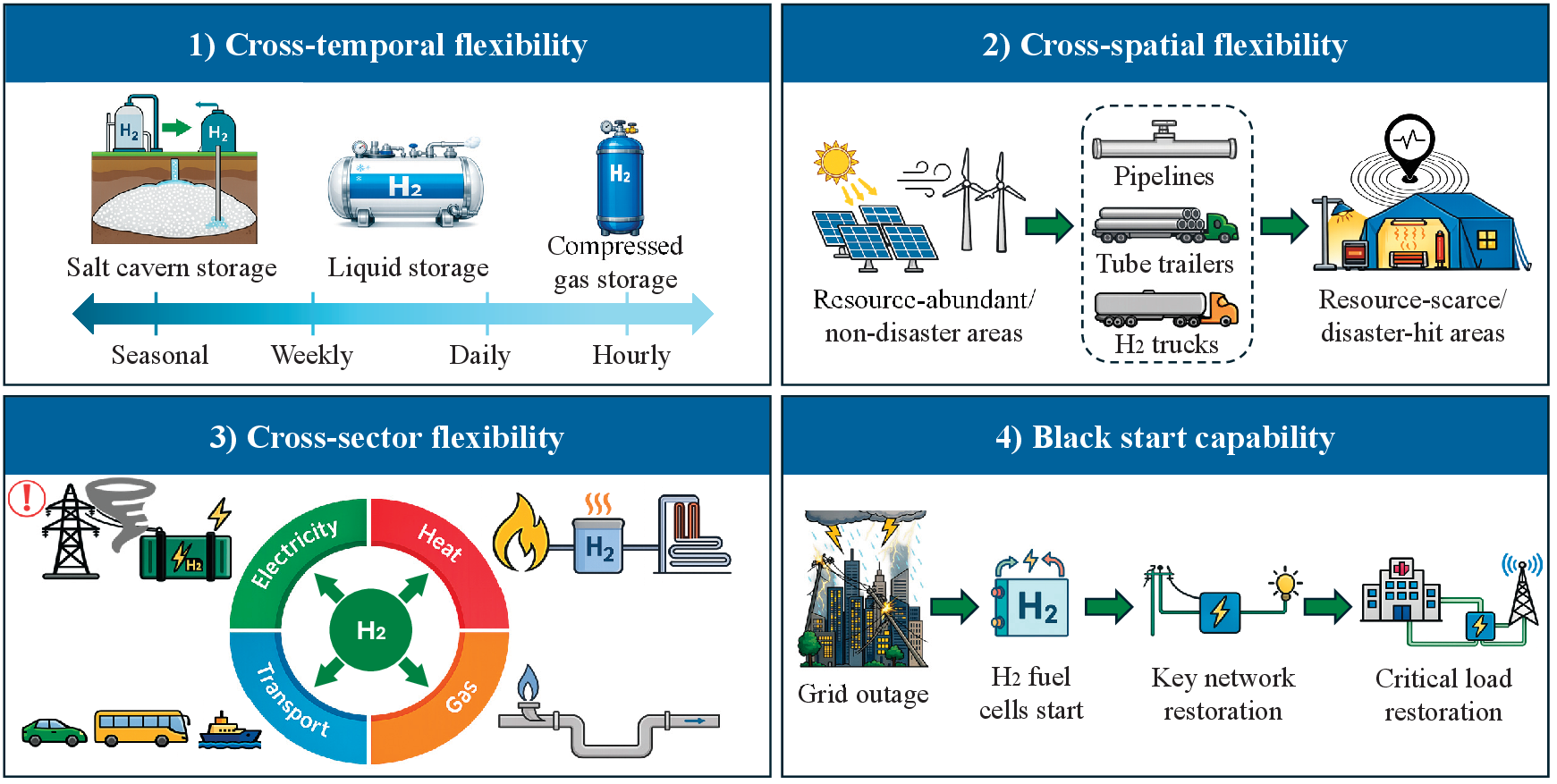}
\caption{Distinguishing features of hydrogen for enhancing MES resilience. Note that cross-temporal flexibility refers to the ability to store hydrogen over long durations and shift energy use from periods of surplus to periods of shortage; cross-spatial flexibility refers to the capability to transport hydrogen across regions to balance spatially uneven supply and demand; cross-sector flexibility refers to the interoperability among the electricity, heat, gas, and transport sectors through hydrogen conversion and blending; and black-start capability refers to the ability of an HMES system to self-start and restore electricity supply to critical infrastructure (e.g., hospitals, emergency shelters, and communication networks) without relying on external grid support during extreme events.}\label{fig_1}
\end{figure}

Among the measures of enhancing MES resilience, the integration of hydrogen energy resources (HERs) has shown exceptional potential. To be specific, HERs refer broadly to the various technologies and systems that produce, store, transport, and utilize hydrogen as an energy carrier, including stationary HERs (e.g., electrolyzers, hydrogen tanks, hydrogen pipelines, hydrogen fuel cell-based conversion systems, and hydrogen refueling stations (HRSs)) and mobile HERs (MHERs), which include hydrogen tube trailers (HTTs)\cite{Yang2024,SunC2025,MaY2025,JiaW2025,ReddiK2014,XiaW2024,Lv2024}, mobile hydrogen trucks\cite{LiB2023,QianH2024,SongJ2024}, mobile fuel cell trucks\cite{ZhangP2024}, hydrogen fuel cell buses (HFCBs)\cite{Dong2023}, hydrogen fuel cell vehicles (HFCVs)\cite{ChenF2024}. The above HERs can offer distinguishing features for enhancing MES resilience in aspects of energy supply and carbon emission as shown in Fig.~\ref{fig_1}\cite{ChenS2024,YuL2024,YueM2021,GenoveseM2023}, i.e., cross-temporal flexibility, cross-spatial flexibility, cross-sector flexibility, and black start capability. Despite the promising potential of hydrogen integration, developing effective resilience enhancement strategies for hydrogen-enabled MESs (HMESs) involves significant complexity across multiple dimensions. First, defining resilience metrics for HMESs is inherently challenging due to the need to capture many aspects, e.g., load loss, carbon resilience, and physical and functional features of hydrogen systems (e.g., hydrogen availability). Second, modeling extreme event-related HESS contingencies requires careful representation of multi-stage processes (such as electrolysis, compression, storage, and reconversion) and each process is subject to distinct cyber and physical operational risks. Third, designing the schemes of hydrogen-related infrastructure planning and HMES operation faces solving challenges caused by uncertainties, mixed integer variables, multi-timescale variables, temporal-spatial coupling constraints, nonlinear and nonconvex constraints, multi-energy coupling constraints, and multi-objective functions.

To overcome the above challenges, researchers have defined many resilience metrics, developed some contingency models, and proposed many planning and operation methods for enhancing the resilience of HMESs\cite{ZhaoH2022,Liu2024,ZhaoY2024,ZhuR2024,Wu2022}. However, there is a lack of a systematic overview of hydrogen-enabled resilience enhancement for MESs. To fill the research gap, we provide a comprehensive overview of leveraging hydrogen to enhance the resilience of MESs. Although there have been many review works related to power grid resilience enhancement\cite{Bie2017,Li2017,Mahzarnia2020,Hossain2021,Xu2021,Ma2021,Shi2022,Younesi2022,Amini2023,Modaberi2023,Huang2024,Zidane2025,Han2023} with different perspectives (e.g., networked microgrids, event timeline of enhancement, cyber-physical interdependence, complex network, and cross-domain multilayer architecture), they neglect to consider the HERs. \cite{Han2023} is the first and only work to review the resilience of a hydrogen-powered smart grid. Although this review paper considers the hydrogen-related facilities, there are many differences between \cite{Han2023} and our work in aspects of objects, perspectives, contents, and research gaps. Firstly, we focus on HMESs, while \cite{Han2023} focuses on hydrogen-enabled smart grid. Secondly, we review the existing works from the perspective of hydrogen-enabled planning and operation. In contrast, \cite{Han2023} mainly focuses on the event timeline of measures, i.e., pre-event, during-event, and post-event. Thirdly, this paper summarizes advantages and challenges of adopting hydrogen in MES resilience enhancement, resilience metrics of HMES and event-oriented contingency models, which are not mentioned in \cite{Han2023}. Fourthly, we classify the existing resilience enhancement measures with the consideration of hydrogen pipeline hardening, more types of HERs, and multi-stage operation coordination, whereas \cite{Han2023} considers measures only within independent stages. Finally, we identify six research gaps and point out future directions, which are neglected in \cite{Han2023}. To the best of our knowledge, our work is the first survey dedicated to hydrogen-enabled resilience enhancement for MESs.

The main contributions of this paper are summarized below.
\begin{enumerate}
  \item We summarize the advantages and challenges of adopting hydrogen in MES resilience enhancement. Specifically, distinguishing features of hydrogen for resilience enhancement, resilience potential and security issues of hydrogen-related technologies, and optimization challenges of hydrogen-related planning and operation are summarized.
  \item We propose a comprehensive resilience enhancement framework for HMESs considering hydrogen-enabled planning and operation measures (including pre-event preventive response, during-event emergency response, and post-event restoration response). Under the proposed framework, the widely used resilience metrics and event-oriented contingency models in existing works are summarized.
  \item We classify the planning enhancement of HMES resilience from two aspects, i.e., single-type hydrogen facility and multi-type hydrogen facility. The hydrogen-related facilities considered include HESSs, hydrogen pipelines, HRSs, HFCBs, fuel cell-driven combined cooling, heat, and power (CCHP), and hydrogen-related production facilities. Moreover, we summarize the classifications of uncertainty handling methods, scenario generation methods, and problem formulation frameworks for HMES planning.
  \item We categorize the operation enhancement of HMES resilience according to three types of operation response stages involved, including preventive response, emergency response, and restoration response. Moreover, the insights about the particular function of hydrogen in each operation stage are summarized.
  \item We identify six research gaps and point out future directions in aspects of comprehensive resilience metric design, advanced extreme-event scenario generation, multi-type temporal-spatial cyber-physical contingency modeling under compound extreme events, multi-network multi-timescale coordinated planning and operation, low-carbon resilient planning and operation, and large language model-assisted whole-process resilience enhancement.
\end{enumerate}

The rest of this paper is organized as follows. In Section~\ref{s2X}, advantages and challenges of adopting hydrogen in MES resilience enhancement are summarized. In Section~\ref{s2}, a comprehensive resilience enhancement framework for HMESs is described. In Section~\ref{s3}, the definitions and metrics of HMES resilience are described. In Section~\ref{s4}, event-oriented contingency modeling is summarized. In Section~\ref{s5}, planning enhancement measures for HMES resilience are classified, and some insights are provided. In Section~\ref{s6}, operation enhancement measures for HMES resilience are categorized. Section~\ref{s7} identifies six research gaps and possible future directions. Finally, conclusions and lessons learned are summarized in Section~\ref{s8}. For better understanding, we provide the list of abbreviations in alphabetical order in Table~\ref{table_1}.

\begin{table}[htbp]
\center
\setlength{\arrayrulewidth}{1pt}
\caption{List of abbreviations in alphabetical order}\label{table_1} \centering
\rowcolors{1}{yellow!12}{yellow!12} 
\begin{tabular}{|c|c|}
\hline
\rowcolor{yellow!25} 
\textbf{Abbreviation} &\textbf{Description}\\
\hline
C\&CG & Column and constraint generation \\
\hline
CCHP & Combined cooling, heat and power \\
\hline
CRI & Comprehensive resilience index \\
\hline
ELNS &Expected load not served \\
\hline
ERL & Expected restored load \\
\hline
ESS  &Energy storage system\\
\hline
GAN & Generative adversarial networks \\
\hline
GSHP & Ground-source heat pump \\
\hline
H2P & Hydrogen-to-power \\
\hline
HESS & Hydrogen energy storage system \\
\hline
HER & Hydrogen energy resource \\
\hline
HFCB & Hydrogen fuel cell bus \\
\hline
HFCV & Hydrogen fuel cell vehicle \\
\hline
HMES & Hydrogen-enabled multi-energy system\\
\hline
HRS & Hydrogen refueling station \\
\hline
HTT & Hydrogen tube trailer \\
\hline
LLM  & Large language model\\
\hline
LL & Load loss \\
\hline
LSR & Load served ratio \\
\hline
MHER & Mobile hydrogen energy resource\\
\hline
MILP & Mixed integer linear programming \\
\hline
MINLP & Mixed integer nonlinear programming \\
\hline
MISOCP & Mixed-integer second-order cone programming \\
\hline
PV & Photovoltaic \\
\hline
SHS & Seasonal hydrogen storage \\
\hline
TES  &Thermal energy storage\\
\hline
WSLL &Weighted sum of load loss \\
\hline
\end{tabular}
\end{table}

\section{Advantages, Vulnerabilities, and Challenges Related to Hydrogen-enabled MES Resilience Enhancement}\label{s2X}

In this section, the distinguishing features of adopting hydrogen for MES resilience enhancement are firstly provided. Then, from the perspective of the hydrogen supply chain, potentials and vulnerabilities related to hydrogen technologies in resilience enhancement are summarized. Finally, optimization challenges associated with resilience-oriented planning and operation of hydrogen-enabled MESs are analyzed.

\subsection{Distinguishing Features of Hydrogen for MES Resilience Enhancement}
Among various measures of enhancing MES resilience, the integration of HERs has shown exceptional potential due to their distinctive multi-functional attributes as shown in Fig.~\ref{fig_1}, which are described below.

\subsubsection{\textbf{Cross-temporal Flexibility}} This feature refers to the ability to store hydrogen over long durations and shift energy use from periods of surplus to periods of shortage\cite{Gu2024,WenZ2024,Hassan2023}. Unlike traditional battery energy storage systems, hydrogen can be stored at a much larger scale and over longer durations, e.g., underground salt-cavern hydrogen storage can be used to provide cross-seasonal low-cost and large-scale hydrogen storage\cite{CaoY2021}. The above feature can improve MES energy resilience by providing energy backup in the case of long-duration power outages. For example, electricity and heat/cold demand in the summer and winter are typically higher than those in spring and autumn. To prepare for possible prolonged outages (e.g., lasting for more than 48 hours\cite{Haggi2022,CicekA2025,resiliency}) under extreme events (e.g., heat waves or ice disaster), proactive hydrogen production and storage could be conducted in spring and autumn. In addition, when produced from renewable energy, hydrogen improves the consumption rate of renewable energy. By converting excess electricity into storable hydrogen using seasonal hydrogen storage (SHS) and conventional HESSs, MESs can buffer supply-demand mismatches more effectively and operate with greater carbon resilience (i.e., an energy system's ability to maintain or recover to its current and predicted emission level during and after shocks such as renewable intermittency, public health emergencies, or climate-related disruptions)\cite{WuC2024}. As shown in \cite{WangY2025}, 100\% self-sufficiency and zero carbon emissions can be simultaneously achieved in distant oceanic islands with abundant renewable energy resources by coordinating SHSs and HESSs.

\subsubsection{\textbf{Cross-spatial Flexibility}} This feature refers to the capability to transport hydrogen across regions to balance spatially uneven supply and demand\cite{Gu2024,WenZ2024,Hassan2023}. For example, hydrogen can be flexibly transported across regions through hydrogen pipelines, HTTs, or liquid hydrogen trucks, making it possible to allocate energy geographically according to regional demand and supply conditions. This spatial transfer capability becomes especially valuable in resilience enhancement when local renewable resources are insufficient or disrupted. For instance, during an earthquake, hydrogen can be transported from unaffected industrial production bases to disaster-hit areas to power temporary shelters via fuel cells, ensuring heating and lighting during prolonged grid outages. Similarly, mobile hydrogen refueling trucks can be deployed to support remote or damaged areas after extreme weather events, such as floods, by providing temporary hydrogen fuel for microgrids and emergency vehicles. By bridging spatial gaps between supply and demand, hydrogen transportation infrastructure allows MESs to dynamically reallocate resources, maintain critical services, and prevent cascading failures in interconnected energy networks, resulting in improved energy resilience.

\subsubsection{\textbf{Cross-sector Flexibility}} This feature refers to the interoperability among the electricity, heat, gas, and transport sectors through hydrogen conversion and blending\cite{ZhuR2024}. For example, the power-to-hydrogen technologies enable surplus electricity from renewable sources like wind and solar to be converted into hydrogen, which can then be stored and later used for electricity and heat generation or as a fuel for HFCVs in the transportation sector. In addition, hydrogen can be converted to synthetic methane or blended with natural gas for transportation\cite{ZhuR2024},\cite{YangT2025},\cite{Shahid2025}, which can reduce dependence on pure natural gas. The above interoperability not only enhances system efficiency, but also supports more robust and adaptive energy supply chains. Based on dynamic reconfiguration, hydrogen can facilitate resource sharing across different energy vectors, especially during extreme events. In such cases, if one subsystem experiences a failure, hydrogen can provide an alternative energy pathway, thereby ensuring the continuous operation of essential services and improving MES resilience.

\subsubsection{\textbf{Black Start Capability}} This feature refers to the ability of an HMES system to self-start and restore electricity supply to critical infrastructure (e.g., hospitals, emergency shelters, and communication networks) without relying on external grid support during extreme events\cite{LuJ2025}. In the event of prolonged power outages or natural gas supply failures, hydrogen-based infrastructure (such as proton exchange membrane fuel cells and hydrogen tanks) can operate off-grid and provide autonomous ``black start" capabilities\cite{LuJ2025}, significantly enhancing the system's ability to self-recover without external intervention during extreme events. According to \cite{LuJ2025}, black start should be implemented before network reconfiguration and load restoration are performed. Compared with black-start diesel generators, hydrogen fuel cells produce zero local emissions, operate quietly, and offer higher reliability with fewer maintenance requirements.

In summary, hydrogen has some distinguishing features and its role in enhancing energy resilience and carbon resilience of MESs is multi-dimensional, which can be summarized as shown in Table~\ref{table_2}.

\begin{table*}[htbp]
\centering
\setlength{\arrayrulewidth}{1pt}
\caption{Distinguishing features of hydrogen and its role in MES resilience enhancement}
\label{table_2}
\renewcommand{\arraystretch}{1.2}
\rowcolors{1}{yellow!12}{yellow!12} 
\begin{tabular}{|>{\color{black}}m{1.4cm}<{\centering}|>{\color{black}}m{2.1cm}<{\centering}|>{\color{black}}m{4.8cm}<{\centering}|>{\color{black}}m{3.8cm}<{\centering}|>{\color{black}}m{3.8cm}<{\centering}|}
\hline
\rowcolor{yellow!25} 
\textbf{Index} & \textbf{Distinguishing Feature} & \textbf{Role in MES Resilience Enhancement} & \textbf{Energy Resilience} & \textbf{Carbon Resilience}
\\
\hline
1 & Cross-temporal flexibility & Ability to store hydrogen over long durations (e.g., seasonal storage) and shift energy use from surplus periods to shortage periods & Long-duration backup and seasonal balancing reduce risk of supply shortages & Enables higher renewable integration and reduces curtailment \\
\hline
2 & Cross-spatial flexibility & Capability to transport hydrogen across regions via pipelines, HTTs, or liquid hydrogen trucks to match spatially uneven supply-demand patterns & Allows resource reallocation during local outages or regional shortages & Enables renewable-rich regions to export clean energy to high-demand or low-renewable areas \\
\hline
3 & Cross-sector flexibility & Interoperability between electricity, heat, gas, and transport sectors via hydrogen conversion and blending & Provides alternative energy pathways when one sector fails & Facilitates green fuel substitution and decarbonization in multiple sectors \\
\hline
4 & Black start capability & Off-grid operation of hydrogen fuel cells and tanks to restart or sustain critical infrastructure during outages & Ensures rapid recovery and maintains essential services in blackout scenarios & Zero carbon emission compared with diesel generators \\
\hline
\end{tabular}
\end{table*}

\subsection{Resilience Potential of Hydrogen-related Technologies}
In the hydrogen supply chain, various hydrogen production pathways (e.g., electrolyzers and steam methane reforming \cite{Le2024}), storage options (e.g., salt-cavern hydrogen storage and hydrogen tanks), transportation modes (e.g., pipelines, trucks, and tube trailers), and hydrogen-to-power technologies (e.g., proton exchange membrane fuel cells, solid oxide fuel cells, and hydrogen turbines) provide different resilience capabilities in terms of cross-temporal flexibility, cross-spatial flexibility, cross-sector flexibility, and black-start capability, as summarized in Table \ref{table_3}. The basis for the qualitative ratings (i.e., high, medium, and not obvious) is explained in the last column of the table. For example, salt-cavern hydrogen storage is rated as high in cross-temporal flexibility because it offers very large-scale and long-duration storage, making it particularly effective for bridging prolonged supply-demand mismatches and supporting resilience during extended disturbances. By contrast, hydrogen tanks are rated as medium because they mainly provide short-term to medium-term on-site buffering with more limited storage scale and duration. Hydrogen pipelines are also rated as medium, as they can provide a certain degree of temporal balancing through linepack; however, this capability is generally much more limited than that of dedicated geological storage, and pipelines are primarily designed for transport rather than long-duration storage. In addition, HTT is mainly characterized by cross-spatial flexibility because its principal function is to transport hydrogen from one location to another. Its cross-temporal flexibility is rated as not obvious for two reasons. First, the transportation time associated with HTT is essentially a logistics-related delivery lag rather than intentional temporal shifting enabled by storage and deferred use. In other words, simply moving hydrogen from one location to another over several hours does not provide the same type of temporal flexibility as storing hydrogen in tanks, pipelines, or salt caverns for later use. Second, although HTT may provide limited and scenario-dependent temporal support in practice (e.g., the HTT itself can serve as a form of mobile hydrogen inventory and careful dispatch may allow some degree of deferred delivery under emergency conditions), this effect is indirect, logistics-dependent, and generally weaker than the intrinsic time-shifting capability of dedicated storage technologies. It should also be noted that additional resilience benefits can be achieved through the coordinated use of different hydrogen-related technologies. For example, combining proton exchange membrane fuel cells with on-site hydrogen tanks can enable fast, clean, and autonomous black start for critical infrastructure such as hospitals. Similarly, integrating electrolyzers, salt-cavern storage, pipelines, and hydrogen turbines can support large-scale grid restoration. Therefore, careful selection and coordination of hydrogen-related technologies according to specific application scenarios are essential for the design of hydrogen-enabled MESs.

\begin{table*}[t]
\centering
\setlength{\arrayrulewidth}{1pt}
\caption{Resilience Potential of Hydrogen-Related Technologies in MESs (H=High, M=Medium, $-$=Not obvious)}
\label{table_3}
\rowcolors{1}{yellow!12}{yellow!12} 
\begin{tabular}{|>{\color{black}}m{2.5cm}<{\centering}|>{\color{black}}m{2.3cm}<{\centering}|>{\color{black}}m{1.3cm}<{\centering}|>{\color{black}}m{1.3cm}<{\centering}|>{\color{black}}m{1.3cm}<{\centering}|>{\color{black}}m{1.3cm}<{\centering}|>{\color{black}}m{5.0cm}<{\centering}|}
\hline
\rowcolor{yellow!25} 
\textbf{Category} & \textbf{Technology} & \textbf{Cross-temporal flexibility} & \textbf{Cross-spatial flexibility} & \textbf{Cross-sector flexibility} & \textbf{Black-start capability} & \textbf{Resilience Potential Explanations} \\
\hline
\centering Hydrogen production & Electrolyzer & - & - & H & - & Converts surplus power to hydrogen; siting is flexible but not mobile; not a generator \\
\hline
\centering Hydrogen production & Steam methane reforming & - & - & H & - & Centralized, utility-dependent hydrogen production; strong industrial fuel support, but weak emergency agility \\
\hline
\centering Hydrogen storage & Salt-cavern hydrogen storage & H & - & - & - & Seasonal, very-large storage; immobile; withdrawal needs powered compressors \\ \hline
\centering Hydrogen storage & Hydrogen tank & M & - & - & - & Hours--days buffer onsite; fixed asset; enables generators but does not generate \\
\hline
\centering Hydrogen transport & Hydrogen pipelines & M & H & - & - & Pipeline linepack; cross-spatial transport; not a generator \\
\hline
\centering Hydrogen transport & HTT & - & H & - & - & Mobile hydrogen supply to critical loads; not a generator \\
\hline
\centering Hydrogen utilization & Proton exchange membrane fuel cell & - & - & H & H & Combined heat and power
generation; Fast black-start for critical loads \\
\hline
\centering Hydrogen utilization & Solid oxide fuel cell & - & - & H & - & Combined heat and power generation; long warm-up; poor for rapid/black-start use \\
\hline
\centering Hydrogen utilization & Hydrogen turbines & - & - & H & H & Combined heat and power
generation; configured units can black-start and support restoration \\
\hline
\end{tabular}
\end{table*}

\subsection{Vulnerabilities of Hydrogen-related Technologies}
Although resilience potential can be offered by hydrogen-related technologies in the chain of production, storage, transportation, and utilization, security vulnerabilities also exist in each stage, which can create new disturbance propagation paths and may weaken rather than strengthen system resilience under adverse conditions. To be specific, in the production stage, especially in electrolysis and steam methane reforming, the risk of hydrogen leakage and explosion is significant due to high operating pressures and temperatures. In the storage stage, compressed hydrogen tanks face threats such as high-pressure rupture, while cryogenic liquid hydrogen presents risks of rapid venting. Long-term storage methods like salt caverns must address potential leakage and environmental contamination. In the transportation stage, hydrogen's small molecular size increases the likelihood of leaks in pipelines or tube trailers, and mobile transport systems are vulnerable to traffic accidents or mechanical damage. Additionally, improper handling during transfer can lead to static ignition or pressure imbalances. In the utilization stage, inadequate ventilation may cause hydrogen accumulation and fire hazards. Moreover, failures in electrolyzers, hydrogen storage tanks, hydrogen pipelines, or fuel cell systems may compromise local energy supply and even lead to load shedding. In addition, the growing digitalization of hydrogen infrastructures introduces cybersecurity threats\cite{RiziD2025}, \cite{HuaD2025}, such as false data injection attacks on tank levels and demand information\cite{HuaD2025}. Therefore, hydrogen should be regarded as a technology with both resilience benefits and resilience risks, and its positive contribution to MES resilience depends on proper design, operation, protection, and cybersecurity measures.

\begin{table*}[htbp]
\centering
\setlength{\arrayrulewidth}{1pt}
\footnotesize
\caption{Optimization Challenges Introduced by Adopting Hydrogen in MES for Resilience Enhancement}\label{table_4}
\begin{tabular}{|>{\color{black}}m{0.8cm}<{\centering}|>{\color{black}}m{1.6cm}<{\centering}|>{\color{black}}m{4.3cm}<{\centering}|>{\color{black}}m{5.0cm}<{\centering}|>{\color{black}}m{4.0cm}<{\centering}|}
\hline
\rowcolor{yellow!25} 
\textbf{Index} & \textbf{Challenges} & \textbf{Hydrogen-related Example} & \textbf{Formulas / Model Implications} & \textbf{Formulations and Optimization Methods/Techniques} \\
\hline
\rowcolor{yellow!12} 
1 & Uncertainties & Hydrogen demand varies with mobility or industrial loads & $\min_{x \in X} c^\top x + \mathbb{E}_{\xi}[Q(x,\xi)]$ & Stochastic programming, robust optimization, scenario-based MILP \\
\hline
\rowcolor{yellow!12} 
2 & Mixed integer variables & Investment locations of electrolyzers and hydrogen tanks & $y_i^{\text{electrolyzer}} \in \{0,1\}, \quad z_j^{\text{tank}} \in \{0,1\}$ & MILP, MINLP, branch-and-bound, Lagrangian relaxation \\
\hline
\rowcolor{yellow!12} 
3 & Multi-timescale variables & Slow hydrogen storage level update and fast battery dispatch  & $H_t^{\text{stored}}$ (in hours), $B_{\tau}^{\text{stored}}$ (in 5 minutes) & MILP, Nested model predictive control, hierarchical deep reinforcement learning  \\
\hline
\rowcolor{yellow!12} 
4 & Temporal coupling constraints& Hydrogen produced at off-peak hours and used later during peak demand, slow dynamics of hydrogen pipeline pressure & $H_t^{\text{stored}} = H_{t-1}^{\text{stored}} + H_t^{\text{prod}} - H_t^{\text{used}}$ & MILP, model predictive control, dynamic programming, deep reinforcement learning \\
\hline
\rowcolor{yellow!12} 
5 & Spatial coupling constraints& Hydrogen produced in region A, stored in region B, and used in region C via pipelines & $\sum_{j \in \mathcal{N}, j \ne i} h_{ji,t} - \sum_{j \in \mathcal{N}, j \ne i} h_{ij,t} = h^{\text{net}}_{i,t},~\forall i \in \mathcal{N}$ & MILP, Benders decomposition, distributed optimization \\
\hline
\rowcolor{yellow!12} 
6 & Nonlinear \& nonconvex constraints & Load-dependent electrolyzer/fuel cell efficiencies, part-load operating regimes, nonlinear hydrogen pipeline hydraulics, compression/liquefaction energy consumption, and nonlinear hydrogen storage state relations & $\eta_t^{\mathrm{EL}} = f(P_t^{\mathrm{EL}})$, $\eta_t^{\mathrm{FC}}$ = $g(P_t^{\mathrm{FC}})$, $H_{t}^{\mathrm{prod}}$ =$\frac{\eta_{t}^{\mathrm{EL}} P_{t}^{\mathrm{EL}} \Delta t}{\omega^{\mathrm{L}}}$, $P_{t}^{\mathrm{FC,out}}$ =$\frac{\eta_{t}^{\mathrm{FC}} H_{t}^{\mathrm{FC,in}} \omega^{\mathrm{L}}}{\Delta t}$, $h_{ij,t}$=$k_{i,j}(p_{i,t},~p_{j,t})$, $E_t^{\mathrm{c}}$=$\varphi(q_t^{\mathrm{H_2}},p_t^{\mathrm{in}},p_t^{\mathrm{out}})$,~$p_t^{\mathrm{tk}}$=$\phi(m_t^{\mathrm{H_2}},T_t,V_t)$ & MINLP, NLP, piecewise linear approximation, convex/cone relaxation, sequential linearization, McCormick envelopes, metaheuristics, and deep reinforcement learning \\
\hline
\rowcolor{yellow!12} 
7 & Multi-energy coupling constraints & Hydrogen enables electricity-gas-heat-transport sector integration & $
P_{t}^{\mathrm{FC,e}} =\frac{\eta_{t}^{\text{FC,e}} H_{t}^{\mathrm{FC,in}} \omega^{\mathrm{L}}}{\Delta t}$,\;
$Q_{t}^{\mathrm{FC,h}} =\frac{\eta_{t}^{\text{FC,h}} H_{t}^{\mathrm{FC,in}} \omega^{\mathrm{L}}}{\Delta t}$ & Integrated MILP/MINLP, Benders decomposition  \\
\hline
\rowcolor{yellow!12} 
8 & Multi-objective functions & Tradeoff between cost minimization and resilience/load recovery maximization  & $\min~\text{Cost}$; $\max~\text{Load~Recovery}$ & MILP/MINLP, Multi-objective optimization/reinforcement learning \\
\hline
\end{tabular}
\end{table*}

\subsection{Optimization Challenges Related to Hydrogen-enabled Planning and Operation Enhancement}
Integrating hydrogen into the resilience-oriented planning and operation of MESs introduces a range of complex optimization challenges. Firstly, uncertainty in hydrogen demand complicates the optimization process, often requiring stochastic or robust optimization frameworks. Secondly, many hydrogen-related decisions, including infrastructure investments and on/off operational states of electrolyzers or compressors, involve binary or integer variables, resulting in mixed-integer programming problems. Thirdly, there are multi-timescale decision variables. Due to the physical limits on gas flow, thermal dynamics, and control systems, hourly updates of hydrogen storage level are sufficient for modeling storage dynamics. In contrast, batteries could be dispatched every 5 minutes to capture the fluctuations of renewable generation and load demand. When conducting operation optimization on an hourly basis, multi-timescale decision variables would arise. Fourthly, hydrogen enables cross-temporal operations (e.g., producing and storing energy in low-demand periods for later use), which creates strong inter-temporal coupling and increases the dimensionality of the optimization problem. Fifthly, hydrogen transport via pipelines or trailers introduces spatial coupling, requiring network-based flow models with additional constraints. Sixthly, there are many nonlinear or even nonconvex constraints involved in the hydrogen production and the associated compression, storage, transport, and utilization processes, which complicate solution tractability. Taking electrolyzers, fuel cells, hydrogen pipelines, hydrogen compressors/liquefiers, and hydrogen storage tanks into consideration, there are nonlinear or nonconvex constraints incurred by load-dependent electrolyzer/fuel cell efficiencies, part-load operating regimes of electrolyzers/fuel cells, nonlinear hydrogen pipeline hydraulics, energy consumption of hydrogen compressors/liquefiers, and nonlinear hydrogen storage state relations. Seventhly, the integration of hydrogen also deepens the coupling between electricity, gas, heat, and mobility sectors, leading to multi-energy coupling constraints. Finally, there are multi-objective functions in HMESs, e.g., system cost minimization and operation risk minimization, operation cost minimization and load recovery maximization. It is worth noting that the mathematical models of hydrogen equipment do not necessarily introduce features that are entirely absent from all other MES components, since nonlinearities, mixed-integer decisions, and temporal couplings may also arise in other devices such as CCHP units, batteries, and gas networks. However, compared with conventional MES equipment, hydrogen-related devices tend to exhibit these characteristics in a more pronounced and integrated manner, especially when hydrogen production, compression, storage, transport, and utilization are modeled simultaneously. In order to address the above challenges, more advanced solution methods should be adopted, such as mixed-integer nonlinear programming, decomposition techniques, scenario-based planning, robust optimization\cite{ZhouY2024}, metaheuristics\cite{HuaD2025}, convex relaxation, model predictive control\cite{ChenTSG2025}, and deep reinforcement learning\cite{Yu2023}.

For the ease of understanding, we summarize the optimization challenges introduced by adopting hydrogen in MESs for resilience enhancement in Table~\ref{table_4}. In line 1, $x$ denotes the first-stage decision variables (e.g., hydrogen storage or electrolyzer sizing);
$\xi$ represents the uncertainty parameters (e.g., hydrogen demand, renewable availability); $X$ is the uncertainty set; $c^\top x$ is the deterministic first-stage investment cost; $Q(x, \xi)$ is the recourse cost in the second stage given scenario $\xi$. In line 2, $y_i^{\text{electrolyzer}}$ is a binary variable indicating whether an electrolyzer is installed at location $i$, $z_j^{\text{tank}}$ is a binary variable representing whether a hydrogen storage tank is constructed at node $j$, $y, z = 1$ implies the component is built/active, while $y, z = 0$ implies it is not. In line 3, $H_t^{\text{stored}}$ and $B_{\tau}^{\text{stored}}$ denote the energy level in hydrogen tanks and batteries at time $t$ and $\tau$, respectively. In line 4, $H_t^{\text{prod}}$ and $H_t^{\text{used}}$ denote the produced hydrogen and the consumed hydrogen at time $t$, respectively. In line 5, $\mathcal{N}$ is the set of all nodes (e.g., cities, regions, or hydrogen hubs) in the hydrogen network, $\mathcal{T}$ is the set of time periods in the optimization horizon, $h_{ij,t}$ is the amount of hydrogen flow from node $i$ to node $j$ at time $t$, $h_{ji,t}$ is the amount of hydrogen flow from node $j$ to node $i$ at time $t$, $h^{\text{net}}_{i,t}$ is the net hydrogen injection at node $i$ at time $t$, which is positive if hydrogen is injected (e.g., from an electrolyzer) and negative if it is consumed (e.g., by a fuel cell or local demand). In line 6, $\eta_t^{\mathrm{EL}}$ and $\eta_t^{\mathrm{FC}}$ denote the electrolyzer and fuel cell efficiencies at time $t$, respectively; $f$, $g$, $k_{i,j}$, $\varphi$ and $\phi$ are nonlinear or even nonconvex functions. $P_t^{\mathrm{EL}}$ and $P_t^{\mathrm{FC}}$ denote the electrolyzer input power and fuel cell operating power, respectively; $\Delta t$ is the time-step duration; and $\omega^{\mathrm{L}}$ is the lower heating value of hydrogen. $P_t^{\mathrm{FC,out}}$ denotes the fuel cell electric output power, and $H_t^{\mathrm{FC,in}}$ is the hydrogen input to the fuel cell. $p_{i,t}$ and $p_{j,t}$ denote the nodal hydrogen pressures. $E_t^{\mathrm{c}}$ denotes the compressor energy consumption, $q_t^{\mathrm{H_2}}$ is the hydrogen flow through the compressor, and $p_t^{\mathrm{in}}$ and $p_t^{\mathrm{out}}$ denote the inlet and outlet pressures, respectively. $p_t^{\mathrm{tk}}$ is the hydrogen tank pressure, while $m_t^{\mathrm{H_2}}$, $T_t$, and $V_t$ denote the stored hydrogen mass, storage temperature, and tank volume at slot $t$, respectively. In line 7, $\eta_t^{\mathrm{FC,e}}$ and $\eta_t^{\mathrm{FC,h}}$ are the electrical and thermal conversion efficiencies of the fuel cell, respectively. $P_t^{\mathrm{FC,e}}$ and $Q_t^{\mathrm{FC,h}}$ denote the electrical and thermal outputs of the fuel cell, respectively.

\section{Hydrogen-enabled Resilience Enhancement Framework for MESs}\label{s2}

\begin{figure*}[!htb]
\centering
\includegraphics[scale=0.64]{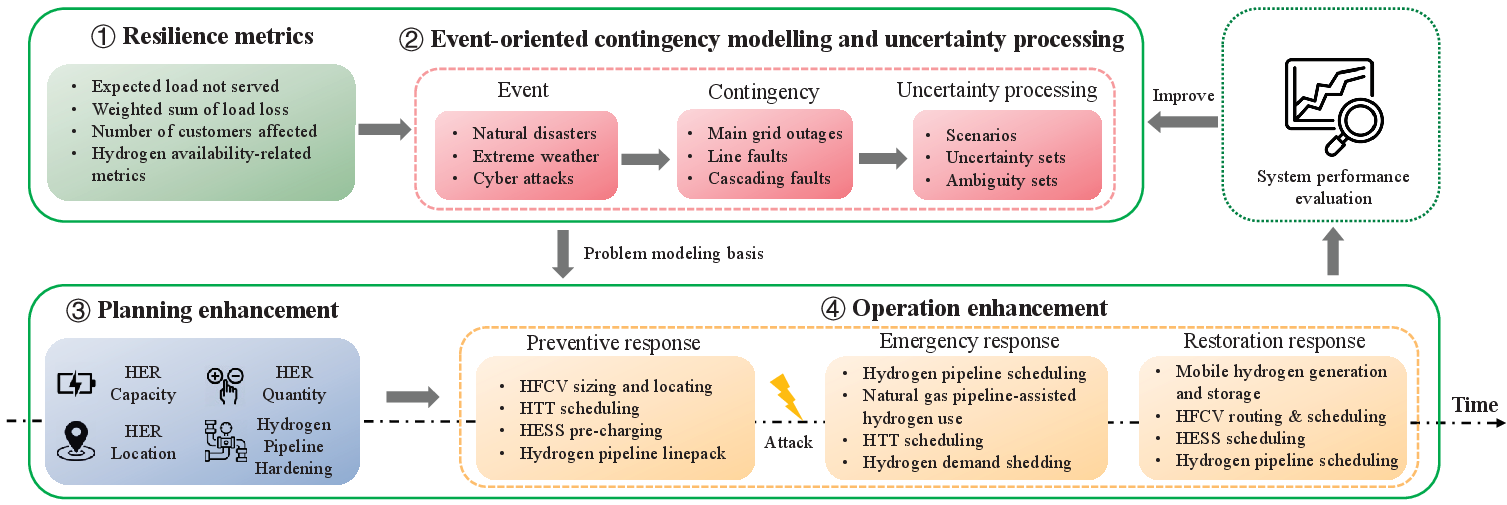}
\caption{The framework of hydrogen-enabled resilience enhancement for MESs}\label{fig_2}
\end{figure*}

As shown in Fig.~\ref{fig_2}, the framework of hydrogen-enabled resilience enhancement for MESs is provided, where four steps could be identified, i.e., resilience metrics, event-oriented contingency modeling and uncertainty processing, planning enhancement, and operation enhancement. To be specific, a resilience metric related to the investigated HMESs should be first defined, which would be incorporated into the objectives or constraints of planning or operation problem models. Then, according to the type of extreme events (e.g., hurricane, flood, ice disaster, earthquake, and cyber attacks), the corresponding event models could be developed, which consist of event-related physical attributes\cite{ZhangY2022}, e.g., movement speed and landfall point of typhoons, and the intensity of earthquakes. Since such attributes depend on some unknown factors, the occurrence of an extreme event is full of uncertainties in the planning stage, e.g., there are many possible landfall scenarios when typhoons are considered\cite{WenZ2024}. After obtaining the physical attributes of extreme events, the contingencies of MESs incurred by extreme events need to be modeled. Typically, stochastic multi-type contingencies are considered in existing works\cite{Dong2023}. In order to cope with the above-mentioned event uncertainties and contingency uncertainties, as well as parameter uncertainty (e.g., renewable generation and multi-type loads), uncertainty processing methods will be adopted, e.g., scenario-based methods, uncertainty-set-based methods, and ambiguity-set-based methods. Such methods can provide scenario parameter information, feasible decision space, and probability distribution sets for planning and operation optimization problems. More details about uncertainty processing methods and extreme-event scenario generation methods will be introduced in the Table~VIII and Table~X in Section~VI, respectively. In the planning stage, some permanent facilities will be deployed, e.g., SHSs, HESSs, hydrogen pipelines, and HRSs. During the operation stage, several kinds of responses can be made, e.g., preventive response, emergency response, and restoration response. During the preventive response, measures such as MHER location allocation, hydrogen procurement, and HESS pre-charging can be taken. During the emergency response, some measures are taken to minimize load shedding, including MHER scheduling, natural gas pipeline-assisted hydrogen use, HESS discharging, hydrogen demand adjustment, and hydrogen transformation. In the restoration stage, mobile hydrogen generation and storage, MHER routing and scheduling, HESS scheduling, hydrogen fuel cell scheduling, and hydrogen pipeline scheduling could be adopted to restore loads. By evaluating the practical system performance in the aspect of resilience, the rationality of the developed resilience metrics and event-oriented contingency modeling could be further improved.

%

\section{The Definitions and Metrics of HMES Resilience}\label{s3}

Before enhancing the resilience of an HMES, its definitions and metrics should be established, since resilience metrics will be incorporated into the objectives or constraints of planning/operation problem models. In the following parts, HMES definitions and architectures are introduced. Then, the definitions of HMES resilience are described. Next, typical resilience metrics in existing works are classified. Finally, a summary of the performance metrics is provided.

\subsection{HMES Definitions and Architectures}\label{s31}
\begin{figure}[!htb]
\centering
\includegraphics[scale=0.48]{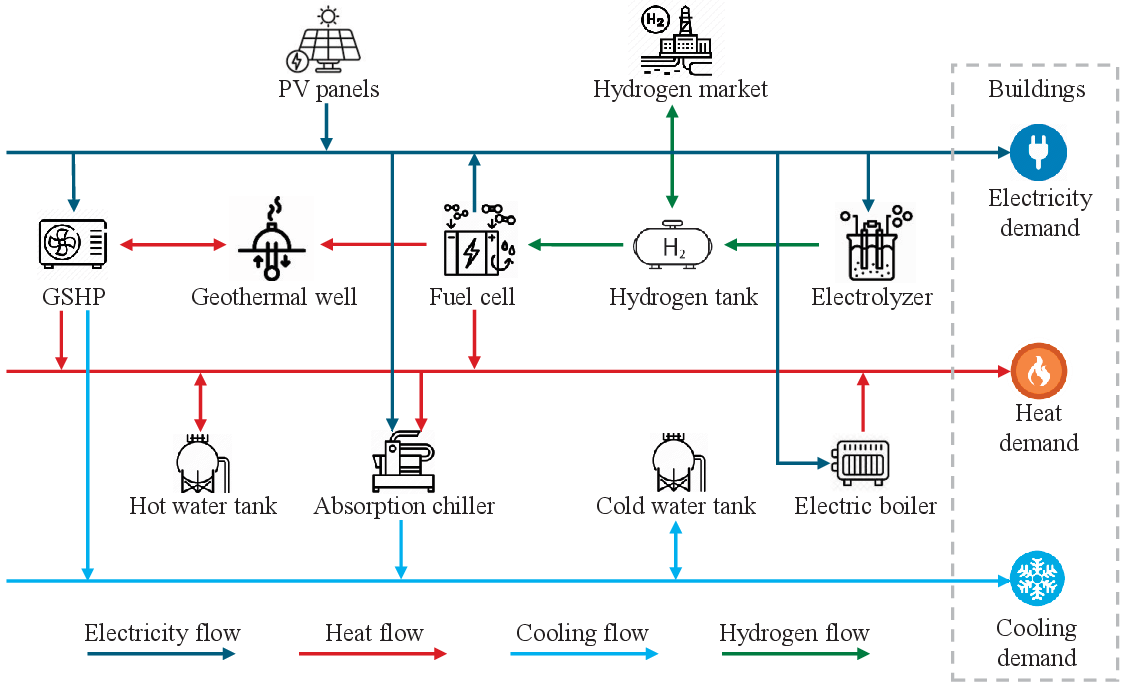}
\caption{Typical HMES architecture (building-level)}\label{fig_3}
\end{figure}

\begin{figure}[!htb]
\centering
\includegraphics[scale=0.46]{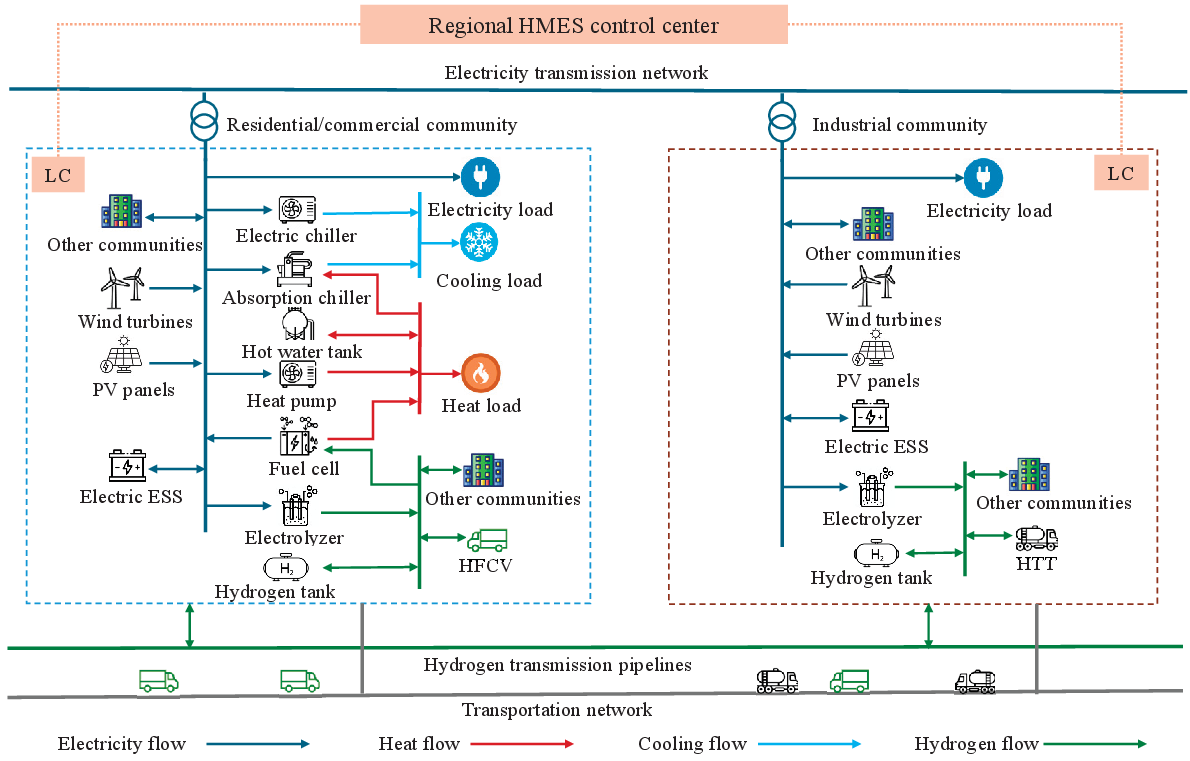}
\caption{Typical HMES architecture (region-level)\textcolor[rgb]{0.0,0.0,0.00}{\cite{FanG2024}}}
\label{fig_4}
\end{figure}

Similar to the MES concept\cite{Mancarella2014}, an HMES is a system in which hydrogen, electricity, heat, cooling, gas, and transport interact optimally with each other at various levels to increase technical, economic, and environmental performance relative to ``classical" energy systems whose sectors are treated separately. According to the spatial perspectives, HMESs can be divided into building-level, district-level, region-level, and nation-level. For simplicity, two typical architectures of HMESs are introduced, i.e., building-level HMESs and region-level HMESs. As shown in Fig.~\ref{fig_3}, the framework of a zero-carbon building-level HMES can be observed, and the corresponding practical system picture can be found in \cite{WangX2024}. In Fig.~\ref{fig_3}, there are four kinds of energy flows, i.e., electricity flow, heat flow, cooling flow, and hydrogen flow. As far as electricity flow is concerned, the electricity generated from photovoltaic (PV) panels and the fuel cell is used to satisfy the electricity demand of an electric boiler, a ground-source heat pump (GSHP), an absorption chiller, an electrolyzer, and electric demand in a commercial building. As to heat flow, the heat demand is satisfied by the electric boiler, GSHP, fuel cell, and hot water tank. Similarly, cooling demand is satisfied by the GSHP, absorption chiller, and cold water tank in the cooling flow. As to hydrogen flow, PV generation can be used for the electrolyzer to produce hydrogen, which will be stored in the hydrogen tank for future use by driving fuel cells. When the hydrogen stored in the hydrogen tank is not enough, hydrogen purchasing from the market is required. In Fig.~\ref{fig_4}, the architecture of a region-level HMES can be observed\cite{FanG2024}, which consists of three types of communities, e.g., residential communities, commercial communities, and industrial communities. In residential and commercial communities, electricity/cold/heat/hydrogen demands can be identified. In industrial communities, electricity/hydrogen demands can be observed. All communities are connected through electricity distribution networks, hydrogen transmission pipelines, and road transportation networks. With the help of transportation networks, MHERs (e.g., HTTs and HFCVs) can be redistributed among multiple communities for improving network resilience, e.g., providing emergent power supply or hydrogen supply to a community when it suffers from an extreme event (e.g., power outages or hydrogen pipeline leakage).

\subsection{Definitions of HMES Resilience}\label{s31}

\begin{figure}[!htb]
\centering
\includegraphics[scale=0.42]{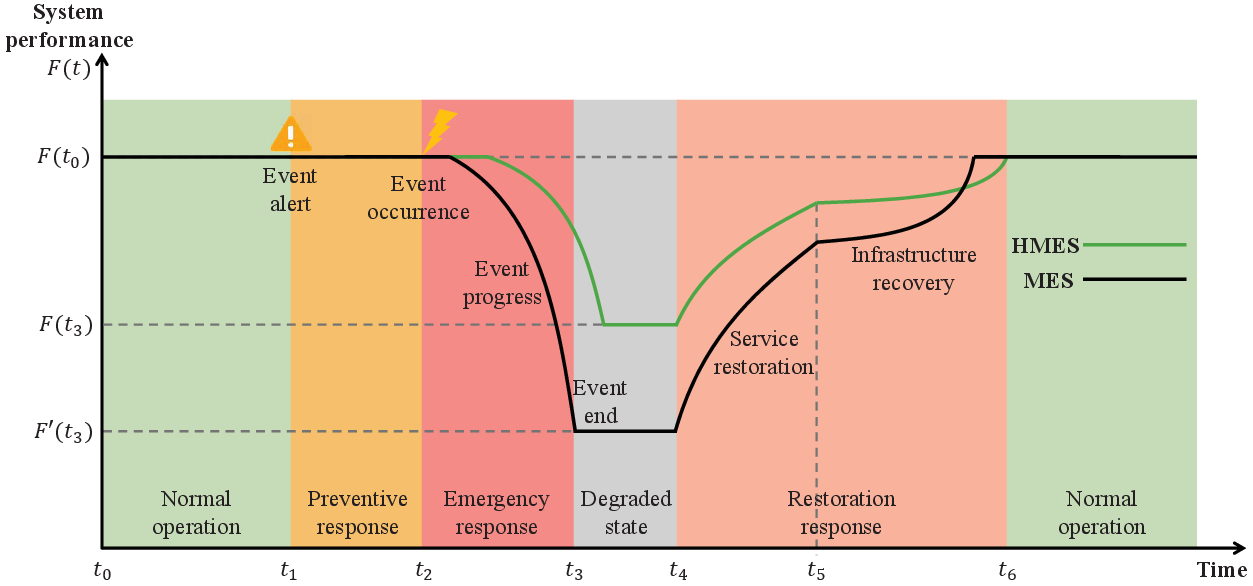}
\caption{System performance curves related to MES and HMES operation}\label{fig_5}
\end{figure}

Before describing the definitions of HMES resilience, we first provide two system performance curves related to MES and HMES under extreme events in Fig.~\ref{fig_5}, respectively. Although the typical curve under MES cannot perfectly capture the realistic system performance trajectories (e.g., local performance fluctuations) as illustrated in \cite{Carrington2021,Dobson2024,Ahmad2025}, it can reflect the system's ability to prepare for, absorb, and recover from disturbances and has been widely used in existing works\cite{Haggi2022},\cite{Lin2024,Cai2023,Jordehi2024,Afsari2024,Xie2024,MehrjerdiH2022}. In addition, while it is not tailored to hydrogen-enabled MESs, the performance trajectory can be applied to hydrogen-enabled MESs after minor modifications. By comparing the differences between two curves, the physical and functional features of hydrogen can be observed. Firstly, we introduce the whole process of resilience enhancement under MES. In existing works, load loss is a widely used metric of system performance. Under this premise, the curve under MES consists of normal operation stages and three kinds of responses (including pre-event preventive response, during-event emergency response, and post-event restoration response). In the normal operation stage, the optimal economic operation is typically conducted from time $t_0$ to time $t_1$. After detecting the extreme event information, the preventive response will be carried out from time $t_1$ to time $t_2$, e.g., purchasing mobile electric vehicle resources. During the extreme event, the emergency response is adopted to minimize the load shedding from time $t_2$ to time $t_3$ so that the extent of system performance degradation could be alleviated (note that the system performance can be maintained for a short time due to the deployed mobile resources in the preventive response). Then, the degraded state could be observed from time $t_3$ to time $t_4$ at the end of the event. From time $t_4$ to time $t_5$, the restoration response is achieved to restore critical loads \cite{Afsari2024} like hospitals and water pumps by adopting temporary measures, such as grid-island operation using remaining storage and mobile energy storage units (e.g., electric vehicles). These temporary measures restore partial or prioritized functionality, even if the underlying infrastructure (e.g., transmission lines and distribution equipment) is not yet fully repaired. After stabilizing essential services, the system proceeds to recover full infrastructure, e.g., repairing lines and equipment.

Similarly, after integrating hydrogen into MESs, the system performance curve should reflect both the physical and functional characteristics of hydrogen. In the preventive stage, additional mobile energy resources (e.g., HTTs for gas delivery and liquid-hydrogen tankers, as well as HFCVs/HFCBs used as mobile generators) can be deployed, so that system performance can be sustained longer in the emergency period. During the emergency response, if a grid outage coincides with hydrogen-pipeline leakage, part of critical demand can be supplied by fuel cells. Owing to pipeline linepack and pressure regulation, fuel-cell stack inlet pressure/flow conditions can be temporarily maintained, allowing the fuel cell to sustain its set power and thereby reduce load shedding as long as the available upstream pressure/flow remains above the requirements of hydrogen regulators and fuel cell stacks. If the leakage reduces the available supply below those requirements, hydrogen starvation occurs\cite{JiaF2017}, leading to a rapid voltage drop and possible protective shutdown, which increases load shedding. Although linepack buffering may delay this adverse impact, fuel-cell power output may continue to decline even after the initiating event ends once the buffer is depleted. Since more resources are available for deployment, an HMES can exhibit a higher performance level in the degraded state, and an example of this phenomenon can be seen in Fig. 7 of \cite{ChangS2025}. Utilizing mobile resources or conducting pipeline reconfiguration, the recovery trajectory of HMES can also be raised. Compared with MESs, hydrogen-related equipment failures in HMESs may involve longer restoration time, as illustrated in Fig.~5. This extended recovery is influenced by stochastic transportation network states and hydrogen flow dynamics as supported by Fig. 6 of \cite{Wangz2021}.

Based on the above description, HMES resilience can be defined as the capacity of an HMES to withstand the disturbance from an extreme event (i.e., event progress during $t_2$ and $t_3$), and to rebuild and renew it afterwards\cite{Afgan2012}. Similarly, HMES resilience can also be described as the capacity to limit the system performance degradation and the duration of the degraded state in order to maintain critical services following an extreme event\cite{Stankovi2023}. Although many definitions related to HMES resilience are proposed, they are originally borrowed from power grids and MESs, which can not truly reflect the physical and functional features of hydrogen\cite{LiB2023}\cite{ZhaoH2022}\cite{Lin2024,Cai2023,Jordehi2024,Afsari2024,Xie2024}. For example, when an extreme event (e.g., ice disaster) causes simultaneous power outages and hydrogen pipeline leakage, even if no load loss is incurred due to backup sources, hydrogen fuel cells may not operate normally due to the loss of pressurized hydrogen supply. At this time, no resilience enhancement measure is required since there is no system performance degradation (i.e., load loss). In fact, hydrogen-related systems need immediate repair so that further pressure reduction can be avoided. Otherwise, once backup sources are unavailable, the whole HMES may collapse and result in larger negative social and economic impacts. Therefore, the definition of HMES resilience should take any negative impacts (e.g., hydrogen availability and hydrogen-demand loss) of extreme events on hydrogen energy resources into consideration.

\subsection{Metrics of HMES Resilience}\label{s32}
In existing works, resilience metrics related to HMESs can be divided into three types, i.e., time indexes, load loss indexes, and load rate indexes\cite{Yang2022}. To be specific, time indexes are used to describe the duration of each operation stage, while load loss indexes are adopted to describe the extent of load shedding in different stages. Smaller time and load-loss indices are preferred. Load rate indexes are used to evaluate the rates of load loss or load recovery. Thus, a smaller load loss rate and a larger load recovery rate are expected. In existing works related to HMES resilience, the most widely used resilience metrics for HMES operation are summarized in Table~\ref{table_5}, and detailed descriptions are given below.

\begin{table}[htbp]
\center
\setlength{\arrayrulewidth}{1pt}
\caption{Summary of widely-used resilience metrics for HMES operation}\label{table_5} \centering
\begin{tabular}{|m{0.8cm}<{\centering}|m{3.1cm}<{\centering}|m{3.2cm}<{\centering}|}
\hline
\rowcolor{yellow!25} 
\textbf{Metrics}&\textbf{Definitions}&\textbf{References} \\
\hline
\rowcolor{yellow!12} 
LSR  & Load served ratio   &     \cite{Lin2024}, \cite{Cao2023}, \cite{Haggi2022}  \\
\hline
\rowcolor{yellow!12} 
ELNS & Expected load not served   &  \cite{Liu2024}, \cite{Cai2023,Jordehi2024,ZhaoY2023}, \cite{Yuan2024}\\
\hline
\rowcolor{yellow!12} 
ERL  & Expected restored load & \cite{Afsari2024}, \cite{ZhuS2024,Su2024,ZouX2024}  \\
\hline
\rowcolor{yellow!12} 
WSLL & Weighted sum of load loss   &    \cite{WenZ2024}, \cite{LiB2023}, \cite{WangYuze2024,Cicek2024,Shahbazbegian2023} \\
\hline
\rowcolor{yellow!12} 
LL   & Load loss   &    \cite{ZhuR2024}, \cite{Xie2024}, \cite{Sharifpour2023,Haggi2022,Liu2021,Tang2022,Yang2024,AminHashemifar2022,Huangchunjun2023} \\
\hline
\rowcolor{yellow!12} 
CRI  & Comprehensive resilience index & \cite{ZhaoH2022} \\
\hline
\end{tabular}
\end{table}

\subsubsection{\textbf{LSR}}
The load served ratio (\textbf{LSR}) is adopted to describe the operation resilience of hydrogen-powered microgrids\cite{Lin2024}, which is defined by
\begin{gather}\label{f_1}
\textbf{LSR}=\sum\nolimits_{i\in \mathbf{N}}\lambda_{i,t}P_{i,t}/\sum\nolimits_{i\in N}P_{i,t},
\end{gather}
where $\mathbf{N}$ denotes the set of buses, $\lambda_{i,t}$ is a binary variable to indicate whether the load at bus $i$ at slot $t$ is served or not. When the corresponding bus is served, the value of $\lambda_{i,t}$ is one. Otherwise, its value is zero.

\subsubsection{\textbf{ELNS}}
In \cite{Cai2023} and \cite{Jordehi2024}, Cai \emph{et al.} and Jordehi \emph{et al.} proposed the same resilience metric in the preventive stage for hydrogen fuel station-integrated power systems and industrial energy hubs with electric, thermal and hydrogen demands, respectively. The resilience metric is the expected load not served (\textbf{ELNS}), which can be described by
\begin{gather}\label{f_2}
\textbf{ELNS}=\mathbb{E}_s\{\sum\limits_{i}\sum\limits_{t}(w_eP_{i,t,s}^{\text{shed}}+w_hH_{i,t,s}^{\text{shed}}+w_{g}G_{i,t,s}^{\text{shed}})\},
\end{gather}
where $s$ denotes the index of uncertainty scenarios; $t$ denotes the time slot index; $i$ denotes the bus index; $w_e$, $w_h$, and $w_{g}$ denote the weight coefficients related to electric load shedding, hydrogen load shedding, and heat load shedding, respectively; $P_{i,t,s}^{\text{shed}}$, $H_{i,t,s}^{\text{shed}}$, and $G_{i,t,s}^{\text{shed}}$ denote the quantity of electric, hydrogen, and heat load shedding, respectively. Similarly, expected thermal load not served and expected electric load not served are considered for hydrogen-penetrated distribution systems in the restoration response stage\cite{ZhaoY2023}.

\subsubsection{\textbf{ERL}}
In the restoration response stage, expected restored load (\textbf{ERL}) maximization was considered for hydrogen-based microgrids in \cite{Afsari2024}, which is given by
\begin{gather}\label{f_3}
\textbf{ERL}=\sum\nolimits_{\theta}\varrho_{\theta}\Phi_{\theta},
\end{gather}
where $\theta$ denotes the uncertainty scenario index, $\varrho_{\theta}$ denotes the scenario probability, $\Phi_{\theta}$ denotes the served weighted load of heating, cooling, electricity, gas, and hydrogen demands\cite{ZhuS2024,Su2024,ZouX2024}.

\subsubsection{\textbf{WSLL}}
In \cite{LiB2023}, Li \emph{et al.} proposed a resilience metric in the emergency response stage for a hydrogen-penetrated multi-energy supply microgrid, which is the weighted sum of load loss (\textbf{WSLL}) related to electricity, heat, and gas, i.e.,
\begin{gather}\label{f_4}
\textbf{WSLL}=\sum\nolimits_{t}(\gamma L_{t}^{e}+\delta L_{t}^{h}+\epsilon L_{t}^{g}),
\end{gather}
where $\gamma$, $\delta$, and $\epsilon$ denote the weights related to electric load shedding, heat load shedding, and gas load shedding, respectively. $L_{t}^{e}$, $L_{t}^{h}$, and $L_{t}^{g}$ are the quantities of electric load shedding, heat load shedding, and gas load shedding, respectively. When just electric load shedding is involved, \textbf{WSLL} is reduced to load loss (\textbf{LL}), which was adopted in many works\cite{Xie2024}.

\subsubsection{\textbf{CRI}}
Note that the above indexes mainly focus on load loss, restored load, or served load ratio, they neglect the temporal performance that load can be restored. To overcome the challenge, Zhao \emph{et al.} proposed a comprehensive resilience index (\textbf{CRI}) to reflect the temporal performance and system overall performance\cite{ZhaoH2022}, i.e.,
\begin{gather}\label{f_5}
\textbf{CRI}=\beta R_0(t_5)+(1-\beta)R_1(t_5), \\
R_{\sigma}(t_5)=\frac{\int_{t_2}^{t_5}F(t_0)^{\sigma}[F(t)>F^{\min}]dt}{\int_{t_2}^{t_5}F(t_0)^{\sigma}dt},
\end{gather}
where $t_2$ and $t_5$ denote the event happening time and full recovery time as shown in Fig.~\ref{fig_3}, $F^{\min}$ denotes the minimum system performance requirement, and $[F(t)>F^{\min}]=1$ if $F(t)>F^{\min}$. Otherwise, $[F(t)>F^{\min}]=0$. $\sigma\in\{0,1\}$. Thus, $R_0(t_5)$ can reflect the time ratio that can satisfy the minimum system performance requirement, and $R_1(t_5)$ can reflect the overall service performance.

\text{Summary}: In existing works, many load-based metrics have been widely adopted, which often translate into economic damage or loss. However, they often overlook the social impacts of service disruption. In practice, the number and types of customers affected are also critical indicators of resilience, especially during extreme events~\cite{AlMuhaini2024},\cite{Kaloti2023}. For instance, shedding the same amount of load for a hospital and a residential building may lead to totally different consequences. The reason is that the former relies on power for life-saving operations, while the latter may experience only reduced comfort. Therefore, to develop a comprehensive resilience metric, it is necessary to combine load-based and customer-based metrics, as mentioned in \cite{Amiri2024}. In addition, existing metrics can not reflect the degraded state (e.g., HESS failure or hydrogen pipeline leakage) of HERs during an extreme event. Therefore, more comprehensive metrics are expected.

\section{Event-oriented Contingency Modeling in HMESs}\label{s4}

\begin{figure}[!htb]
\centering
\includegraphics[scale=0.4]{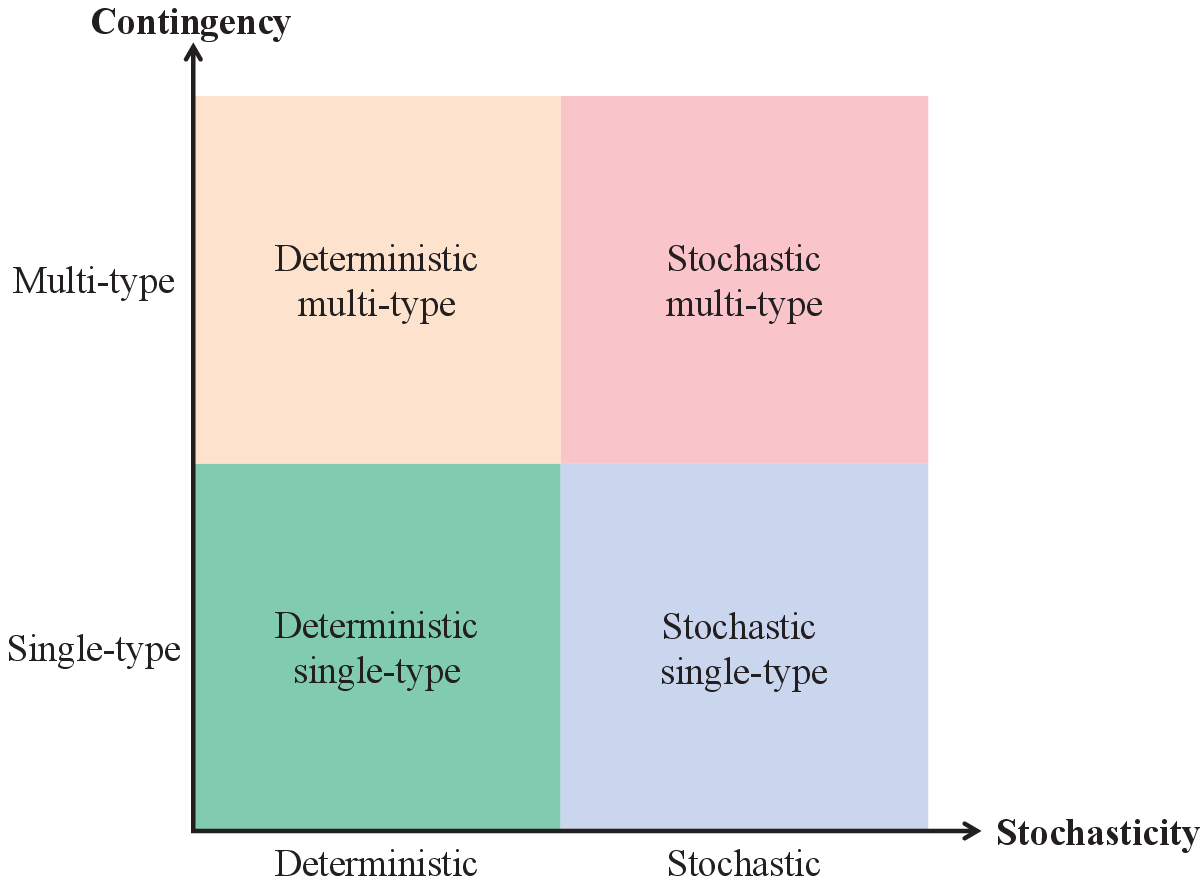}
\caption{The classification of event-oriented contingency modeling}\label{fig_6}
\end{figure}

Different high-impact and low-probability events would incur different types of contingencies. According to the stochasticity and contingency type, existing contingency modeling could be classified into four types as shown in Fig.~\ref{fig_6}, i.e., deterministic single-type contingencies\cite{ZhaoH2022,Liu2024,ZhaoY2024},\cite{Xie2024},\cite{Haggi2022},\cite{ZhuS2024},\cite{Su2024},\cite{Cicek2024},\cite{Shahbazbegian2023},\cite{Amini2024},\cite{Haggi2021}, deterministic multi-type contingencies\cite{Yan2025}, stochastic single-type contingencies\cite{WenZ2024},\cite{Yang2024},\cite{Dong2023},\cite{ChenF2024},\cite{Cai2023},\cite{Jordehi2024},\cite{Yuan2024},\cite{Cao2023},\cite{Tang2022},\cite{MaH2025,Gao2024,SunYong2023},\cite{Bennett2021}, and stochastic multi-type contingencies\cite{Shao2023},\cite{LiuKun2023},\cite{Gu2024},\cite{LiB2023}. Since deterministic contingencies can not represent the uncertain characteristics of high-impact and low-probability events, this paper mainly focuses on stochastic contingencies.

\subsection{Stochastic Single-type Contingency}\label{s41}

\subsubsection{Main grid disconnection model}
There are two types of main grid disconnection models, i.e., the $T-\tau$ criterion emergency islanding model and the stochastic grid outage model. The former is widely used for the microgrid islanding model, which means that a microgrid can operate in an islanding mode for continuous $\tau$ periods that could occur at any period of scheduling horizon $T$, which can be described by\cite{MaH2025}
\begin{gather}\label{f_6}
y_{h,t}=0,~\forall~t\in~[h:\min(T,h+\tau-1)],~\\
\sum\nolimits_{t\in [\gamma:T]}y_{h,t}=T-\gamma-\tau+1, h=\gamma,
\end{gather}
where $\gamma$ denotes the current scheduling slot, $y_{h,t}$ denotes the connection state between a microgrid and the main grid at slot $t$ under the emergency islanding scenario $h$. $h=r$ denotes the scenario $h$ with starting disconnection slot $\gamma$. When $y_{h,t}=1$, it means that the microgrid is connected to the main grid at slot $t$ under the scenario $h$. Otherwise, $y_{h,t}=0$. The stochastic grid outage model is used to generate main grid power outage scenarios with different starting times and ending times according to a given random distribution\cite{Yuan2024}. Compared with the $T-\tau$ criterion emergency islanding model, the stochastic grid outage model can generate power outage profiles with different durations, instead of a constant duration.

\subsubsection{Transmission line fault model}
The transmission line failure model is used to generate a large number of line failure scenarios, where each scenario consists of combinations of all random line states\cite{Gao2024}. To be specific, when the randomly generated values are smaller than the corresponding line failure probability, the considered line is broken. Otherwise, the considered line is still in a normal state. As shown in \cite{SunYong2023}, the failure probability of line $z$ can be described by
\begin{gather}\label{f_7}
P_{k}=1-(1-P_0)^{l_z},
\end{gather}
where $P_0$ denotes the failure probability of unit length transmission line and $l_z$ denotes the length of line $z$. Note that $P_0$ is the function of attributes of ice disasters, e.g., ice load and wind speed. In addition, the $N-k$ line contingency model is adopted in many existing works. To be specific, let $v_{i,j}$ be the connection state of line $(i,j)$, which is a binary variable. Moreover, $v_{i,j}=1$ if the line $(i,j)$ operates under normal state, and $v_{i,j}=0$ if it is damaged. Then, we have $\sum\nolimits_{(i,j)\in \mathcal{L}}(1-v_{i,j})\leq k$. Note that such a model is considered in many existing works\cite{Yang2024},\cite{Cao2023}.

When the starting period is considered, the $N-k$ contingency model can be reformulated as follows, i.e.,
\begin{gather}\label{f_8}
\sum\nolimits_{(i,j) \in \mathcal{L}} (1-v_{i,j})\leq k,~\sum\nolimits_{t\in \mathbf{T}} \varepsilon_t\leq 1,
\end{gather}
where $\varepsilon_t=1$ means that the contingency begins at period $t$, and $\mathbf{T}=\{1,2,\cdots,24\}$\cite{Dong2023}.

Since many extreme events have spatial features (e.g., typhoons), a location-aware line outage uncertainty model can be adopted, e.g., a line outage uncertainty set under a typhoon scenario $s$ can be defined by\cite{WenZ2024}
\begin{gather}\label{f_8}
\mathbb{U}_s=\Big\{\boldsymbol{\vartheta}_s|K_s\geq \sum\limits_{\{i,j\}\in \Omega_{\text{PLrisk}}}(\nu_{i,j}-\vartheta_{i,j})\Big\},~\forall~s \in \boldsymbol{S},
\end{gather}
where $\nu_{i,j}=1$ if $\{i,j\}\in \Omega_{\text{PLrisk}}$; $\vartheta_{i,j}=0$ indicating the line outage, while $\vartheta_{i,j}=1$ means that there is no line outage; $K_s$ denotes the allowable maximum number of line outages; $\boldsymbol{\vartheta}_s=(\vartheta_{i,j})|_{\{i,j\}\in \Omega_{\text{PLrisk}}}$.

\subsubsection{Hydrogen pipeline fault model}
According to \cite{LiuS2026}, the dynamic probability model related to hydrogen pipeline faults can be described by a lognormal cumulative distribution function $\psi(\cdot)$ as follows,
\begin{gather}\label{f_9}
\text{Pr}_{m,n,k,t}=\psi\Bigg(\frac{\ln(\zeta_{m,n,k}\sum\nolimits_{\tau\leq t}r_{m,n,k,\tau})}{\chi_{m,n,k}}\Bigg),
\end{gather}
where $r_{m,n,k,\tau}$ denotes the rainfalls suffered by the segment $k$ of hydrogen pipeline $mn$ at slot $\tau$, $\zeta_{m,n,k}$ and $\chi_{m,n,k}$ are two parameters estimated from empirical data. Then, the failure probability of the hydrogen pipeline $mn$ at slot $t$ is give by
\begin{gather}\label{f_10}
\text{Pr}_{m,n,t}=1-\Pi_{k=1}^{N_s}(1-\text{Pr}_{m,n,k,t}),
\end{gather}
where $N_{s}$ is the number of segments in hydrogen pipeline $mn$.

\subsubsection{Single-type device impact model}
Device fragility models related to poles and towers, insulators, transformers, and photovoltaic panels under extreme weather can be calculated according to the related physical parameters. More details could be found in \cite{ChenF2024},\cite{Tang2022},\cite{Bennett2021}. For example, photovoltaic panel icing has a large impact on the intensity of insolation, which can lead to a decrease in photovoltaic generation power. Let $P_{t}^{\text{fault}}$ be the maximum generation output when panels experience freezing rain disaster at slot $t$. Then, we have\cite{ChenF2024}
\begin{gather}\label{f_11}
P_{t}^{\text{fault}}=\Gamma\Big(G\exp\big(-\frac{1.5\rho_{\text{frz}}H_t^{\text{PV}}}{\rho_{\text{ic}}r_{\text{ef}}}\big)\Big),
\end{gather}
where $\Gamma(\cdot)$ denotes a function; $G$ is the insolation intensity on the surface of the ice cover; $\rho_{\text{frz}}$ and $\rho_{\text{ic}}$ are the intensities of freezing rain and ice, respectively; $r_{\text{ef}}$ denotes the effective grain radius; $H_t^{\text{PV}}$ denotes the ice thickness on the panel surface.

\subsubsection{Repairable equipment failure model}
Since some devices caused by extreme events (e.g., hurricanes) are repairable, a Markov two-state transition model was adopted to describe the ``operation-failure-operation" cycle process of wind turbine or photovoltaic equipment in \cite{ChenZhe2024}. To be specific, the transition from normal operation state to the failure state in a continuous-time Markov process happens according to a transition rate $\lambda_{o}$. Similarly, the transition from the failure state to normal operation state happens according to a transition rate $\lambda_{f}$. Let $T_{otf}$ and $T_{fto}$ denotes the average sojourn time of operation state and failure state, respectively. Then, $T_{otf}=1/\lambda_{o}$ and $T_{fto}=1/\lambda_{f}$, which also represent the mean time to failure and mean time to repair, respectively. Moreover, the above-mentioned sojourn time follows an exponential distribution.

\subsection{Stochastic Multi-type Contingency}\label{s42}
\subsubsection{Multi-device fault model}
Multi-device fault model was adopted in \cite{Gu2024} to capture the uncertain contingencies of electrolyzer units, fuel cells, and gas turbine units caused by disastrous events. Let $N_g$ be the minimum number of components under normal operation. Then, the multi-device fault model can be described by an uncertainty contingency set as follows,
\begin{gather}\label{f_12}
\Omega=\{\boldsymbol{u}|\sum\nolimits_{e}u_{\text{EL},e}+\sum\nolimits_{f}u_{\text{FC},f}+u_{\text{GT}}\geq N_g\},
\end{gather}
where $\boldsymbol{u}=(u_{\text{EL},e},~u_{\text{FC},e},~u_{\text{GT}})$. Moreover, $u_{\text{EL},e}$,~$u_{\text{FC},e}$,~$u_{\text{GT}}$ are binary variables related to electrolyzers, fuel cells, and a gas turbine, respectively. When there is a fault in a component, the corresponding value is zero. Otherwise, its value is one. Similar to the above multi-device fault model, $N-k$ contingency model was considered in \cite{Shao2023} for the stochastic failures of critical components (e.g., fuel cell stacks).

\subsubsection{Temporal-spatial destructive model}
In \cite{LiB2023}, a temporal-spatial destructive model was adopted to describe the contingencies of electricity/gas/heat networks under the impact of the hurricane, which is varying and highly related to spatial locations of components in electricity/gas/heat networks. Firstly, the grid division method is used to describe the damage of hurricanes at slot $t$. Specifically, a three-dimensional matrix $G_{\text{state}}^{r,c,t}$ is constructed and its element is equal to one if the rectangle area with location $(r,c)$ belongs to the hurricane area at slot $t$. Otherwise, its element is equal to zero. Secondly, a two-dimensional matrix $G_{\text{loc}}^{r,c}$ is created to describe which components belong to the rectangle area $(r,c)$. Based on the above matrixes, the damaged level could be obtained. Continually, a node export ability in the network can be calculated by the percentage of survival paths from all energy sources to the current node.

\emph{Summary:} Compared with stochastic single-type contingencies, stochastic multi-type contingency modeling has a stronger representation ability. Even so, the multi-device fault model still cannot capture the dynamics of contingencies related to extreme events, since the sequences and starting times of fault appearances are not considered. In fact, extreme events (e.g., hurricanes and floods) have temporal and spatial attributes, which may lead to multi-type contingencies with different sequences and starting times. In addition, some cascading contingencies incurred by cyber-physical attacks are not considered in existing works\cite{Artime2024}. Overall, existing works rarely consider hydrogen-specific vulnerabilities. In fact, there are many physical vulnerabilities and cyber vulnerabilities in HMESs. In physical aspects, hydrogen tends to cause material embrittlement, particularly in metals such as steel, which can lead to long-term degradation of pipelines, storage tanks, and connectors. These structural vulnerabilities increase the likelihood of mechanical failure under normal operations or stress conditions such as temperature fluctuations and pressure surges during extreme events. In cyber aspects, a hacker manipulating the pressure control in a hydrogen storage tank could trigger an overpressure event, potentially causing mechanical damage or explosion. Similarly, false data injection into hydrogen demand forecasts or electrolyzer schedules could destabilize coordinated operations across the electricity-hydrogen interface\cite{HuaD2025}, leading to misallocation of resources or even blackouts. However, most contingency models treat hydrogen assets as idealized components with fixed performance characteristics in existing works, ignoring the risks posed by leakage, material fatigue, and cyber intrusion. Therefore, future research should develop risk-aware, vulnerability-inclusive contingency modeling that captures both the physical fragilities and cyber risks related to hydrogen.

\section{Hydrogen-enabled Planning Enhancement for MESs}\label{s5}

In the stage of planning, many hydrogen resources will be deployed permanently to enhance the HMES resilience, such as SHSs, HESSs, hydrogen pipelines, HRSs, fuel-cell-driven CCHP, and hydrogen-related production facilities. According to the types of investment-related hydrogen resources, existing works can be generally classified into two types, i.e., single-type hydrogen facilities and multi-type hydrogen facilities. In the following parts, related works will be introduced in detail.

\subsection{Single-type Hydrogen Facilities}\label{s51}
\begin{figure}[!htb]
\centering
\includegraphics[scale=0.42]{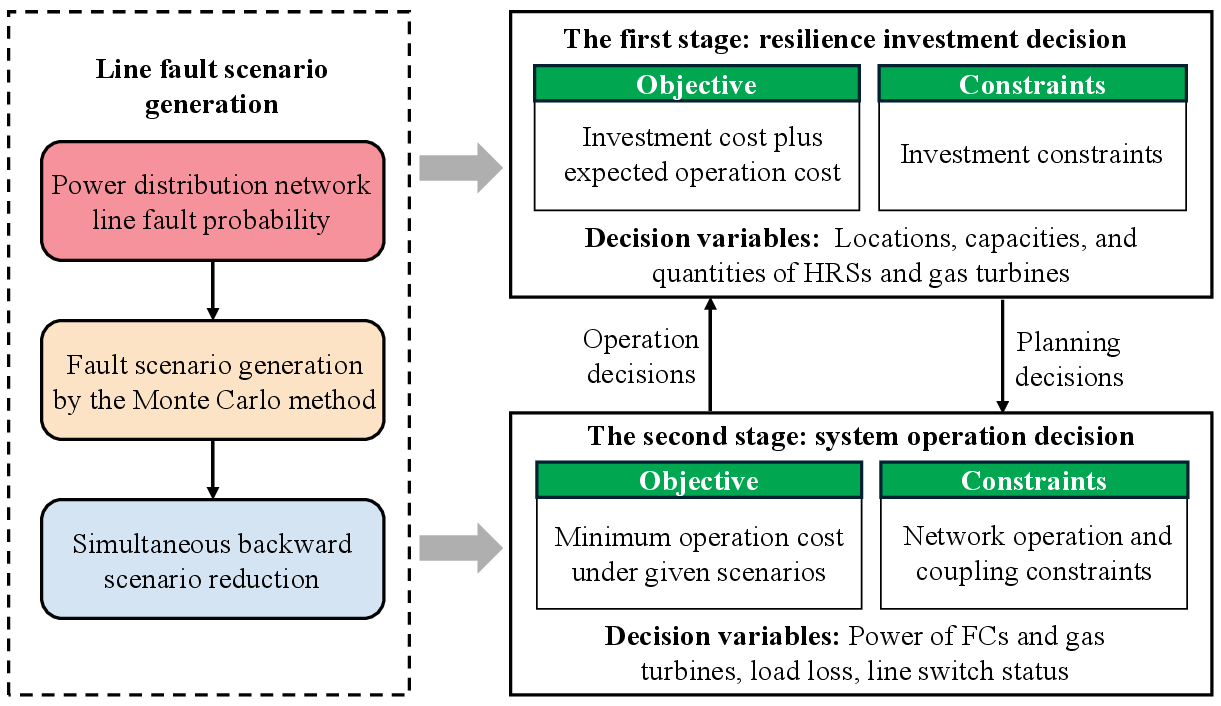}
\caption{The framework of two-stage resilience planning method\cite{Gao2024}}
\label{fig_7}
\end{figure}

Many resilience-oriented planning methods have been developed to decide the capacity, location, and quantity of single-type hydrogen facilities (e.g., HESSs, HRSs, and H2Ps). For example, Yan \emph{et al.}\cite{Yan2025} proposed a resilient planning method for power systems to decide electric ESSs and HESS capacities against heatwave events. Firstly, the impact model of heatwave events on the power demand and photovoltaic generation was developed. Then, a total cost minimization problem was formulated with the consideration of the daily investment cost, operation/maintenance cost, fuel cost, power curtailment penalty cost, and so on. Finally, CPLEX was adopted to solve the formulated problem. The proposed planning method can combine the advantages of electric ESSs and HESSs, i.e., HESSs are adopted for long-term storage before heatwave events happen, whereas electric ESSs are responsible for short-term intra-day power balance. During the heatwave events, HESS adopted a discharging strategy to relieve the negative impact of heatwave events. Similarly, Zhao \emph{et al.}\cite{ZhaoH2022} investigated a resilient planning problem to decide the photovoltaic panels, HESS, and electric ESS capacities for a hydrogen-integrated airport energy system considering main grid power outages incurred by extreme events. To achieve the above aim, a planning method was proposed to minimize the total economic cost that consists of the capital cost, operational cost, and carbon emission cost. Simulation results showed that the comprehensive resilience index can increase by 18\% when the hydrogen-penetration ratio in the airport energy system increases from 20\% to 40\%. In the above works, the deterministic effects of extreme events were considered. In practice, those effects may be stochastic. For example, Gao \emph{et al.}\cite{Gao2024} proposed a two-stage resilience planning method for an integrated electricity-gas energy system with the consideration of stochastic line fault impacts incurred by typhoons. As shown in Fig.~\ref{fig_7}, the first stage solves the investment decision problem to decide the locations and capacities of HRSs and gas-fired distributed generators under the given quantity constraints. The second stage solves the system operation optimization problem in all fault scenarios, which were generated based on the Monte Carlo method and line fragile probability models under typhoons. The decision variables are upstream purchase, electric/gas load shedding, fuel cell generation, and gas-fired distributed generation. Finally, a single-level MISOCP problem was formulated to minimize the sum of investment cost and expected operation cost and solved by the Gurobi solver. Simulation results showed that the proposed planning method can reduce the total cost compared to other four baselines by 6.36\%-50.54\%, which focus on only resilience (load shedding), only investment economics, only HRS planning, and only gas-fired distributed generators.

\begin{table*}[htbp]
\center
\setlength{\arrayrulewidth}{1pt}
\caption{\textcolor[rgb]{0.0,0.0,0.00}{Summary of Planning Enhancement for HMES (Single-type Facility)}}\label{table_6} \centering
\rowcolors{2}{yellow!12}{yellow!12}

\begin{tabular}{|
>{\centering\arraybackslash}m{0.8cm}|
>{\centering\arraybackslash}m{1.9cm}|
>{\centering\arraybackslash}m{2.2cm}|
>{\centering\arraybackslash}m{2.7cm}|
>{\centering\arraybackslash}m{2.4cm}|
>{\centering\arraybackslash}m{2.4cm}|
>{\centering\arraybackslash}m{2.3cm}|}
\hline
\rowcolor{yellow!25} 
\textbf{Ref.} & \textcolor[rgb]{0.0,0.0,0.00}{\textbf{Parameter/event uncertainty}} &\textcolor[rgb]{0.0,0.0,0.00}{\textbf{Contingency uncertainty}}  & \textcolor[rgb]{0.0,0.0,0.00}{\textbf{Uncertainty processing methods}} & \textbf{Hydrogen facility investment decisions}  & \textcolor[rgb]{0.0,0.0,0.00}{\textbf{Optimization objective components}} &\textbf{Problem types and solving methods}\\
\hline
\hline
\cite{Yan2025} & \textcolor[rgb]{0.0,0.0,0.00}{$\times$} & \textcolor[rgb]{0.0,0.0,0.00}{Deterministic impacts}   & \textcolor[rgb]{0.0,0.0,0.00}{$\times$} & HESS (capacity)  & \textcolor[rgb]{0.0,0.0,0.00}{Total cost} & Single-level deterministic MILP, CPLEX\\
 \hline
\cite{ZhaoH2022} & \textcolor[rgb]{0.0,0.0,0.00}{$\times$} & \textcolor[rgb]{0.0,0.0,0.00}{Deterministic impacts}   & \textcolor[rgb]{0.0,0.0,0.00}{$\times$} & HESS (capacity) &\textcolor[rgb]{0.0,0.0,0.00}{Overall economic cost} & Single-level deterministic MILP, CPLEX\\
\hline
\cite{Gao2024} & \textcolor[rgb]{0.0,0.0,0.00}{$\times$} & \textcolor[rgb]{0.0,0.0,0.00}{Line fault uncertainty}   & \textcolor[rgb]{0.0,0.0,0.00}{Line fault scenarios} & HRS (capacity, location, quantity) &\textcolor[rgb]{0.0,0.0,0.00}{Investment cost and expected operation cost} & Two-stage MISOCP, Gurobi \\
\hline
\cite{Qu2024} & \textcolor[rgb]{0.0,0.0,0.00}{Natural disaster event uncertainty} & \textcolor[rgb]{0.0,0.0,0.00}{Component faults and fault durations}  & \textcolor[rgb]{0.0,0.0,0.00}{Uncertainty set, natural disaster scenarios} & HESS (capacity, location, quantity) & \textcolor[rgb]{0.0,0.0,0.00}{Total cost, operation cost, load shedding, and switching cost} & Tri-level robust MILP, mixed-integer L-shaped method\\
\hline
\cite{Gu2024} & \textcolor[rgb]{0.0,0.0,0.00}{$\times$} & \textcolor[rgb]{0.0,0.0,0.00}{Fault seasons and component fault combinations}   & \textcolor[rgb]{0.0,0.0,0.00}{Season scenarios, fault uncertainty sets} & HESS (capacity) &\textcolor[rgb]{0.0,0.0,0.00}{Total profit, load shedding cost} & Tri-level robust MILP, nested C\&CG\\
\hline
\cite{BerneckerM2026} & \textcolor[rgb]{0.0,0.0,0.00}{$\times$} & \textcolor[rgb]{0.0,0.0,0.00}{Low renewable generation}  &  \textcolor[rgb]{0.0,0.0,0.00}{Cardinality-constrained uncertainty sets} & HESS (location, capacity)  & \textcolor[rgb]{0.0,0.0,0.00}{Total annualized costs; operation cost} & Tri-level robust MILP; C\&CG\\
\hline
\cite{WenZ2024} & \textcolor[rgb]{0.0,0.0,0.00}{Typhoon event uncertainty} & \textcolor[rgb]{0.0,0.0,0.00}{Stochastic line outages incurred by typhoons}  & \textcolor[rgb]{0.0,0.0,0.00}{Typhoon scenarios, line outage uncertainty sets} & H2P (capacity, location, quantity) &\textcolor[rgb]{0.0,0.0,0.00}{Investment cost and expected worst-case system cost} & Tri-level stochastic robust optimization, nested C\&CG with hedging\\
\hline
\cite{Oh2024} & \textcolor[rgb]{0.0,0.0,0.00}{$\times$} & \textcolor[rgb]{0.0,0.0,0.00}{$N-k$ contingencies}  & \textcolor[rgb]{0.0,0.0,0.00}{Contingency scenarios} & HESS (location, quantity) & \textcolor[rgb]{0.0,0.0,0.00}{Total cost, load restoration quantity} & Bi-level MILP, CPLEX\\
\hline
\end{tabular}
\end{table*}

\begin{figure}[!htb]
\centering
\includegraphics[scale=0.43]{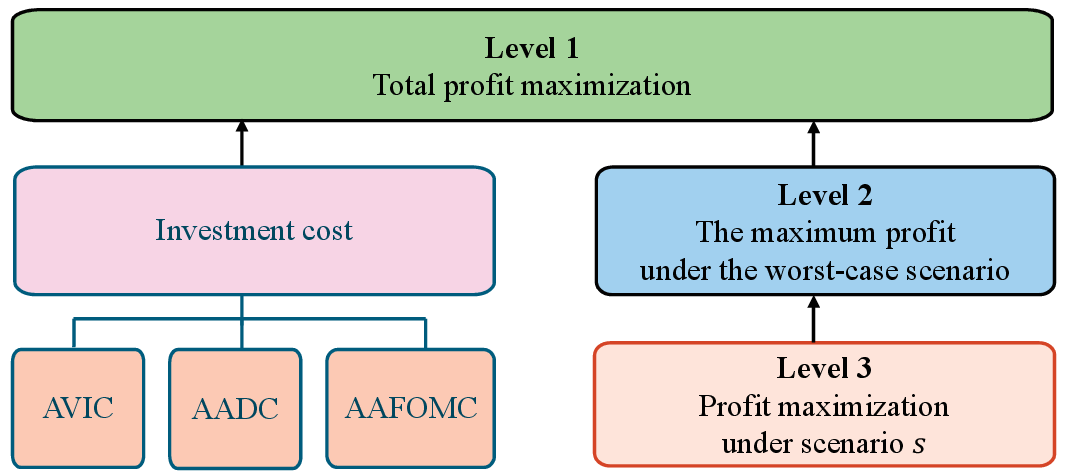}
\caption{The framework of the tri-level robust planning method}
\label{fig_8}
\end{figure}

In \cite{Qu2024}, Qu \emph{et al.} investigated a tri-level robust planning problem for enhancing the resilience of distribution networks considering deterministic line faults. The optimization objective is to minimize the weighted sum of annual investment cost and annual load shedding cost, and decision variables of the first-level problem are the locations, quantities, and capacities of hydrogen microgrids and soft open points. To generate the stochastic line fault scenarios, a multi-type natural disaster model was developed. To solve the formulated tri-level MILP problem, a mixed-integer L-shaped method was proposed. Simulation results indicated that cooperative planning of hydrogen microgrids and soft open points can significantly enhance the resilience of the distribution network and reduce load shedding by 68.4\%. In \cite{Gu2024}, Gu \emph{et al.} proposed a tri-level planning method to determine the capacities of HESSs and gas turbines for an electricity-hydrogen island energy system while taking the uncertainties of wind power, electric load, and line faults into consideration. As shown in Fig.~\ref{fig_8}, three levels could be identified. Note that levels 2 and 3 intend to maximize the profit under the worst-case scenario, while level 1 intends to maximize the sum of the maximum profit obtained from level 2 minus the investment cost, which consists of the annual average investment cost, the annual average depreciation cost, and the annual average fixed operation and maintenance cost. To achieve the above aim, a $\max$-$\min$-$\max$ tri-level problem was formulated under different representative days. Finally, a nested C\&CG algorithm was adopted to solve the formulated problem. Compared with the case that only considers worst-case contingency, the proposed planning method can reduce load shedding by 86\%. Similarly, Bernecker \emph{et al.}\cite{BerneckerM2026} investigated a capacity planning problem for fully decarbonized European electricity systems based on tri-level robust optimization framework considering Dunkelflaute extreme weather events (i.e., prolonged periods of low wind and solar availability). The first level focuses on minimizing investment cost, the second level focuses on maximizing the total system cost by finding the worst-case realizations, and third level focus on minimizing the total system cost under the given uncertainty realization. By merging the dual problem of the third-level subproblem with the second-level subproblem, an equivalent maximization problem can be derived. Finally, the $\min-\max$ problem could be solved by C\&CG algorithm. Simulation results indicated the importance of considering widespread regional weather events when designing a robust Europe electricity system.

\begin{figure}[!htb]
\centering
\includegraphics[scale=0.52]{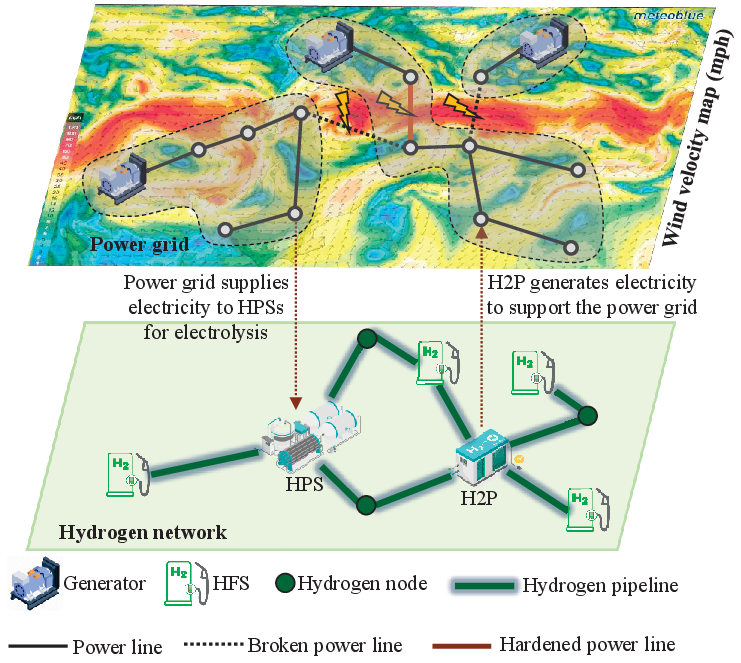}
\caption{Illustration of line outages in an electricity-hydrogen integrated energy system under typhoons\cite{WenZ2024}}
\label{fig_9}
\end{figure}

Although stochastic effects of an event have been considered in the above-mentioned works, they neglect the event uncertainty. For example, when typhoons are considered in the resilient planning of an MHES, there are many possible scenarios related to typhoon spatial distributions, and stochastic effects under each scenario are incurred. As shown in Fig.~\ref{fig_9}, line outages within the high-risk areas (e.g., high wind speed areas) under a typhoon scenario may happen. By taking event uncertainty into consideration, Wen \emph{et al.} proposed a tri-level planning method for a hydrogen-electricity integrated energy system\cite{WenZ2024}. To be specific, a Wasserstein generative adversarial network with a gradient penalty-based method was presented to generate the typical typhoon scenarios. Then, spectral clustering was adopted to reduce typhoon scenarios. Next, a tri-level optimization problem was formulated, where the upper level intends to decide the investment decisions related to power line hardening and H2P facilities, the middle level intends to find the worst-case line outage event under each typhoon scenario, while the lower level intends to minimize the weighted sum of load shedding related to power demand and hydrogen demand under each line outage event. The above tri-level problem was finally solved by nested C\&CG with a progressive hedging algorithm. Simulation results indicated that the proposed planning method can reduce load shedding by 95.7\% compared with existing uncertain planning methods.

Different from above-mentioned works that deal with uncertainty in optimization, leader-follower structure is adopted for modeling the planning problem with the consideration of contingencies. To be specific, Oh \emph{et al.}\cite{Oh2024} proposed a two-level resilient planning approach for gas-electric-based multi-energy systems, including a microgrid, when $N-k$ contingencies are considered. The upper level intends to minimize the total cost of the microgrid by deciding the capacity, location, and quantity of each component (e.g., distributed generators, electric ESSs, HESSs, wind turbines, photovoltaic panels). The lower level intends to maximize the restored load. Simulation results showed that the proposed approach can enhance the system's resilience by the adoption of the microgrid and reduce expected electric load loss by 30\%-34\%.

\begin{table*}[htbp]
\center
\setlength{\arrayrulewidth}{1pt}
\caption{\textcolor[rgb]{0.0,0.0,0.00}{Summary of Planning Enhancement for HMES (Multi-type facilities)}}\label{table_7} \centering

\begin{tabular}{|
>{\centering\arraybackslash}m{0.8cm}|
>{\centering\arraybackslash}m{2.3cm}|
>{\centering\arraybackslash}m{1.8cm}|
>{\centering\arraybackslash}m{2.4cm}|
>{\centering\arraybackslash}m{3.1cm}|
>{\centering\arraybackslash}m{1.8cm}|
>{\centering\arraybackslash}m{2.3cm}|}
\hline
\rowcolor{yellow!25} 
\textbf{Ref.} & \textcolor[rgb]{0.0,0.0,0.00}{\textbf{Parameter/event uncertainty}} & \textcolor[rgb]{0.0,0.0,0.00}{\textbf{Contingency uncertainty}} & \textcolor[rgb]{0.0,0.0,0.00}{\textbf{Uncertainty processing methods}} & \textbf{Hydrogen facility investment decisions}   & \textcolor[rgb]{0.0,0.0,0.00}{\textbf{Optimization objective components}} &\textbf{Problem types and solving methods}\\
\hline
\rowcolor{yellow!12} 
\cite{Ameli2024} & \textcolor[rgb]{0.0,0.0,0.00}{$\times$} & \textcolor[rgb]{0.0,0.0,0.00}{Deterministic impacts}  & \textcolor[rgb]{0.0,0.0,0.00}{$\times$} & HESS, hydrogen production facilities (capacity, location)  & \textcolor[rgb]{0.0,0.0,0.00}{Total cost} & Single-level deterministic MINLP, Xpress solver\\
\hline
\rowcolor{yellow!12} 
\cite{Shao2023} & \textcolor[rgb]{0.0,0.0,0.00}{Renewable generation and demand} &  \textcolor[rgb]{0.0,0.0,0.00}{$N-k$ stochastic device failure}  & \textcolor[rgb]{0.0,0.0,0.00}{Normal operation scenarios, emergency operational scenarios} &  HESS, hydrogen-driven CCHP, SHS (capacity, location, quantity)  & \textcolor[rgb]{0.0,0.0,0.00}{Total cost, operation cost} & Two-stage stochastic MINLP, dual cutting-plane algorithm\\
\hline
\rowcolor{yellow!12} 
\cite{XieC2024} & \textcolor[rgb]{0.0,0.0,0.00}{Renewable generation, demand, and price} & \textcolor[rgb]{0.0,0.0,0.00}{$N-k$ line outage contingencies}  & \textcolor[rgb]{0.0,0.0,0.00}{Normal operation scenarios, emergency operational scenarios} & Fuel cell electric truck, HRS, renewable generator, and remote switch (capacity, location)  & \textcolor[rgb]{0.0,0.0,0.00}{Investment cost, expected operation cost} & Two-stage stochastic MINLP, Gurobi\\
\hline
\rowcolor{yellow!12} 
\cite{WangT2025} & \textcolor[rgb]{0.0,0.0,0.00}{Renewable generation, electrical and hydrogen load} & \textcolor[rgb]{0.0,0.0,0.00}{$N-k$ contingencies}  & \textcolor[rgb]{0.0,0.0,0.00}{Normal operation scenarios, network disruption scenarios} & HESS, MHER (location, capacity)  & \textcolor[rgb]{0.0,0.0,0.00}{Total cost, expected operation cost} & Risk-averse two-stage stochastic MILP, Gurobi\\
\hline
\rowcolor{yellow!12} 
\cite{Dong2023} & \textcolor[rgb]{0.0,0.0,0.00}{$\times$} & \textcolor[rgb]{0.0,0.0,0.00}{$N-k$ line contingencies} & \textcolor[rgb]{0.0,0.0,0.00}{Cardinality-constrained uncertainty sets} &  HRS and hydrogen fuel cell electric bus (capacity, location, quantity)  &\textcolor[rgb]{0.0,0.0,0.00}{Profit, load restoration}  & Tri-level MILP, nested C\&CG\\
\hline
\rowcolor{yellow!12} 
\cite{LiuS2026} &\textcolor[rgb]{0.0,0.0,0.00}{Wind speed and rainfall intensity} & \textcolor[rgb]{0.0,0.0,0.00}{Power line and hydrogen pipeline uncertainties} & \textcolor[rgb]{0.0,0.0,0.00}{Endogenous second-order moment ambiguity set, scenarios} &  Hydrogen pipeline hardening, HESS (capacity)  &\textcolor[rgb]{0.0,0.0,0.00}{Load shedding}  & Distributionally robust MISOCP, C\&CG algorithm\\
\hline
\rowcolor{yellow!12} 
\cite{DongY2026} & \textcolor[rgb]{0.0,0.0,0.00}{Demand, renewable generation, line outages, vessel availability} & \textcolor[rgb]{0.0,0.0,0.00}{$N-k$ contingencies}  & \textcolor[rgb]{0.0,0.0,0.00}{Decision dependent-based ambiguity set, scenarios} & HESS (location, capacity), hydrogen vessel purchasing and routing   & \textcolor[rgb]{0.0,0.0,0.00}{Total costs; operational and maintenance costs} & Distributionally robust MILP, C\&CG with strong cutting planes\\
\hline
\rowcolor{yellow!12} 
\cite{WangHong2025} & \textcolor[rgb]{0.0,0.0,0.00}{$\times$} & \textcolor[rgb]{0.0,0.0,0.00}{Deterministic impacts}  & \textcolor[rgb]{0.0,0.0,0.00}{$\times$} & Fuel cell, battery, electrolyzer, and hydrogen tank (capacity, rated power, location)   & \textcolor[rgb]{0.0,0.0,0.00}{Configuration cost, operation cost} & Bi-level problem, PSO and MILP\\
\hline
\end{tabular}
\end{table*}

\begin{table*}[t]
\centering
\setlength{\arrayrulewidth}{1pt}
\caption{\textcolor[rgb]{0.0,0.0,0.00}{Classification of Uncertainty Processing Methods in HMES Planning Studies}}
\label{table_8}
\renewcommand{\arraystretch}{1.25}
\rowcolors{2}{yellow!12}{yellow!12}
\begin{tabular}{|>{\centering\arraybackslash}m{2.0cm}
                |>{\centering\arraybackslash}m{4.4cm}
                |>{\centering\arraybackslash}m{3.2cm}
                |>{\centering\arraybackslash}m{4.2cm}
                |>{\centering\arraybackslash}m{1.8cm}|}
\hline
\rowcolor{yellow!25} 
\textbf{Category} & \textbf{Key idea} & \textbf{Typical representations} & \textbf{Representative characteristics} & \textbf{Formulation framework} \\
\hline

\textcolor[rgb]{0.0,0.0,0.00}{Scenario-based methods}
&
\textcolor[rgb]{0.0,0.0,0.00}{Uncertainty is represented by a finite number of discrete scenarios, each corresponding to a possible realization of uncertain parameters, extreme events, or contingencies.}
&
\textcolor[rgb]{0.0,0.0,0.00}{Renewable generation scenarios\cite{Shao2023}; load scenarios\cite{XieC2024}; natural disaster scenarios\cite{Qu2024}; line outage scenarios\cite{WenZ2024}}
&
\textcolor[rgb]{0.0,0.0,0.00}{Suitable for explicitly modeling the evolution of uncertain events; may suffer from scenario explosion and high computational burden}
&
\textcolor[rgb]{0.0,0.0,0.00}{Stochastic programming}
\\
\hline

\textcolor[rgb]{0.0,0.0,0.00}{Uncertainty-set-based methods}
&
\textcolor[rgb]{0.0,0.0,0.00}{Uncertainty is characterized by deterministic sets, and decisions are optimized to remain feasible or robust against all realizations within the prescribed sets.}
&
\textcolor[rgb]{0.0,0.0,0.00}{Interval sets; box sets; cardinality-constrained uncertainty sets\cite{Dong2023}\cite{BerneckerM2026}}
&
\textcolor[rgb]{0.0,0.0,0.00}{Does not require exact probability distributions; usually provides strong robustness guarantees, but may be conservative}
&
\textcolor[rgb]{0.0,0.0,0.00}{Robust optimization}
\\
\hline

\textcolor[rgb]{0.0,0.0,0.00}{Ambiguity-set-based methods}
&
\textcolor[rgb]{0.0,0.0,0.00}{Uncertainty is modeled through a family of probability distributions rather than a single known distribution, and the decision is optimized against the worst-case distribution within the ambiguity set.}
&
\textcolor[rgb]{0.0,0.0,0.00}{Moment-based ambiguity sets\cite{LiuS2026}; decision-dependent ambiguity sets\cite{DongY2026}}
&
\textcolor[rgb]{0.0,0.0,0.00}{Suitable when the exact probability distribution is unknown but partial statistical information is available; balances robustness and probabilistic information}
&
\textcolor[rgb]{0.0,0.0,0.00}{Distributionally robust optimization}
\\
\hline

\end{tabular}
\end{table*}

\begin{table*}[htbp]
\centering
\setlength{\arrayrulewidth}{1pt}
\caption{Modeling Frameworks and Solving Methods of Planning Enhancement Problem for HMESs} \label{table_9} \centering
\renewcommand{\arraystretch}{1.3}
\rowcolors{2}{yellow!12}{yellow!12}
\begin{tabular}{|>{\color{black}}m{2.6cm}<{\centering}|>{\color{black}}m{4.4cm}<{\centering}|>{\color{black}}m{4.9cm}<{\centering}|>{\color{black}}m{4.2cm}<{\centering}|}
\hline
\rowcolor{yellow!25} 
\textbf{Modeling framework} & \textbf{Model explanation} & \textbf{Mathematical formulation} & \textbf{Solution method} \\
\hline
Single-level deterministic optimization & Classical optimization with known parameters; no uncertainty or hierarchical decision structure & $\min_{x \in X} c^\top x+f(x)$  & Linear/Nonlinear/Mixed-integer programming solvers (e.g., Gurobi, CPLEX, IPOPT, Xpress)\cite{Yan2025} \\
\hline
Two-stage stochastic programming & First-stage decisions made before uncertainty is realized; second-stage recourse decisions made after uncertainty is revealed & $\min_{x \in X} c^\top x + \mathbb{E}_{\xi}[Q(x,\xi)]$, where $Q(x,\xi) = \min_{y} \{q^\top y : Wy +T x \ge h \}$ & Column-and-Constraint Generation (C\&CG)\cite{LiuS2026}, dual cutting-plane decomposition\cite{Shao2023}, commercial solvers (e.g., Gurobi\cite{XieC2024}\cite{WangT2025}) \\
\hline
Tri-level robust optimization & Hierarchical decision structure with robustness against worst-case uncertainty & $\min_{x \in X} c^\top x+\max_{\xi \in \mathcal{U}} \min_{y \in Y(x,\xi)} f(x,y,\xi)$ & L-shaped method\cite{Qu2024}, nested C\&CG\cite{Gu2024}\cite{Dong2023}, C\&CG\cite{BerneckerM2026} \\
\hline
Tri-level stochastic robust optimization & Outer stage handles here-and-now, middle stage captures uncertainty distribution/worst-case, inner stage executes corrective recourse & $\min_{x} c^\top x+\mathbb{E}_{\xi}[ \max_{u \in \mathcal{U}(\xi)} \min_{y} f(x,y,u)]$ & Nested C\&CG with progressive hedging\cite{WenZ2024} \\
\hline
Distributionally robust optimization & Decisions are robust against the worst-case probability distribution within an ambiguity set, rather than a single known distribution & $\min_{x \in X} c^\top x+\sup_{P \in \mathcal{P}} \mathbb{E}_{\xi \sim P}[f(x,\xi)]$ s.t. $f(x,\xi)=\min_{y\in Y(x,\xi)}\{q^\top y: Wy +T x \ge h \}$ & C\&CG with strong cutting planes\cite{DongY2026} \\
\hline
Bi-level optimization & Leader--follower structure where the upper level (leader) decision depends on the optimal response of the lower level (follower) & $\min_{x \in X} c^\top x+F(x,y)$ s.t. $y \in \arg\min_{y \in Y(x)} f(x,y)$ & Reformulation via KKT conditions\cite{Oh2024}, evolutionary algorithms\cite{WangHong2025} \\
\hline
\end{tabular}
\end{table*}

\subsection{Multi-type Hydrogen Facilities}\label{s52}
\begin{figure}[!htb]
\centering
\includegraphics[scale=0.38]{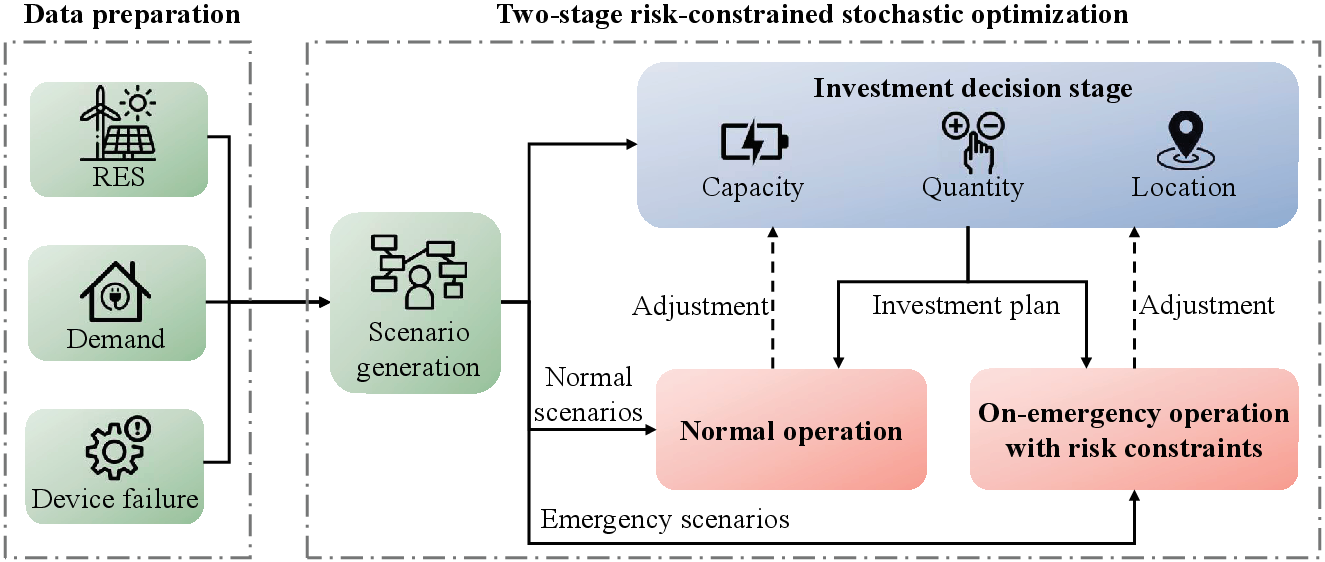}
\caption{The framework of the two-stage planning method\cite{Shao2023}}
\label{fig_10}
\end{figure}

To maximize the potential of utilizing hydrogen-related facilities against extreme events, multi-type hydrogen facilities could be considered. For example, Ameli \emph{et al.}\cite{Ameli2024} proposed a resilient planning method to decide the capacity and location of HESS and hydrogen production facilities for integrated gas and electricity systems with HESSs considering deterministic scenarios under extreme weather events. Different from the above studies, Shao \emph{et al.}\cite{Shao2023} proposed a planning method to decide the capacity, location, and quantity of HESS, CCHP, and SHS in hydrogen-based microgrids with the consideration of uncertainties related to power supply, load demand, and device failure caused by disastrous events. The framework of the proposed planning method is shown in Fig.~\ref{fig_10}, where data sources related to renewable energy sources, load demand, and device failure are first prepared. Then, scenarios used for planning are generated, which consist of normal scenarios and emergency scenarios. Next, a two-stage risk-constrained stochastic optimization framework is adopted for problem formulation, where the investment decision stage belongs to the first stage, and the operation stage belongs to the second stage. After the investment plan is determined, it will be transmitted to the second stage. According to the generated scenarios, the corresponding normal-status operation or on-emergency operation is carried out, and the resulting performance metrics are used for the adjustment of the investment plan of the first stage. Note that risk constraints are considered to reduce the conservatism of the investment plan. Simulation results indicated that the proposed planning method can implement flexible tradeoff between economics and resilience. In \cite{XieC2024}, Xie \emph{et al.} developed a planning problem model related to hydrogen-based seaport MESs, where normal operation scenarios and contingency operation scenarios are simultaneously considered. The model is formulated as a two-stage stochastic programming, where the first-stage decision is to decide the quantity and location of HRSs, fuel cell electric trucks, renewable generators, and remote control switches, and the second-stage decision is to decide scheduling power of each equipment under the given scenario. Simulation results indicated that the investment of resilience and low-carbon energy infrastructure can bring three aspects of benefits, i.e., power outage risk, system cost, and environmental conservation. In \cite{WangT2025}, Wang \emph{et al.} proposed a risk-averse two-stage stochastic programming framework for the economic–resilient planning of hydrogen-enriched distribution networks with MHERs, where the first stage minimizes investment cost by co-siting and sizing stationary distributed energy resources and MHERs. To hedge long-run operational risk, a conditional-value-at-risk-based objective is introduced, and on-emergency corrective actions (including energy storage dispatch, dynamic re-routing of MHERs, and distribution feeder reconfiguration) are modeled to enhance system resilience in the second stage. Case studies demonstrated that proper configuration of stationary and mobile distributed energy resources reduces load-curtailment risk by more than 75\% under comparable expenditure levels.

In order to deal with the worst-case cost under extreme events, Dong \emph{et al.}\cite{Dong2023} proposed a co-planning method for hydrogen-based microgrids and fuel-cell bus operation centers to decide the capacity, location, and quantity of HRS and FC electric bus considering stochastic line outages incurred by extreme weather events. Note that the proposed planning method is based on a nested C\&CG algorithm, which can solve the formulated $\max-\min-\max$ tr-level optimization problem efficiently. In \cite{LiuS2026}, Liu \emph{et al.} investigated a hydrogen pipeline hardening and HESS allocation problem for electric-hydrogen network against extreme weather. Then, the above problem was formulated based on two-stage distributionally robust optimization framework, where the first stage focuses on pre-disaster decision under the given budget and the allowed hydrogen leakage risk, the second stage focuses on minimizing load shedding cost after events happen. The formulated problem was transformed into a two-stage MISOCP problem and was solved by C\&CG problem. Simulation results indicated that the proposed strategy can minimize load shedding and control hydrogen pipeline leakage risk. Similar works could be observed in \cite{DongY2026}. In \cite{WangHong2025}, Wang \emph{et al.} developed a resilient planning method for hybrid hydrogen and battery energy storage systems under extreme events based on particle swarm optimization and MILP, where particle swarm optimization is adopted to decide the system configuration (including installed location of fuel cells and batteries; the rated power of fuel cells, batteries, electrolyzers, and hydrogen tanks; the capacity of batteries) and the MILP related to the system operation cost minimization is solved by commercial solvers. In this work, parameter uncertainty and event uncertainty are not considered, while contingency modeling related to renewable energy reduction, load increase, and line capacity reduction is considered. Simulation results indicated that the hybrid energy system can improve the resilience index (i.e., load loss ratio) by 23.8\% and 0.7\%, compared with single battery energy storage or hydrogen energy storage, respectively.

\emph{Summary}: To show the differences of existing works more clearly, their main features are summarized in Table~\ref{table_6} and Table~\ref{table_7}. \textcolor[rgb]{0.0,0.0,0.00}{It can be observed that there are typically three kinds of uncertainty processing methods in existing works. To better understand the differences among these methods, we provide a clear classification of them, as shown in Table~\ref{table_8}. Among them, scenario-based methods are suitable for explicitly modeling the evolution of uncertain events, and related planning problems typically follow the framework of stochastic programming. Instead of representing uncertainty by a finite number of scenarios, uncertainty-set-based methods adopt deterministic sets to describe uncertainties, e.g., interval sets and cardinality-constrained uncertainty sets. These methods do not require exact probability distributions, and related planning problems typically follow the robust optimization or stochastic robust optimization framework. When the exact probability distribution is unknown but partial statistical information is available, ambiguity-set-based methods can be adopted, which model uncertainty through a family of probability distributions. In this case, related planning problems typically follow the framework of distributionally robust optimization.} Moreover, when more complex HMESs are considered (e.g., electricity-hydrogen-gas systems), the number of decisions related to different components becomes larger \textcolor[rgb]{0.0,0.0,0.00}{and more nonlinear or nonconvex constraints incurred by hydrogen equipment (e.g., electrolyzers, fuel cells, hydrogen pipelines, hydrogen compressors/liquefiers, and hydrogen storage tanks) are introduced. Consequently, the formulated problems have more complex forms (e.g., a MINLP problem is incurred due to the nonlinear hydrogen pipeline hydraulics (i.e., equation (28) in \cite{Ameli2024})) and advanced solution methods are expected}. Furthermore, there are six kinds of modeling frameworks for planning enhancement problem of HMESs and we summarize their differences in \textcolor[rgb]{0.0,0.0,0.00}{Table~\ref{table_9}}. In row 1, $x$ denotes all decision variables, $c^\top x$ denotes the investment cost, $f(x)$ denotes other cost except investment cost; In row 2, $y$ denotes decision variables of lower-level optimization problem with objective function $f(x,y)$, $F(x,y)$ denotes the operation cost in the upper-level optimization problem; In row 3, $Q(x,\xi)$ denotes the operation cost under the scenario $\xi$; In row 4, $f(x,y,\xi)$ denotes the operation cost under the scenario $\xi$; In row 5, $f(x,y,u)$ denotes the operation cost under a decision $u$, which belongs to $\mathcal{U}(\xi)$; In row 6, $p$ denotes a probability distribution, $f(x,\xi)$ denotes the optimal operation cost under the scenario $\xi$ and decision $x$. Finally, the above works mainly focus on simultaneous line outages and/or device faults incurred by natural extreme events, while cascading faults and cyber attacks are seldom considered\cite{Sturmer2024}. \textcolor[rgb]{0.0,0.0,0.00}{In addition, a few of them considered scenario generation for extreme events based on generative adversarial networks\cite{WenZ2024} and natural disaster models\cite{Qu2024}. Although such methods have some advantages, they face some limitations in aspects of physical implausibility and bias correction. To show the advantages and disadvantages of other scenario generation methods, we provide a classification of extreme-event scenario generation methods in Table~\ref{tab_extreme_scenario_methods}, where representative methods, key idea, advantages, and limitations could be observed}.

\begin{table*}[htbp]
\centering
\setlength{\arrayrulewidth}{1pt}
\caption{\textcolor[rgb]{0.0,0.0,0.00}{Classification of extreme-event scenario generation methods considering the uncertainty of extreme events themselves}}
\label{tab_extreme_scenario_methods}
\rowcolors{2}{yellow!12}{yellow!12}
\renewcommand{\arraystretch}{1.3}
\begin{tabular}{|>{\color{black}}m{1.5cm}<{\centering}|>{\color{black}}m{3.4cm}<{\centering}|>{\color{black}}m{3.6cm}<{\centering}|>{\color{black}}m{3.6cm}<{\centering}|>{\color{black}}m{3.6cm}<{\centering}|}
\hline
\rowcolor{yellow!25} 
\textbf{Category} & \textbf{Representative methods} & \textbf{Key idea} & \textbf{Advantages} & \textbf{Limitations} \\
\hline

\textcolor[rgb]{0.0,0.0,0.00}{Statistical sampling-based scenario generation}
&
\textcolor[rgb]{0.0,0.0,0.00}{Historical event selection\cite{Arjomandi2020}; bootstrap resampling\cite{LongoE2026}; block bootstrapping\cite{HuJ2026}; parametric distribution fitting with random sampling\cite{WenZ2024}; representative-period extraction\cite{ZhangJ2025}}
&
\textcolor[rgb]{0.0,0.0,0.00}{Generate extreme-event scenarios directly from historical observations or their statistical properties, such as empirical frequencies, fitted marginal distributions, or resampling rules}
&
\textcolor[rgb]{0.0,0.0,0.00}{Simple and transparent; easy to implement; low computational burden; easy to integrate into stochastic planning models; preserves some historical temporal characteristics}
&
\textcolor[rgb]{0.0,0.0,0.00}{Difficult to generate unprecedented or very rare events; limited ability to represent compound extremes and high-dimensional dependence}
\\
\hline

\textcolor[rgb]{0.0,0.0,0.00}{Dependence-aware probabilistic modeling}
&
\textcolor[rgb]{0.0,0.0,0.00}{Copula-based models\cite{WuX2026}\cite{HuaD2026}; multivariate parametric distributions\cite{OrtegaJ2009}; Bayesian networks\cite{ZhaoZ2026}; spatio-temporal probability modeling\cite{SerreC2026}}
&
\textcolor[rgb]{0.0,0.0,0.00}{Construct explicit joint probabilistic relationships among multiple variables, so that generated scenarios preserve inter-variable, spatial, and temporal dependence structures}
&
\textcolor[rgb]{0.0,0.0,0.00}{Better captures spatial correlation and temporal clustering; more suitable than simple resampling for multi-variable extreme-event modeling}
&
\textcolor[rgb]{0.0,0.0,0.00}{Model estimation becomes difficult in high-dimensional settings; tail dependence is hard to identify under limited extreme samples}
\\
\hline

\textcolor[rgb]{0.0,0.0,0.00}{Climate- and weather-model-driven scenario generation}
&
\textcolor[rgb]{0.0,0.0,0.00}{Regional climate models\cite{Qu2024}; global climate model outputs\cite{LongoE2026}; numerical weather prediction models\cite{GuJ2025}; bias-corrected climate projections\cite{IizumiT2012}; weather generators\cite{NajibiN2024}}
&
\textcolor[rgb]{0.0,0.0,0.00}{Generate extreme-event scenarios from physically based simulation or projection of meteorological and climate processes, rather than only from historical observations}
&
\textcolor[rgb]{0.0,0.0,0.00}{Physically grounded; suitable for future climate conditions and long-return-period events; capable of representing persistent and evolving meteorological extremes}
&
\textcolor[rgb]{0.0,0.0,0.00}{Requires downscaling, bias correction, and large computational effort; outputs may contain structural uncertainty from the underlying climate/weather models}
\\
\hline

\textcolor[rgb]{0.0,0.0,0.00}{Artificial intelligence-enabled synthetic scenario generation}
&
\textcolor[rgb]{0.0,0.0,0.00}{Generative adversarial networks\cite{WenZ2024}\cite{WangH2025}\cite{GuJ2025}; variational autoencoders\cite{WuX2026}; diffusion models\cite{YangR2025}\cite{FuX2026}; hybrid deep generative models\cite{LiY2025EI}}
&
\textcolor[rgb]{0.0,0.0,0.00}{Learn complex nonlinear spatio-temporal patterns from data and generate synthetic extreme-event trajectories beyond direct historical resampling}
&
\textcolor[rgb]{0.0,0.0,0.00}{Flexible in handling high-dimensional and nonlinear patterns; potentially better at capturing complex spatio-temporal structures than classical approaches}
&
\textcolor[rgb]{0.0,0.0,0.00}{Usually data-hungry; may generate physically implausible scenarios without domain constraints; interpretability is weak}
\\
\hline

\end{tabular}
\end{table*}

\section{Hydrogen-enabled Operation Enhancement for MESs}\label{s6}

As mentioned above, three types of HMES operation response stages under extreme events are involved. According to the measures adopted in the above-mentioned three stages, existing works can be divided into two types, i.e., single-stage operation strategy and multi-stage operation strategy. The former means that the designed strategy is involved in one operational stage, while the latter means that the designed strategy is involved in multiple operational stages. In the following parts, the related works are introduced in detail.

\subsection{Single-stage operation strategy}\label{s61}
Many single-stage operation strategies have been developed in existing works. For example, Sharifpour \emph{et al.}\cite{Sharifpour2023} proposed a preventive response strategy for a networked microgrid by integrating HESSs and demand response programs, which encourages replenishing HESSs and restricting non-essential loads before the impending main grid power outages incurred by extreme events. Similarly, Shahbazbegian \emph{et al.}\cite{Shahbazbegian2023} proposed a preventive response approach to optimize operation cost and resilience metrics (e.g., electricity purchasing quantity, hydrogen tank level, and power loss) related to a microgrid with power-to-hydrogen systems when line outages caused by extreme events are considered. Although some advances have been made, the above methods assume that the future event attributes (e.g., power outage duration and starting time) are known. Moreover, parameter uncertainties related to load and renewable generation were neglected. To overcome such challenges, many preventive response strategies are designed with the consideration of uncertainty modeling\cite{Yuan2024}\cite{Liu2024}. For example, in \cite{Yuan2024}, Yuan \emph{et al.} proposed a risk-constrained day-ahead preventive response method for photovoltaic-penetrated power distribution networks equipped with power-to-hydrogen systems based on two-stage stochastic programming considering uncertainties in wind speed, solar radiation, active and reactive loads. Liu \emph{et al.}\cite{Liu2024} proposed a resilient day-ahead scheduling method for power distribution networks in the presence of uncertainties in renewable output, active/reactive demands, and disaster duration, which intends to prepare HRSs for electricity generation during power outage emergencies. In \cite{ZhaoY2024}, Zhao \emph{et al.}  proposed a robust hydrogen-penetrated formation strategy against emergent power outages caused by extreme events with the help of HRSs and hydrogen transit while taking uncertainties in load and hydrogen demands, renewable generators, and temporal distances between HRSs into consideration.

\begin{figure}[!htb]
\centering
\includegraphics[scale=0.55]{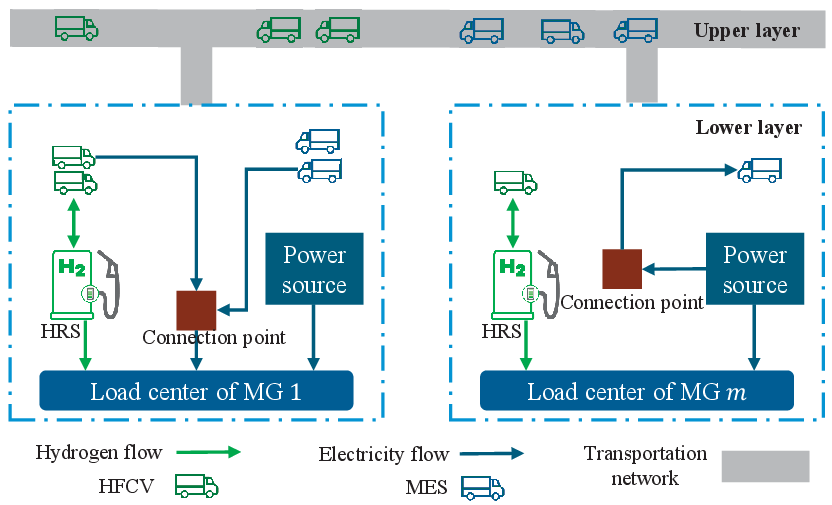}
\caption{The key idea of the proposed scheduling method for the hydrogen-based multi-energy supply microgrid\cite{LiB2023}}
\label{fig_11}
\end{figure}

\begin{figure}[!htb]
\centering
\includegraphics[scale=0.39]{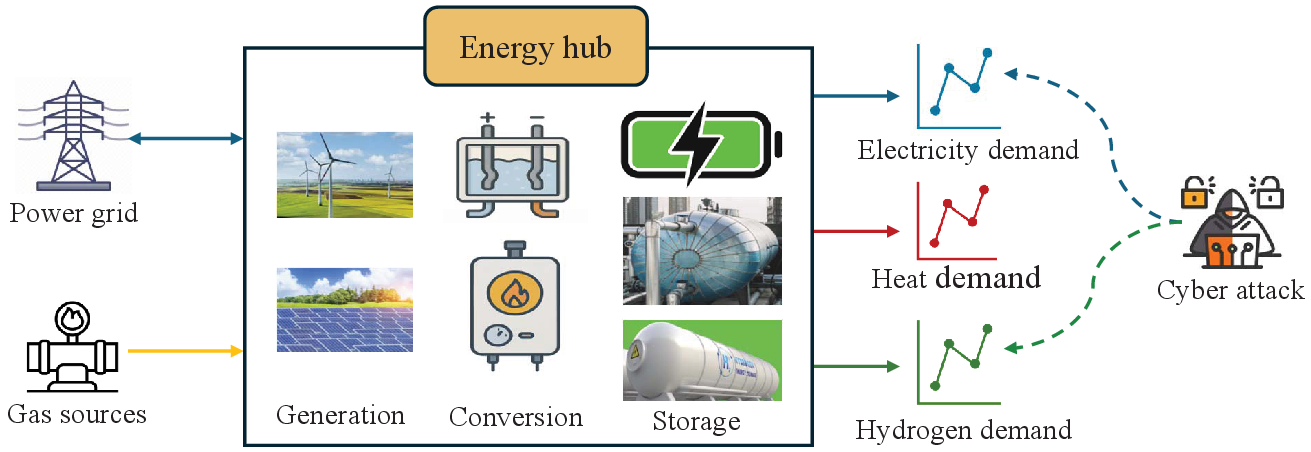}
\caption{The illustration of an energy hub under threats of false data injection attacking}
\label{fig_12}
\end{figure}

When extreme events happen, an emergency response can be adopted to minimize load shedding in existing works. For example, $\c{C}$i$\c{c}$ek \emph{et al.}\cite{CicekA2025} investigated an economic operation problem of a hydrogen-based country house with the consideration of power outages and uncertainties in renewable energy and electricity prices. When power outages occur, fuel cells will provide the electricity supply. In \cite{TangD2025}, Tang \emph{et al.} investigated an economic and resilient operation problem for a hydrogen-electricity coupled system in ports considering main grid outages caused by extreme events. Moreover, a resilience enhancement method based on hydrogen and electricity energy storage was proposed. Simulation results indicated the effectiveness of the proposed method in reducing load shedding. In \cite{ZhuR2024}, Zhu \emph{et al.} developed an operation strategy for the integrated electricity and natural gas system by utilizing hydrogen-enriched compressed natural gas to generate electricity during emergent power outages. In \cite{LiB2023}, Li \emph{et al.} proposed an emergency scheduling method for a hydrogen-based multi-energy supply microgrid by utilizing mobile hydrogen storage during the event of hurricanes. The key idea of the proposed scheduling method is illustrated in Fig.~\ref{fig_11}, where four steps could be identified. Firstly, the multi-energy network supply ability is estimated based on the developed temporal-spatial destructive model. Then, according to the accepted waiting time related to hydrogen energy delivery from the hydrogen company, the optimal traffic flow and the maximum mobile hydrogen energy are decided. Next, mobile hydrogen energy is delivered through the transportation network. Finally, based on the above information, a load shedding minimization-based scheduling method is designed. Simulation results showed that delivering mobile hydrogen tanks to the end-user microgrid can reduce heat load shedding and electric load shedding effectively. In \cite{MehrjerdiH2022}, Mehrjerdi \emph{et al.} investigated a multicarrier microgrid operation problem to minimize the expected annual operating cost considering component outages. Based on the cross-sector flexibility, the model considered the use of hydrogen generated from renewables for natural gas generation to reduce load shedding. In \cite{GuZhong2025}, extreme renewable generation caused by extreme events was considered in the resilient operation of HMESs. Under the emergency mode, hydrogen-related measures were used, e.g., hydrogen pipeline scheduling, non-critical hydrogen load shedding, and HESS discharging. However, the above works mainly focus on non-hydrogen contingencies, e.g., power outages, non-hydrogen component faults, load surge, and renewable generation dropping. In \cite{RiziD2025}, Rizi \emph{et al.} proposed a robust K-nearest neighbors (KNN)-based data modification approach for an energy hub as shown in Fig.~\ref{fig_12} to mitigate the threats caused by false data injection attacks towards electricity demand and hydrogen demand. Simulation results highlighted the need for developing advanced detection and mitigation strategies for protecting energy hubs against evolving cyber attacks.


\begin{table*}[htbp]
\center
\setlength{\arrayrulewidth}{1pt}
\caption{Summary of Operation Enhancement for HMES (Single-stage)}\label{table_10} \centering
\rowcolors{2}{yellow!12}{yellow!12}
\begin{tabular}{|
>{\centering\arraybackslash}m{0.9cm}|
>{\centering\arraybackslash}m{2.5cm}|
>{\centering\arraybackslash}m{3.2cm}|
>{\centering\arraybackslash}m{2.1cm}|
>{\centering\arraybackslash}m{1.4cm}|
>{\centering\arraybackslash}m{3.7cm}|
>{\centering\arraybackslash}m{1.0cm}|}
\hline
\rowcolor{yellow!25}
\textbf{Ref.}&\textcolor[rgb]{0.0,0.0,0.00}{\textbf{Parameter/event uncertainty}}&\textcolor[rgb]{0.0,0.0,0.00}{\textbf{Contingency uncertainty}} & \textcolor[rgb]{0.0,0.0,0.00}{\textbf{Uncertainty processing methods}} & \textbf{Operation stages} & \textbf{Problem types and solving methods} &\textbf{Measure types}\\
\hline
\hline
\cite{Sharifpour2023}  & \textcolor[rgb]{0.0,0.0,0.00}{$\times$} & \textcolor[rgb]{0.0,0.0,0.00}{Deterministic grid outage}  & \textcolor[rgb]{0.0,0.0,0.00}{$\times$}  & Preventive & MILP; CPLEX & A\\
\hline
\cite{Shahbazbegian2023}& \textcolor[rgb]{0.0,0.0,0.00}{$\times$} & \textcolor[rgb]{0.0,0.0,0.00}{Deterministic grid outage} & \textcolor[rgb]{0.0,0.0,0.00}{$\times$} & Preventive & MINLP; Generalized benders decomposition & A\\
\hline
\cite{Haggi2021}& $\times$ & \textcolor[rgb]{0.0,0.0,0.00}{Deterministic line and generator outages} &  \textcolor[rgb]{0.0,0.0,0.00}{$\times$} & Preventive  & MINLP   & A \\
\hline
\cite{Yuan2024}& \textcolor[rgb]{0.0,0.0,0.00}{Wind speed, solar radiation, active and reactive loads} & \textcolor[rgb]{0.0,0.0,0.00}{Deterministic grid outage}  & \textcolor[rgb]{0.0,0.0,0.00}{Scenarios} & Preventive  &Stochastic MILP, fuzzy method & A\\
\hline
\centering \cite{Liu2024}& \textcolor[rgb]{0.0,0.0,0.00}{Demands and renewable generation} & \textcolor[rgb]{0.0,0.0,0.00}{Deterministic grid outage} & \textcolor[rgb]{0.0,0.0,0.00}{Scenarios} &Preventive &Stochastic MILP; CPLEX & A\\
\hline
\cite{ZhaoY2024}& \textcolor[rgb]{0.0,0.0,0.00}{Demands, renewable generators, HRS distances} & \textcolor[rgb]{0.0,0.0,0.00}{Deterministic grid outage}  & \textcolor[rgb]{0.0,0.0,0.00}{Scenarios} &Preventive & Stochastic MILP; Gurobi & B \\
 \hline
 \cite{CicekA2025} & \textcolor[rgb]{0.0,0.0,0.00}{Renewable energy and electricity prices} & \textcolor[rgb]{0.0,0.0,0.00}{Deterministic grid outage} & \textcolor[rgb]{0.0,0.0,0.00}{Scenarios} & Emergency  & Stochastic MILP; CPLEX &D\\
 \hline
   \cite{TangD2025} & \textcolor[rgb]{0.0,0.0,0.00}{$\times$} & \textcolor[rgb]{0.0,0.0,0.00}{Deterministic grid outage}  & \textcolor[rgb]{0.0,0.0,0.00}{$\times$} & Emergency & Fuzzy multi-objective MISOCP; Gurobi & D \\
\hline
\cite{ChangS2025} & \textcolor[rgb]{0.0,0.0,0.00}{$\times$} & \textcolor[rgb]{0.0,0.0,0.00}{Deterministic line outages and generator trips}  & \textcolor[rgb]{0.0,0.0,0.00}{$\times$} & Emergency & MILP, CPLEX & E \\
  \hline
\cite{ZhuR2024} & \textcolor[rgb]{0.0,0.0,0.00}{$\times$} & \textcolor[rgb]{0.0,0.0,0.00}{Deterministic grid outage and gas shortage} & \textcolor[rgb]{0.0,0.0,0.00}{$\times$}
 & Emergency & MILP; Alternating direction method of multipliers &E\\
 \hline
\cite{LiB2023} & \textcolor[rgb]{0.0,0.0,0.00}{$\times$} & \textcolor[rgb]{0.0,0.0,0.00}{Deterministic component failures} & \textcolor[rgb]{0.0,0.0,0.00}{$\times$}
 &Emergency  & MILP, Frank-Wolfe algorithm &F\\
 \hline
\cite{MehrjerdiH2022} & \textcolor[rgb]{0.0,0.0,0.00}{Loads, renewables, electric vehicle charging behaviors} & \textcolor[rgb]{0.0,0.0,0.00}{Deterministic component outages}  & \textcolor[rgb]{0.0,0.0,0.00}{Scenarios} & Emergency  & Stochastic MILP, GAMS solver &H\\
 \hline
\cite{GuZhong2025} & \textcolor[rgb]{0.0,0.0,0.00}{PV and wind turbine generation} & \textcolor[rgb]{0.0,0.0,0.00}{Extreme PV and wind turbine generation}  & \textcolor[rgb]{0.0,0.0,0.00}{Scenarios} & Emergency  & Stochastic MILP, Gurobi/CPLEX &E, G\\
 \hline
\cite{RiziD2025}  & \textcolor[rgb]{0.0,0.0,0.00}{$\times$} & \textcolor[rgb]{0.0,0.0,0.00}{Cyber-attacks on demand} & \textcolor[rgb]{0.0,0.0,0.00}{$\times$} & Emergency  & MILP; CPLEX &G\\
 \hline
\cite{ZhuS2024} & \textcolor[rgb]{0.0,0.0,0.00}{$\times$} & \textcolor[rgb]{0.0,0.0,0.00}{Deterministic line and pipeline outages}  & \textcolor[rgb]{0.0,0.0,0.00}{$\times$}  & Restoration & MISOCP; Relaxation inducement & K\\
\hline
\cite{Afsari2024}& \textcolor[rgb]{0.0,0.0,0.00}{Load demand} & \textcolor[rgb]{0.0,0.0,0.00}{Line outages}  & \textcolor[rgb]{0.0,0.0,0.00}{Scenarios, ambiguity sets}  & Restoration  & Chance-constrained adaptive distributionally robust MILP; CPLEX & K \\
 \hline
\cite{LuJ2025} & \textcolor[rgb]{0.0,0.0,0.00}{$\times$} & \textcolor[rgb]{0.0,0.0,0.00}{Deterministic large-scale blackouts} & \textcolor[rgb]{0.0,0.0,0.00}{$\times$} & Restoration & MILP
 & L \\
\hline
\cite{Su2024}& \textcolor[rgb]{0.0,0.0,0.00}{Wind forecast} & \textcolor[rgb]{0.0,0.0,0.00}{Deterministic line outages} & \textcolor[rgb]{0.0,0.0,0.00}{Scenarios} & Restoration   & Stochastic MILP; Gurobi  & I \\
\hline
\cite{ZhaoY2023}& \textcolor[rgb]{0.0,0.0,0.00}{Renewable outputs, electric and thermal loads} & \textcolor[rgb]{0.0,0.0,0.00}{Deterministic line outages} & \textcolor[rgb]{0.0,0.0,0.00}{Scenarios} & Restoration & Stochastic MILP; Gurobi
 & J \\
  \hline
\cite{Wangz2021} & \textcolor[rgb]{0.0,0.0,0.00}{$\times$} & \textcolor[rgb]{0.0,0.0,0.00}{Power line and pipeline outages}  & \textcolor[rgb]{0.0,0.0,0.00}{$\times$}  & Restoration & MILP, Gurobi
 & M \\
\hline
\end{tabular}
\end{table*}

\begin{figure}[!htb]
\centering
\includegraphics[scale=0.45]{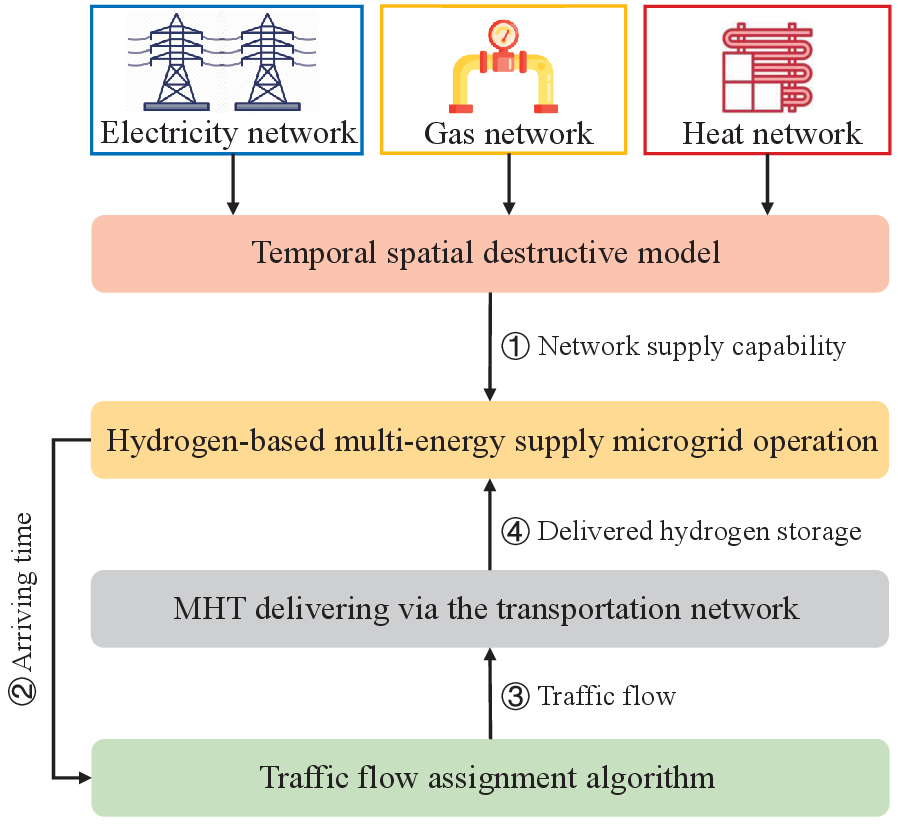}
\caption{The framework of the two-level restoration strategy for a distribution network with HRSs\cite{ZhaoY2023}}
\label{fig_13}
\end{figure}

After the extreme events, some measures related to system topologies or resources could be adopted to restore load. For example, Zhu \emph{et al.}\cite{ZhuS2024} developed a resilience-oriented service restoration strategy for a power-hydrogen distribution system by taking some measures, e.g., network topology reconfiguration, scheduling of dispatchable generators and HESSs, and load prioritization. Similarly, Afsari \emph{et al.}\cite{Afsari2024} proposed a service restoration strategy for microgrids with hydrogen storage based on network topology reconfiguration and the coordination of HESSs and hydrogen-based micro-gas turbines. Before network reconfiguration is performed, an HMES black-start is required. In \cite{LuJ2025}, Lu \emph{et al.} developed a black start strategy for hydrogen-integrated renewable grids with energy storage systems, where hydrogen fuel cells are used as black start resources. In addition to stationary HERs, MHERs could also be scheduled for restoration response. For example, Su \emph{et al.}\cite{Su2024} proposed a restoration strategy for hydrogen-accommodated microgrids based on a two-stage stochastic programming framework, where the location allocation of mobile wind turbines is decided in the first stage according to the shortest-path information in the transportation network considering uncertainties in wind generation. Then, the expected costs of power outages will be minimized by scheduling mobile wind turbines and HESSs in microgrids jointly. In \cite{ZhaoY2023}, Zhao \emph{et al.} proposed a two-level restoration strategy for a distribution network with HRSs. As shown in Fig.~\ref{fig_13}, the distribution system operator schedules mobile energy sources (i.e., HTTs and HFCVs) to minimize the power imbalance among microgrids and reduce the unserved load in the upper level. Then, based on the allocation results, a microgrid energy management strategy is designed to minimize the system operation cost of each microgrid. Simulation results showed that the coordination of mobile energy sources and HFCVs can reduce the power imbalance and improve the self-healing ability of hydrogen-electricity systems. Different from above works, Wang \emph{et al.}\cite{Wangz2021} proposed a restoration strategy for an integrated power and hydrogen distribution system with the consideration of power line and hydrogen pipeline outages caused by extreme events. When power outages occur, reconfigurations of transmission line and hydrogen pipeline are first adopted to avoid islanding. Then, repair crews are dispatched to repair the fault lines and hydrogen pipelines. In addition, mobile battery-carried vehicles are dispatched to cooperate with repair crews to restore critical loads.

\begin{figure}[!htb]
\centering
\includegraphics[scale=0.38]{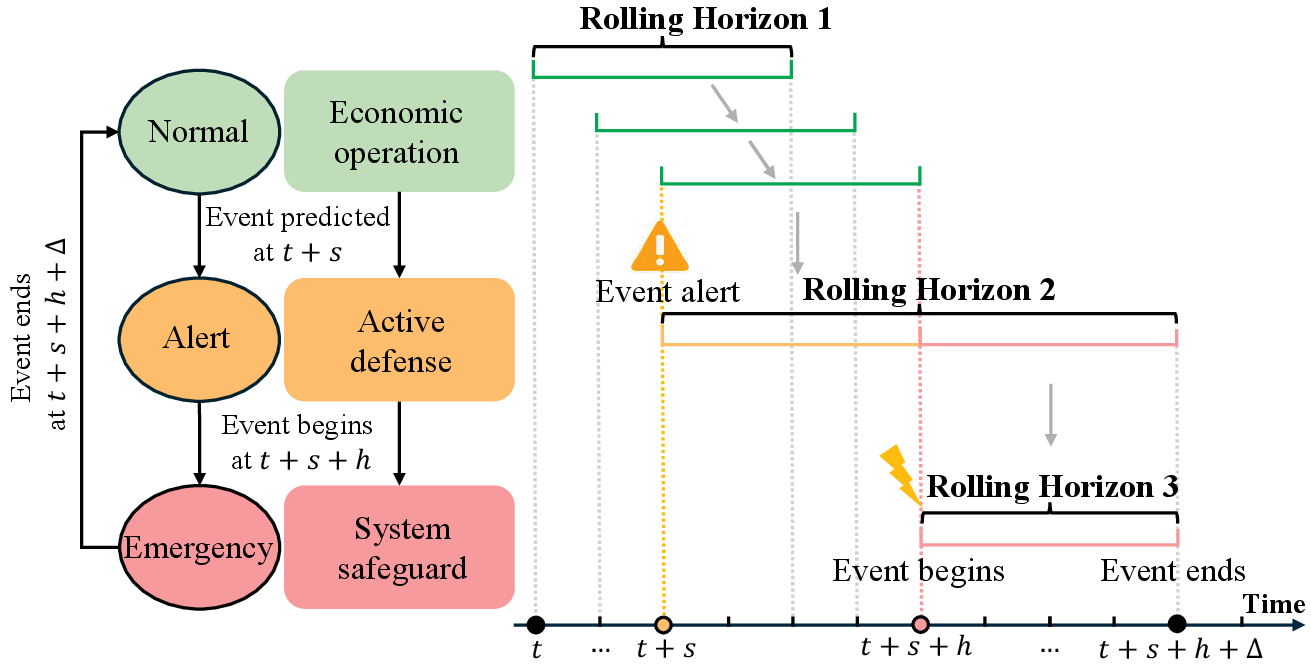}
\caption{The framework of the rolling dispatch method with an alert mechanism\cite{WangYuze2024}}
\label{fig_14}
\end{figure}

\subsection{Multi-stage operation strategy}\label{s62}
Although the above single-stage operation strategies can enhance HMES resilience, they cannot implement the coordination between different stages. To this end, some multi-stage operation strategies for HMESs have been proposed. For example, Haggi \emph{et al.} proposed a proactive rolling horizon optimization-based method for hydrogen systems to enhance the power system resilience\cite{Haggi2022}, which can implement the seamless switching between normal and emergency modes. Under normal mode, hydrogen systems assist the power grid by acting as a load. Once the extreme event time is known based on the forecast, the hydrogen tank is filled to make preparation for emergency mode. In \cite{Liu2021}, Liu \emph{et al.} developed a risk-averse receding horizon optimization method to support the proactive and emergency operation of the industrial park based on integrated hydrogen-electricity-heat microgrids, where the measure of proactive operation includes increasing the reserve capacity of FCs and that of emergency operation is to reschedule components for the electricity and heat load survivability. In \cite{WangYuze2024}, Wang \emph{et al.} proposed a rolling dispatch method to enhance the resilience of hydrogen-penetrated Antarctic energy systems under extreme weather. The framework of the rolling dispatch method can be observed in Fig.~\ref{fig_14}, where three stages could be identified, i.e., normal scheduling, alert scheduling, and emergency scheduling. Moreover, the above-mentioned stages will transfer from one to another according to different event states. To be specific, when the extreme event information is known based on forecasting techniques at slot $t+s$, the normal scheduling stage with horizon length $h$ will transfer to the alert scheduling stage with rolling horizon length $h+\tau$. Here, $\tau$ is the extreme event duration. When the event beginning is detected, the alert scheduling stage will transfer to the emergency scheduling stage with a rolling horizon length $\tau$. Once the event ending is detected, the alert scheduling stage will transfer to the normal scheduling stage. Since the function of alert scheduling is similar to that of preventive response (i.e., making preparation for emergency response), their roles are regarded as equivalent in Table~\ref{table_10}. Similar rolling horizon optimization-based methods considering economic operation, proactive preparation, and emergency operation can be observed in \cite{Huangchunjun2023}.

\begin{table*}[htbp]
\center
\setlength{\arrayrulewidth}{1pt}
\caption{\textcolor[rgb]{0.0,0.0,0.00}{Summary of Operation Enhancement for HMES (Multi-stage)}}\label{table_11} \centering
\rowcolors{2}{yellow!12}{yellow!12}
\begin{tabular}{|m{1.0cm}<{\centering}|m{2.4cm}<{\centering}|m{2.7cm}<{\centering}|m{2.1cm}<{\centering}|m{1.5cm}<{\centering}|m{3.5cm}<{\centering}|m{1.8cm}<{\centering}|}
\hline
\rowcolor{yellow!25}
\textbf{Ref.}&\textcolor[rgb]{0.0,0.0,0.00}{\textbf{Parameter/event uncertainty}}&\textcolor[rgb]{0.0,0.0,0.00}{\textbf{Contingency uncertainty}} & \textcolor[rgb]{0.0,0.0,0.00}{\textbf{Uncertainty processing methods}} & \textbf{Operation stages} & \textbf{Problem types and solving methods} &\textbf{Hydrogen-related measure types}\\
\hline
\hline
\cite{Liu2021}& \textcolor[rgb]{0.0,0.0,0.00}{$\times$} & \textcolor[rgb]{0.0,0.0,0.00}{$N-k$ line contingencies} & \textcolor[rgb]{0.0,0.0,0.00}{Scenarios} & Preventive, emergency  & Stochastic MISOCP; Receding horizon method & A, D\\
\hline
\cite{WangYuze2024}& \textcolor[rgb]{0.0,0.0,0.00}{Renewable energy} & \textcolor[rgb]{0.0,0.0,0.00}{Outage of renewable energy units} & \textcolor[rgb]{0.0,0.0,0.00}{Box uncertainty set} & Preventive, emergency & MILP; Gurobi & A, D \\
\hline
\cite{Huangchunjun2023}& \textcolor[rgb]{0.0,0.0,0.00}{Extreme event starting time}  & \textcolor[rgb]{0.0,0.0,0.00}{Deterministic wind generation curtailment} & \textcolor[rgb]{0.0,0.0,0.00}{Scenarios} & Preventive, emergency & Rolling-horizon MILP; Gurobi & A, D \\
\hline
\cite{Cai2023}& \textcolor[rgb]{0.0,0.0,0.00}{$\times$} & \textcolor[rgb]{0.0,0.0,0.00}{Failed transmission lines, their failure, and repair times}  & \textcolor[rgb]{0.0,0.0,0.00}{Scenarios} &Preventive, emergency & Stochastic multi-objective MILP & D \\
\hline
\cite{ChenF2024} & \textcolor[rgb]{0.0,0.0,0.00}{$\times$} & \textcolor[rgb]{0.0,0.0,0.00}{Component outages} & \textcolor[rgb]{0.0,0.0,0.00}{Scenarios} & Preventive, emergency & MILP; Gurobi & A, D \\
\hline
\cite{Yang2024}& \textcolor[rgb]{0.0,0.0,0.00}{$\times$} & \textcolor[rgb]{0.0,0.0,0.00}{Line faults} & \textcolor[rgb]{0.0,0.0,0.00}{Uncertainty sets}  & Preventive, emergency & Two stage robust optimization; C\&CG
 & B, F\\
\hline
\cite{Jordehi2024}& \textcolor[rgb]{0.0,0.0,0.00}{$\times$} & \textcolor[rgb]{0.0,0.0,0.00}{Line outages} & \textcolor[rgb]{0.0,0.0,0.00}{Scenarios}  & Preventive, emergency & Stochastic MILP & C, D \\
\hline
\cite{Cao2023}& \textcolor[rgb]{0.0,0.0,0.00}{$\times$}  & \textcolor[rgb]{0.0,0.0,0.00}{$N-k$ line contingencies} & \textcolor[rgb]{0.0,0.0,0.00}{Uncertainty sets}
 &Preventive, emergency & Tri-level MILP; Nested C\&CG  & C, F \\
\hline
\cite{ZouX2024}& \textcolor[rgb]{0.0,0.0,0.00}{Load, renewable energy, and travelling time} &  \textcolor[rgb]{0.0,0.0,0.00}{Deterministic grid outages} & \textcolor[rgb]{0.0,0.0,0.00}{Scenarios} & Preventive, restoration & Stochastic MILP; Gurobi & D\\
\hline
\cite{Tang2022}& \textcolor[rgb]{0.0,0.0,0.00}{$\times$} & \textcolor[rgb]{0.0,0.0,0.00}{Lines/component failure} & \textcolor[rgb]{0.0,0.0,0.00}{Scenarios} & Preventive, restoration & Stochastic multi-objective programming; NSGA-II & C, J\\
\hline
\cite{Xie2024}& \textcolor[rgb]{0.0,0.0,0.00}{$\times$} & \textcolor[rgb]{0.0,0.0,0.00}{Deterministic line outage} & \textcolor[rgb]{0.0,0.0,0.00}{$\times$} & Preventive, restoration  & MISOCP; CPLEX & A, J, K \\
\hline
\cite{LiL2025} & \textcolor[rgb]{0.0,0.0,0.00}{$\times$} & \textcolor[rgb]{0.0,0.0,0.00}{Deterministic grid outage} & \textcolor[rgb]{0.0,0.0,0.00}{$\times$} & Emergency, restoration & MILP; CPLEX & G, J\\
\hline
\cite{ZhangP2024} & \textcolor[rgb]{0.0,0.0,0.00}{Renewable generation, load demands, EV charging level} & \textcolor[rgb]{0.0,0.0,0.00}{Deterministic line outage}  & \textcolor[rgb]{0.0,0.0,0.00}{Scenarios} & Emergency, restoration & Bi-level MILP; Adaptive ADMM algorithm & G, J\\
\hline
\cite{ChenJ2025} & \textcolor[rgb]{0.0,0.0,0.00}{$\times$} & \textcolor[rgb]{0.0,0.0,0.00}{Deterministic line outages and insufficient gas supply}  & \textcolor[rgb]{0.0,0.0,0.00}{$\times$} & Emergency, restoration & Stochastic MIQCP; Augmented ADMM algorithm & F, D, J\\
\hline
\end{tabular}
\end{table*}

\begin{table*}[htbp]
\centering
\setlength{\arrayrulewidth}{1pt}
\caption{Representative Hydrogen-related Operation Measures for HMES resilience enhancement}
\label{table_12}
\renewcommand{\arraystretch}{1.15}
\setlength{\tabcolsep}{4pt}

\begin{tabular}{|>{\centering\arraybackslash}m{1.7cm}|>{\centering\arraybackslash}m{1.2cm}|>{\centering\arraybackslash}m{4.1cm}|>{\centering\arraybackslash}m{10cm}|}
\hline
\rowcolor{yellow!25}
\textbf{Stages} & \textbf{Measure types} & \textbf{Hydrogen-related measures} & \textbf{Examples of specific measures} \\
\hline

\cellcolor{yellow!12}
& \cellcolor{yellow!12}Type-A
& \cellcolor{yellow!12}HESS pre-charging
& \cellcolor{yellow!12}Increasing HESS level for reserve\cite{Haggi2022}\cite{Xie2024}\cite{WangYuze2024}\cite{Shahbazbegian2023,Sharifpour2023,Liu2021}\cite{ChenF2024} \\
\cline{2-4}

\cellcolor{yellow!12}Preventive
& \cellcolor{yellow!12}Type-B
& \cellcolor{yellow!12}HTT scheduling
& \cellcolor{yellow!12}HRS-based microgrid formation assisted by transport-based measures (e.g., HTTs and mobile hydrogen trucks) \cite{ZhaoY2024}, pre-disaster HTT scheduling\cite{Yang2024} \\
\cline{2-4}

\cellcolor{yellow!12}
& \cellcolor{yellow!12}Type-C
& \cellcolor{yellow!12}MHER location allocation
& \cellcolor{yellow!12}Location allocation for HTTs\cite{Cao2023}, sizing and locating for HFCVs\cite{Tang2022} \\
\hline

\cellcolor{yellow!12}
& \cellcolor{yellow!12}Type-D
& \cellcolor{yellow!12}HESS discharging
& \cellcolor{yellow!12}Hydrogen tank-driven fuel cell for independent power supply during outages\cite{CicekA2025}, discharging batteries and HESS\cite{Huangchunjun2023}, HESS and CHP work at maximum power\cite{WangYuze2024}, discharging HESS for serving HFCVs\cite{ChenF2024}, discharging HESS for power supply\cite{Liu2021} \\
\cline{2-4}

\cellcolor{yellow!12}
& \cellcolor{yellow!12}Type-E
& \cellcolor{yellow!12}Natural gas/hydrogen pipeline-assisted energy supply
& \cellcolor{yellow!12}Utilizing hydrogen enriched natural gas for fuel support\cite{ZhuR2024}, using pipelines as large-scale energy storage\cite{GuZhong2025}, using hydrogen pipeline for power support\cite{ChangS2025} \\
\cline{2-4}

\cellcolor{yellow!12}Emergency
& \cellcolor{yellow!12}Type-F
& \cellcolor{yellow!12}HTT/hydrogen truck scheduling
& \cellcolor{yellow!12}Using mobile hydrogen trucks for critical load support\cite{LiB2023}\cite{QianH2024}, fuel cell power supply supported by hydrogen from HTTs\cite{Yang2024}, dynamic re-routing of HTTs\cite{Cao2023} \\
\cline{2-4}

\cellcolor{yellow!12}
& \cellcolor{yellow!12}Type-G
& \cellcolor{yellow!12}Hydrogen demand adjustment
& \cellcolor{yellow!12}Shedding non-critical hydrogen demand\cite{GuZhong2025}, restoring the integrity of compromised hydrogen demand data \cite{RiziD2025}, HRSs adjust the charging schedule of HFCVs\cite{ZhangP2024}\cite{LiL2025} \\
\cline{2-4}

\cellcolor{yellow!12}
& \cellcolor{yellow!12}Type-H
& \cellcolor{yellow!12}Hydrogen transformation
& \cellcolor{yellow!12}Using hydrogen generated from renewables for synthetic natural gas production to satisfy demand\cite{MehrjerdiH2022} \\
\hline

\cellcolor{yellow!12}
& \cellcolor{yellow!12}Type-I
& \cellcolor{yellow!12}Mobile hydrogen storage
& \cellcolor{yellow!12}Combined operation of mobile wind turbines and HESSs\cite{Su2024} \\
\cline{2-4}

\cellcolor{yellow!12}
& \cellcolor{yellow!12}Type-J
& \cellcolor{yellow!12}HFCV scheduling
& \cellcolor{yellow!12}HFCV routing and dispatch\cite{Xie2024}\cite{ZhaoY2023}\cite{Tang2022}, HFCV for power support\cite{Xie2024} \\
\cline{2-4}

\cellcolor{yellow!12}Restoration
& \cellcolor{yellow!12}Type-K
& \cellcolor{yellow!12}HESS scheduling
& \cellcolor{yellow!12}Scheduling HESS with other components\cite{ZouX2024}, HESS discharging\cite{Afsari2024}\cite{Xie2024} \\
\cline{2-4}

\cellcolor{yellow!12}
& \cellcolor{yellow!12}Type-L
& \cellcolor{yellow!12}Hydrogen fuel cell scheduling
& \cellcolor{yellow!12}Hydrogen fuel cells as black-start resources\cite{LuJ2025} \\
\cline{2-4}

\cellcolor{yellow!12}
& \cellcolor{yellow!12}Type-M
& \cellcolor{yellow!12}Hydrogen pipeline scheduling
& \cellcolor{yellow!12}Hydrogen pipeline reconfiguration and repair\cite{Wangz2021} \\
\hline
\end{tabular}
\end{table*}

\begin{figure}[!htb]
\centering
\includegraphics[scale=0.4]{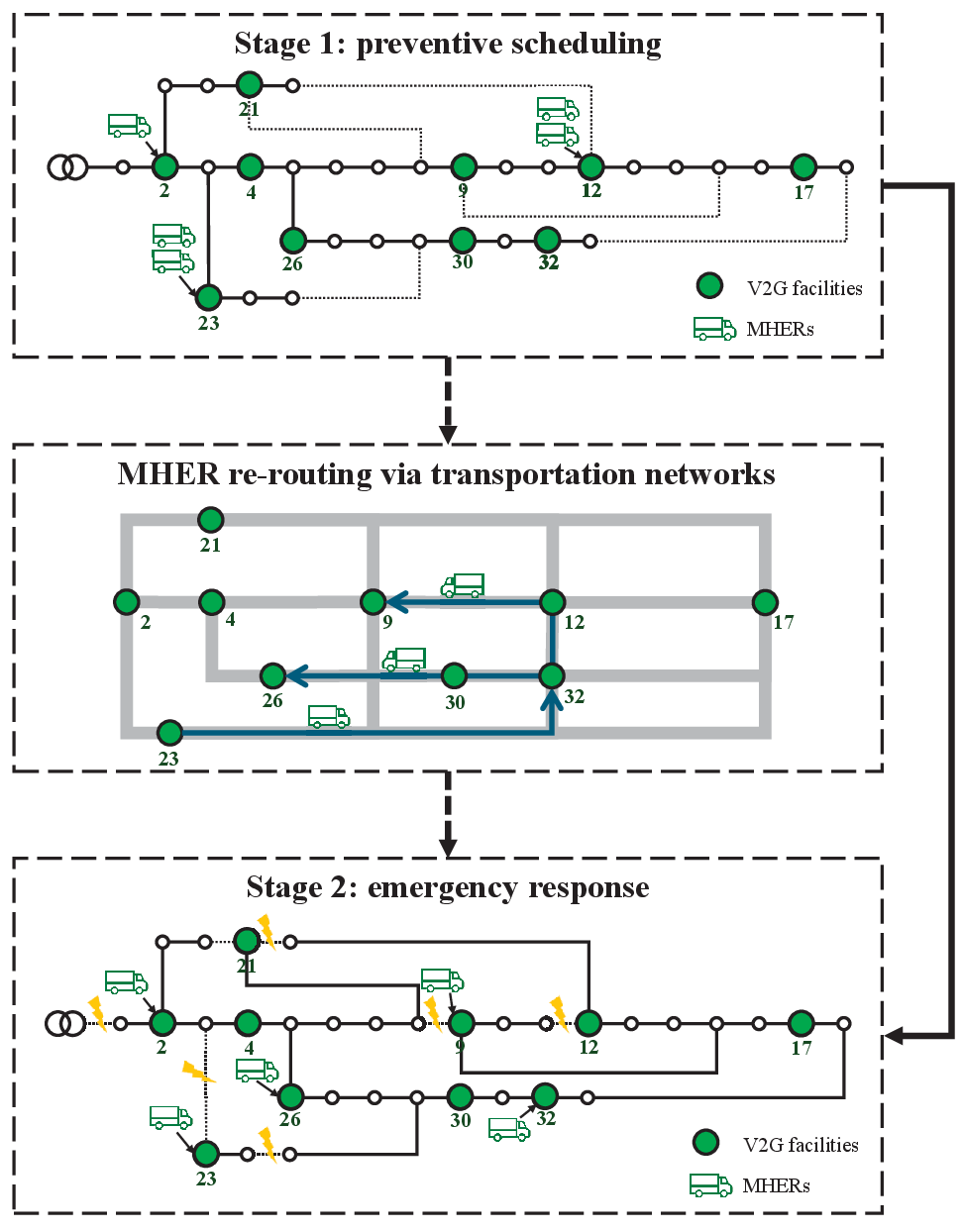}
\caption{The framework of resilient scheduling for electricity-hydrogen distribution networks\cite{Cao2023}}
\label{fig_15}
\end{figure}

Different from the above works, some studies considered the use of mobile energy sources for multi-stage operation. For example, Cai \emph{et al.}\cite{Cai2023} proposed a preventive response strategy to decide the location of mobile batteries in hydrogen fuel station-integrated power systems considering the uncertainties of failed transmission lines, their failure, and repair times. In \cite{ChenF2024}, Chen \emph{et al.} proposed a two-stage resilience enhancement method against the freezing rain disaster for the highway transportation energy system. In the first stage, preventive response is conducted to pre-charge storage systems (including heating ESS, cooling ESS, electric ESS, and HESS) so that stored energy can be discharged to satisfy multiple demands during both emergency and response stages. The optimization objective in this stage is to minimize economic operation costs and maximize the stored energy in all kinds of ESSs. In the second stage, emergency response is adopted to minimize multi-energy load shedding amount and operation cost. In \cite{Yang2024}, Yang \emph{et al.} proposed a two-stage robust operation strategy to decide the location and hydrogen weight of HTTs considering the uncertainties in line outages. Once the uncertain contingency is discovered, HTTs and HRSs will be scheduled to minimize load shedding of electricity and hydrogen demand. In \cite{Jordehi2024}, Jordehi \emph{et al.} considered using mobile energy sources and demand response programs to improve the resilience of the power system under extreme events. To be specific, the location of mobile energy sources was first decided considering the uncertainties in damaged lines, hurricane time, and repair time. Then, the joint scheduling of mobile energy sources, demand response programs, and energy hubs was conducted. In \cite{Cao2023}, Cao \emph{et al.} investigated a resilient scheduling problem for electricity-hydrogen distribution networks considering line fault uncertainties. Then, a tri-level optimization-based scheduling strategy was proposed. The key idea of the proposed operation strategy is shown in Fig.~\ref{fig_15}, where two stages could be identified. In the preventive response stage, the locations of MHERs at different electricity-hydrogen nodes are decided. Then, the rerouting of MHERs with the help of the transportation network and network reconfiguration under the worst-case damage scenario are optimized jointly in the emergency response stage. Given guaranteed load survivability, the proposed scheduling strategy effectively reduces unit reinforcement cost. In addition to the combination of preventive response and emergency response, some studies considered the coordination of preventive response and restoration response. For example, \cite{ZouX2024} \emph{et al.} investigated a resilience enhancement problem of an electricity-transport coupled system considering power outages caused by extreme events under the uncertainties in load, renewable energy, and electric vehicle commuting time. In the preventive stage, a pre-location of electric vehicle was conducted. In the restoration stage, the scheduling of electric vehicles and HESS was implemented. Simulation results showed that the proposed approach offers benefits in aspects of resilience enhancement and supplying critical loads. Similarly, Tang \emph{et al.}\cite{Tang2022} proposed a sizing and location allocation strategy for electric emergency power supply vehicles and hydrogen fuel cell emergency power supply vehicles so as to achieve a balance between resilience and economics. In \cite{Xie2024}, Xie \emph{et al.} proposed a method to enhance power distribution network resilience based on HESSs and HFCVs. In the preventive stage, pre-deployment of HESSs and HFCVs within the electric-road network system was conducted. In the restoration stage, HESSs and HFCVs are scheduled to minimize load power losses and economic costs. Simulation results indicated that network resilience and recovery performance can be enhanced by adopting hydrogen technologies. In order to enhance the resilience of system resilience after natural disasters, Li \emph{et al.} developed a two-layer economic resilience model for distribution network. Specifically, mobile battery storage units are considered and HRSs adjusted the charging schedule of HFCVs during emergency response. In the restoration response, hydrogen fuel cell trucks were dispatched to connection points for electricity supply. Similar measures could be found in \cite{ZhangP2024}. In \cite{ChenJ2025}, Chen \emph{et al.} proposed a self-healing approach for coupled electricity-gas-heat-hydrogen networks, which consists of two parts, i.e., hydrogen fuel cell trucks with storage were deployed to increase hydrogen energy storage level under emergency conditions, and such trucks were dispatched to connection points for electricity supply in restoration response. Simulation results indicated that the proposed approach can harness maximum capacities from smart prosumers for reducing load shedding and improving the infrastructure resilience.

\emph{Summary}: For easy understanding of existing works related to resilient operation for HMESs, we summarize their characteristics in Table~\ref{table_10} and Table~\ref{table_11}, and list hydrogen-related measures in each response stage in Table~\ref{table_12}. It can be observed that hydrogen plays an important role in every response stage, such as pre-charging HESS in the preventive response stage, hydrogen for fuel support via natural gas pipeline in the emergency response stage, and scheduling MHERs in the restoration response stage. Moreover, existing works mainly focus on single-stage or two-stage operation methods. Few of them consider the coordination operation among three kinds of response stages. In addition, hydrogen pipeline-related operational measures in emergency response and restoration response could be observed. The reason is that hydrogen pipelines can serve as large-scale energy storage solutions and play a key role in resilience enhancement. As mentioned in existing work\cite{GuZhong2025},\cite{FanG2025}, the pressure of a hydrogen pipeline can be adjusted dynamically by changing the quantity of hydrogen within it. By exploiting such elasticity, many resilience enhancement measures related to hydrogen pipelines could be adopted. For example, MESs can temporarily store hydrogen in the pipeline when there is surplus renewable energy but limited hydrogen tank space. When an extreme event happens, the hydrogen pipeline can release hydrogen to drive fuel cells for energy supply. Such linepack utilization is helpful in enhancing resilience of MESs\cite{Bah2026}. Moreover, during extreme events (such as main grid outages or natural gas supply interruptions), hydrogen pipelines can be directionally dispatched to recover hydrogen supply to critical loads such as hospitals and communication base stations. Finally, most contingencies in hydrogen-enabled resilience enhancement for MESs in existing works are related to main grid power outages, component outages, renewable generation reduction, and load surge, cyber-attack related contingencies should be paid more attention, which are also very common in cyber-physical energy systems\cite{YaoP2025,LiJ2025,FengY2025,LiuF2025}.

\section{Research Gaps and Future Directions}\label{s7}
Although some advances have been made in existing works, several important research gaps are identified and possible directions are provided below.

\subsection{Comprehensive Design of HMES Resilience Metrics}
Current resilience metrics in HMESs are often borrowed from power systems and MESs, such as LSR, ELNS, ERL, and WSLL. These metrics are primarily designed to quantify the extent of unmet energy demand in electricity, heating, and cooling during and after extreme events. However, these metrics fail to capture physical characteristics and functional complexities of hydrogen systems, such as hydrogen availability and sector-coupled interactions. To be specific, hydrogen availability is not binary and it is strongly influenced by pressure levels within pipelines and storage units. For example, fuel cells require hydrogen at a minimum operating pressure (e.g., $>$1.5 bar). Thus, insufficient pressure, even with sufficient quantity of hydrogen present, leads to functional failure. Consequently, such ``hidden" functional failure may result in serious social impacts under some extreme events, e.g., the failed operation of life-support equipment in an emergency hospital supported by hydrogen fuel cells may result in the loss of many patients' lives. Additionally, HMESs rely on sector-coupled interactions (e.g., excess electricity is used to generate hydrogen via electrolysis, which is later used for electricity generation or heating). A failure in the hydrogen link breaks this coupling, degrading system functionality without necessarily causing measurable load loss. Therefore, existing HMES resilience metrics should be extended to incorporate hydrogen-specific attributes. A possible solution is to develop comprehensive resilience metrics by considering hydrogen-specific attributes (e.g., hydrogen system availability ratio and pressure recovery time), load loss, the number of customers affected, and carbon resilience jointly.

\subsection{Extreme-event Scenario Generation for HMES Resilience Planning}
\textcolor[rgb]{0.0,0.0,0.00}{In existing studies on HMES resilience planning, only a limited number of works have explicitly considered extreme-event uncertainty. Among them, extreme-event scenario generation has mainly been conducted using generative adversarial networks (GANs) \cite{WenZ2024} and climate- and weather-model-driven methods \cite{Qu2024}. GAN-based approaches are attractive because they can learn high-dimensional spatio-temporal patterns from data and generate large numbers of synthetic extreme scenarios, which is particularly useful for HMESs where prolonged renewable deficits may coincide with heating or cooling stress. Climate- and weather-model-driven methods, by contrast, can provide physically grounded realizations of persistent meteorological extremes, such as prolonged heatwaves, droughts, and renewable-energy droughts, which are often insufficiently represented in historical observations. Despite these advantages, both approaches have important limitations, as summarized in Table~\ref{tab_extreme_scenario_methods}. In particular, GAN-based methods are typically data-hungry and may generate physically implausible scenarios in the absence of sufficient domain constraints. Climate- and weather-model-driven methods, on the other hand, suffer from high computational burden, the need for downscaling and bias correction, the difficulty of translating raw climate outputs into planning-ready scenarios for HMESs, and structural uncertainty arising from imperfect or simplified model formulations. More importantly, these limitations are not confined to the above two categories, but are also relevant to other extreme-event scenario generation methods listed in Table~\ref{tab_extreme_scenario_methods}. Among them, one of the most critical unresolved issues is extreme-data scarcity. By definition, severe extreme events are rare, while compound extremes are even less frequently observed. As a result, the available data record is often too short to support robust characterization of long-duration, spatially correlated, and concurrent extreme events. This challenge is particularly pronounced for HMESs, because the resilience value of hydrogen is expected to emerge precisely under low-frequency but high-impact events, whereas such conditions are poorly represented in ordinary operational datasets. Therefore, addressing extreme-data scarcity is a key prerequisite for more credible resilience-oriented planning. Several promising directions may help address this challenge. First, extreme-value-theory-assisted deep generative models offer a valuable hybrid framework, in which deep generative models capture high-dimensional dependence structures while extreme value theory improves the representation and extrapolation of tail behavior \cite{GuJ2025}. Second, physics-informed generative models provide a way to enhance the physical plausibility of synthetic extreme scenarios by embedding conservation laws, known meteorological drivers, or other domain constraints into the generation process \cite{GuJ2025}. Third, response-aware scenario generation methods can identify representative climate scenarios according to system response, thereby reducing scenario complexity while preserving the most critical stress patterns for resilience-oriented planning \cite{DeF2025}. These directions suggest that future extreme-event scenario generation for HMES resilience planning should move beyond purely data-fitting approaches toward hybrid frameworks that are tail-aware, physically consistent, and decision-relevant.}

\subsection{Multi-type Temporal-spatial Cyber-physical Contingency Modeling under Compound Extreme Events}
Most existing event-oriented contingency models focus on static, single-type faults, e.g., main grid power outages, simultaneous line failures, and simultaneous device failures. However, due to the coupling features of MESs, multi-type faults are more common under extreme events, e.g., simultaneous electrical network outages, natural gas network outages, and renewable generator failures\cite{Mehrjerdi2021}. In addition, with the progress of extreme events, the component failure probabilities may also change, e.g., with the increase of ice thickness, component failure probabilities under different time intervals are different\cite{ChenF2024}. Consequently, sequential multi-type failures may happen. Similarly, spatial dynamics of contingencies incurred by a hurricane could be observed in practice\cite{LiB2023}. Furthermore, when compound extreme events happen (e.g., typhoons, earthquakes, freezing ice disasters, tropical-cyclone-blackout-heatwave hazards, and cyber attacks)\cite{XuLuo2024}, the resulting temporal and spatial dynamics of faults are more complex\cite{WangH2022}. Above all, hydrogen-specific cyber vulnerabilities (e.g., false data injection attacks) are seldom considered.
Thus, multi-type temporal-spatial cyber-physical contingency modeling under compound extreme events needs further investigation. A possible solution for such modeling is to use spatially-aware dynamic Bayesian networks\cite{Yodo2017}\cite{HuY2024} and Petri-net\cite{BaiK2025} since the former can capture the probabilistic dependencies of spatially-correlated component states under extreme events and the latter can capture the MES operation logic related to energy state and energy transition.

\subsection{Multi-network Multi-timescale Coordination Planning and Operation}
Most existing works mainly focus on enhancing the resilience of HMESs related to electricity demand and hydrogen demand. However, two aspects are neglected. Firstly, multi-timescale feature existing in electric-hydrogen systems is not considered. To be specific, slow dynamics of hydrogen systems and the fast response of electric systems are coupled with each other. For instance, hydrogen pipelines exhibit response delays and they can not respond instantly to changes in flow or pressure, i.e., the pipeline system does not reach a new equilibrium immediately after a disturbance, and it takes time for changes in input or demand to affect the system's operation. Such delay may span from minutes to hours. In addition to slow dynamics, hydrogen pipelines can be used to store hydrogen temporarily due to the compressibility of hydrogen, which can help to absorb fluctuations in renewable energy generation or load demand without the immediate need for additional storage tanks. Compared with hydrogen tanks/pipelines, large-scale hydrogen energy storage systems (e.g., salt-cavern hydrogen storage) have slower dynamics, which are used as SHSs to adjust the weekly/monthly source-load imbalance\cite{WangY2025}. In contrast, to capture the minute-level fluctuations of renewable energy and load demand, electricity system could be scheduled on a minute timescale. Thus, multi-timescale modeling should be considered in the planning and operation of hydrogen-enabled MESs. Secondly, the same attention should be paid to providing heat demand and gas demand with high service quality under extreme events, since such demands may exceed 50\% of total energy demand in a country\cite{Ameli2024}, e.g., almost half of the energy demand in the U.K. is attributed to heat demand. When HMESs with versatile energy demands are considered, more strong couplings of power/gas/heat networks would arise. To manage such systems of systems efficiently, multi-network coordination is required. since enhancing the HMES resilience requires the use of many MHERs (e.g., HTTs, HFCVs, and HFCBs) and other mobile electric energy resources (e.g., mobile wind turbines and electric vehicles), four-network (i.e., power/gas/heat/transportation networks) coordination in planning and operation deserves further investigation. To support multi-timescale coordination operation, a possible solution is to utilize hierarchical MPC or hierarchical deep reinforcement learning for multi-timescale coordination operation, e.g., deciding the quantity of MHERs at day timescale, scheduling MHERs in emergency response at the hourly timescale, and scheduling electric equipment at the 5-minute timescale.


\subsection{Low-carbon and Resilient Planning and Operation}
As highlighted in \cite{Panteli2017,JiaQ2025,Evro2024}, achieving both low-carbon development and resilience is essential for the future of energy networks, including HMESs discussed in this paper. However, most existing studies on HMES resilience enhancement have largely overlooked the low-carbon objective. For example, in \cite{ZhaoH2022}, carbon emission costs were included in the objective function of a planning model for a hydrogen-integrated airport energy system, helping to reduce system-level emissions. Despite such efforts, focusing solely on low-carbon strategies is insufficient, as the ultimate target for future energy systems is to achieve net-zero carbon emissions \cite{IEA2024,ZhangS2024}. Therefore, future planning and operational strategies for HMESs must jointly address both net-zero carbon targets and system resilience. To tackle this dual-objective challenge under uncertainty, advanced artificial intelligence methods can be employed. One promising solution is multi-objective deep reinforcement learning \cite{Nguyen2020,WangZ2023,Coskun2024,HeQing2023}, which enables coordinated decision-making across multiple objectives, timescales, and stages. Another solution is to use deep reinforcement learning-assisted evolutionary algorithms \cite{Wen2025,WuR2025}, where deep reinforcement learning helps dynamically control the crossover probability between individuals at different levels and the mutation intensity for individuals, thereby improving the quality of evolved offspring and enhancing global search performance.

\subsection{LLM-assisted Whole-Process Resilience Enhancement for HMESs}
LLMs (e.g., GPT-series models) have demonstrated remarkable capabilities in understanding, synthesizing, and reasoning across vast domains of unstructured and structured information\cite{Bi2023,Ding2024}. Their contextual awareness, semantic reasoning, and adaptability position them as promising tools for supporting the whole-process resilience enhancement of HMESs. Firstly, the current resilience metrics failed to capture the hydrogen-related attributes, e.g., hydrogen availability, tank/pipeline pressure recovery, and sector-coupled response. In contrast, LLMs can help extract and organize hydrogen-specific resilience attributes from technical literature, safety standards, policy documents, and incident reports, thereby supporting the development of more comprehensive resilience metrics for HMESs. Secondly, contingency modeling under extreme events in HMESs is hindered by the complexity of hydrogen-related subsystems. For example, during a typhoon event, transmission outages may coincide with road blockages, temporary natural gas shortages, and constrained hydrogen delivery, creating cascading impacts that are difficult to model using only predefined structured data. In this context, LLMs can assist contingency modeling by converting natural-language descriptions from operation manuals, fault logs, and emergency response documents into structured fault trees, event chains, and scenario libraries, which can improve the efficiency of constructing probabilistic or rule-based contingency models under diverse stress conditions. Thirdly, from a planning perspective, LLMs can support semantically grounded extreme-event scenario generation since their training are conducted based on massive textual corpora\cite{GuJ2025}. In addition, LLMs can support strategic siting of hydrogen-related assets by transforming multi-source textual knowledge into candidate rules, constraints, and evaluation criteria. For instance, an LLM may infer that a location close to wind power resources is economically attractive for electrolysis, but unsuitable for emergency hydrogen supply if it is far from hospitals, communication hubs, or major transport corridors. Finally, from an operational perspective, LLMs can support the dynamic dispatch of HMESs by integrating heterogeneous real-time information (e.g., sensor readings, traffic updates, weather alerts, and emergency protocols) and assisting decision-making on hydrogen allocation, mobile resource routing, and recovery prioritization. Therefore, rather than replacing physics-based models or optimization tools, LLMs can serve as an intelligent interface that connects fragmented knowledge, planning rules, and real-time operational information, thereby enhancing the resilience of HMESs across assessment, modeling, planning, and operation.

\section{Conclusions}\label{s8}
This article presents the first comprehensive review of hydrogen-enabled resilience enhancement for MESs from the perspective of planning and operation. Firstly, advantages and challenges of adopting hydrogen to enhance resilience of MESs are introduced. Then, it provides a comprehensive resilience enhancement framework for HMESs. Under the proposed framework, the widely used resilience metrics and event-oriented contingency models are summarized. Then, it classifies the hydrogen-enabled planning measures for resilience enhancement of HMESs according to the type of hydrogen-related facilities and summarizes \textcolor[rgb]{0.0,0.0,0.00}{the classifications of uncertainty processing methods and scenario generation methods}, as well as planning problem formulation frameworks. Next, it categorizes the hydrogen-enabled operation measures for the resilience enhancement of HMESs according to three types of operation response stages and summarizes the role of hydrogen in each response stage. Finally, several research gaps and future directions are identified, i.e., (1) Comprehensive HMES resilience metrics with the consideration of hydrogen-specific attributes (e.g., hydrogen system availability ratio and pressure recovery time), load loss, the number of customers affected, and carbon resilience should be designed; (2) \textcolor[rgb]{0.0,0.0,0.00}{Generating tail-extrapolation-aware, physical-informed, response-relevant extreme-event scenarios for HMES resilient planning based on extreme value theory-assisted deep generative models, physics-informed generative models, and response-aware scenario generation methods}; (3) Achieving multi-type temporal-spatial contingency modeling under compound extreme events based on spatially-aware dynamic Bayesian networks and Petri-net; (4) Implementing multi-network multi-timescale coordination planning and operation based on hierarchical MPC or hierarchical deep reinforcement learning; (5) Designing advanced artificial intelligence-based multi-objective optimization methods for the planning and operation of low-carbon and resilient HMES; (6) Adopting LLM-assisted whole-process resilience enhancement for HMESs in aspects of proper resilience metrics, event-oriented contingency modeling, scenario generation and smart decision in planning process, and MHER scheduling and routing in operation process. While the above-mentioned technical strategies are essential for enhancing the resilience of HMESs, business and market models also play a crucial role. Representative models include hydrogen-as-a-service, priority hydrogen supply contracts, and shared hydrogen pipeline infrastructure. For instance, by leasing hydrogen fuel cell trucks under a hydrogen-as-a-service model, MESs can gain black-start capabilities and rapidly restore critical loads during unexpected power outages without requiring large upfront investments. Likewise, priority hydrogen contracts can ensure uninterrupted supply to critical infrastructure (e.g., hospitals or water treatment plants) during extreme events. Moreover, due to the high capital cost of hydrogen pipeline infrastructure, shared investment models among multiple MES operators can promote cost-effective deployment while improving cross-system resilience.

\begin{IEEEbiography}[{\includegraphics[width=1in,height=1.25in,clip,keepaspectratio]{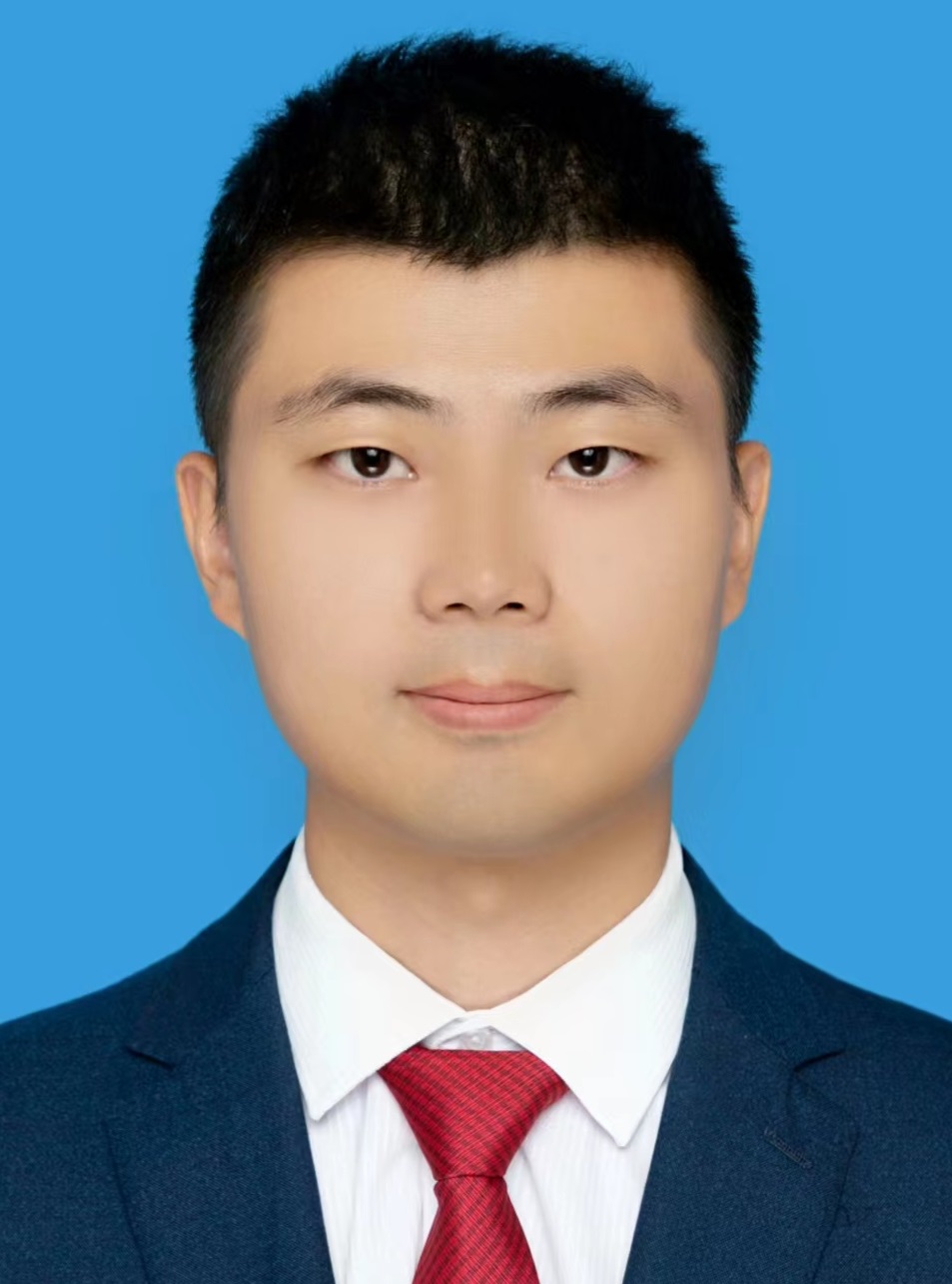}}]{Liang Yu} (Senior Member, IEEE) received the Ph.D. degree in Information and Communication Engineering from Huazhong University of Science and Technology, Wuhan, China, in 2014. Since 2022, he has been a Full Professor and Doctoral Supervisor with the College of Automation, Nanjing University of Posts and Telecommunications, Nanjing, China. From November 2024 to November 2025, he was a Visiting Researcher with Imperial College London, London, U.K. He serves/has served as an Associate Editor for IEEE Transactions on Smart Grid and IEEE Transactions on Industrial Informatics. He has published more than 30 papers in leading IEEE Transactions and Journals. His honors include the First Prize of the Natural Science Award in the 2025 Ministry of Education Outstanding Scientific Research Achievement Award (Natural Science and Engineering Technology), the IEEE Transactions on Smart Grid Best Paper Recognition in 2022, the Award for Hundred Excellent Academic Achievement Papers in Natural Science of Jiangsu Province in 2023, the Higher Education Science and Technology Research Achievement Award of Jiangsu Province in 2023, the Award for Excellent Papers in Natural Science of Nanjing City in 2018, and the Outstanding Doctoral Dissertation Award of Hubei Province in 2015. He also holds 31 authorized Chinese invention patents. His research interests include the planning and operation optimization of hydrogen-enabled multi-energy systems, deep reinforcement learning, and large language models.
\end{IEEEbiography}

\begin{IEEEbiography}[{\includegraphics[width=1in,height=1.25in,clip,keepaspectratio]{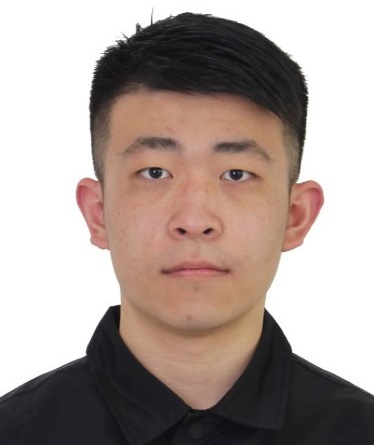}}]{Haoyu Fang} received the B.E. degree in Mechanical Engineering from North China Electric Power University, Baoding, China, in 2022, and the M.S. degree in Future Power Networks from Imperial College London, London, U.K., in 2023. He is currently pursuing the Ph.D. degree in Electrical Engineering at Imperial College London, London, U.K. since 2024. His current research interests include planning and operation of multi-energy systems for resilience enhancement, and distributed optimization.
\end{IEEEbiography}

\begin{IEEEbiography}[{\includegraphics[width=1in,height=1.25in,clip,keepaspectratio]{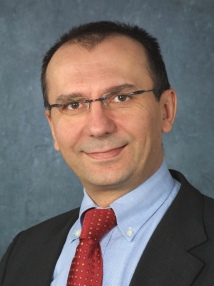}}]{Goran Strbac} (Fellow, IEEE) is Professor of Energy Systems at Imperial College London. He is the Leading Author in IPCC WG 3, member of the European Technology and Innovation Platform for Smart Networks for the Energy Transition, Member of the Joint EU Programme in Energy Systems Integration of the European Energy Research Alliance, Member of UK Energy Regulator Challenge Group, Member of the UK Smart System Forum. He led the development of fundamentally novel Integrated Whole Energy System modeling concept, Advanced Energy Market models for operation and design of low carbon energy systems, that provided critical evidence related to the importance of coordinated operation and planning of multi-vector energy systems, considering extremely granular time, from fractions of seconds to multiple years, while balancing synergies and conflicts between local district and national/international objectives, and achieving cost effective transition to secure, low/zero carbon energy future. He also led the development of advanced models for quantifying security and resilience of energy systems, that were used in the fundamental review of UK power network operation and design standards. He led and contributed to many major and impactful UK and international research projects including EU Horizon 2020. He co-authored 4 books and published over 500 technical papers.
\end{IEEEbiography}

\begin{IEEEbiography}[{\includegraphics[width=1in,height=1.25in,clip,keepaspectratio]{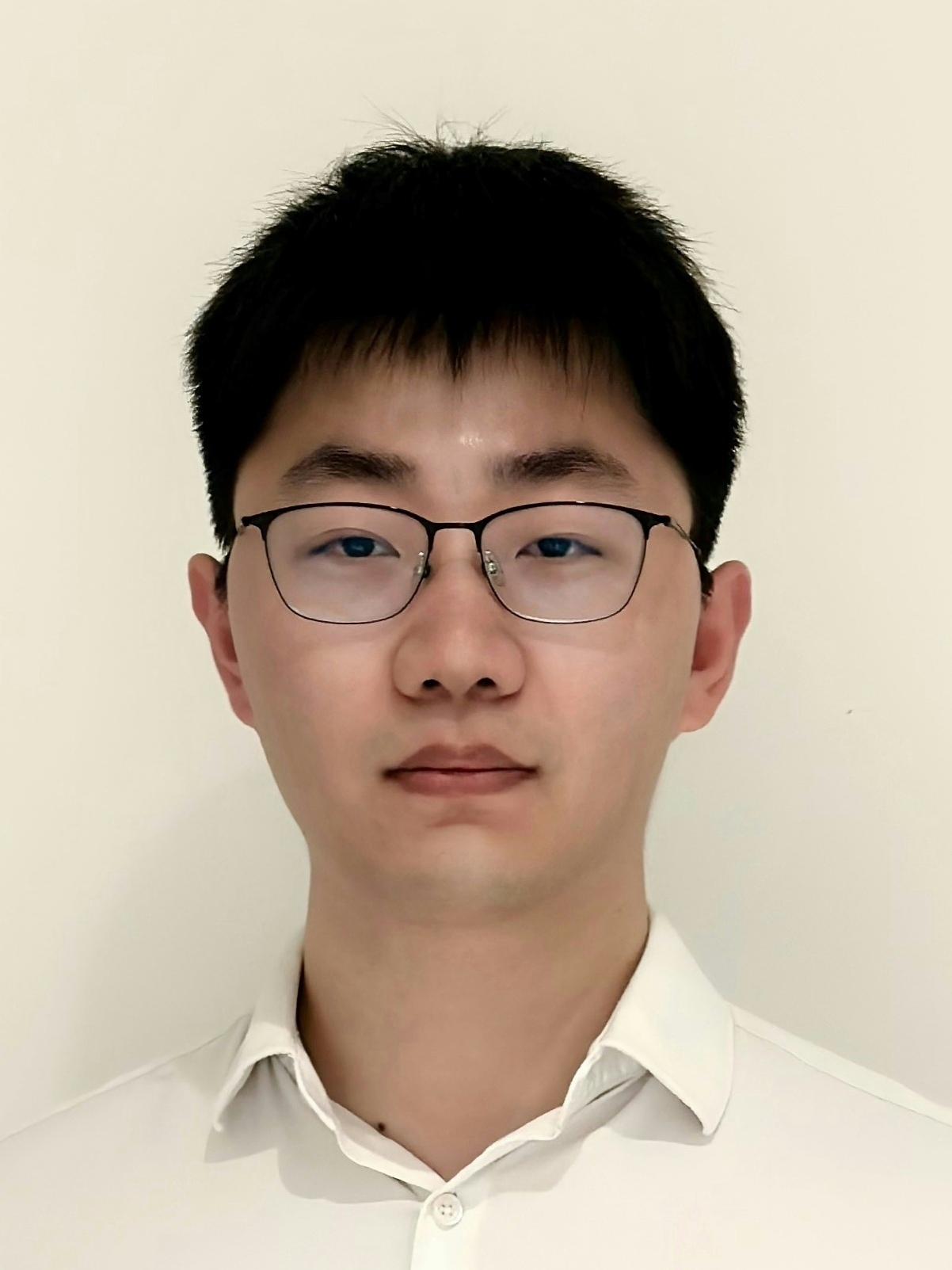}}]{Dawei Qiu} (Senior Member, IEEE) received the Ph.D. degree in Electrical Engineering from Imperial College London in 2020. After graduation, he was employed as a Research Associate and then promoted to be a Research Fellow in Market Design for Low-Carbon Energy Systems at Imperial College London. He is currently a Lecturer in Smart Energy Systems within the Department of Engineering at the University of Exeter, Exeter, U.K. He has published more than 20 papers in leading journals of energy field (e.g., Proceedings of the IEEE, IEEE TSG, IEEE TIA, IEEE TII, IEEE TPWRS, IEEE TSTE, IEEE TVT, and Applied Energy). His research interests include strategic behaviours in electricity markets, market design for peer-to-peer energy trading, and multi-energy system resilience enhancement.
\end{IEEEbiography}

\begin{IEEEbiography}[{\includegraphics[width=1in,height=1.25in,clip,keepaspectratio]{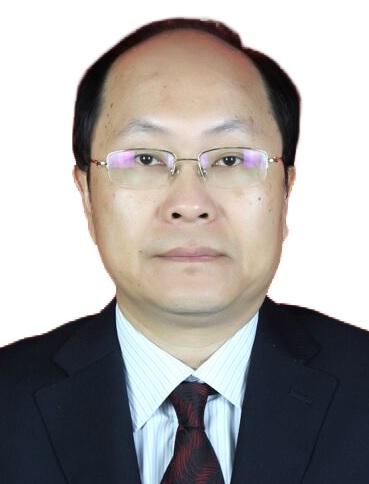}}]{Dong Yue} (Fellow, IEEE) received the Ph.D. degree from the South China University of Technology, Guangzhou, China, in 1995. He is currently a Full Professor and the Dean of the Institute of Advanced Technology for Carbon Neutrality, and Director of Academic Committee of the University at Nanjing University of Posts and Telecommunication. He was the recipient of 2022 IEEE Rudolf Chope Research and Development Award, Norbert Wiener Review Award by IEEE/CAA Journal of Automatica Sinica in 2020 and the Best paper award of IEEE Systems Journal in 2022. He has served as the Chair of IEEE IES Technical Committee on NCS and Applications. He has served as the Co-Editor-in-Chief for IEEE Transactions on Industrial Informatics and the Associate Editor for IEEE Industrial Electronics Magazine, IEEE Transactions on Systems, Man, and Cybernetics:Systems, IEEE Transactions on Industrial Informatics, IEEE Transactions on Neural Networks and Learning Systems, and Journal of the Franklin Institute. His current research interests include networked control, optimization, multiagent systems, and smart grid.
\end{IEEEbiography}

\begin{IEEEbiography}[{\includegraphics[width=1in,height=1.25in,clip,keepaspectratio]{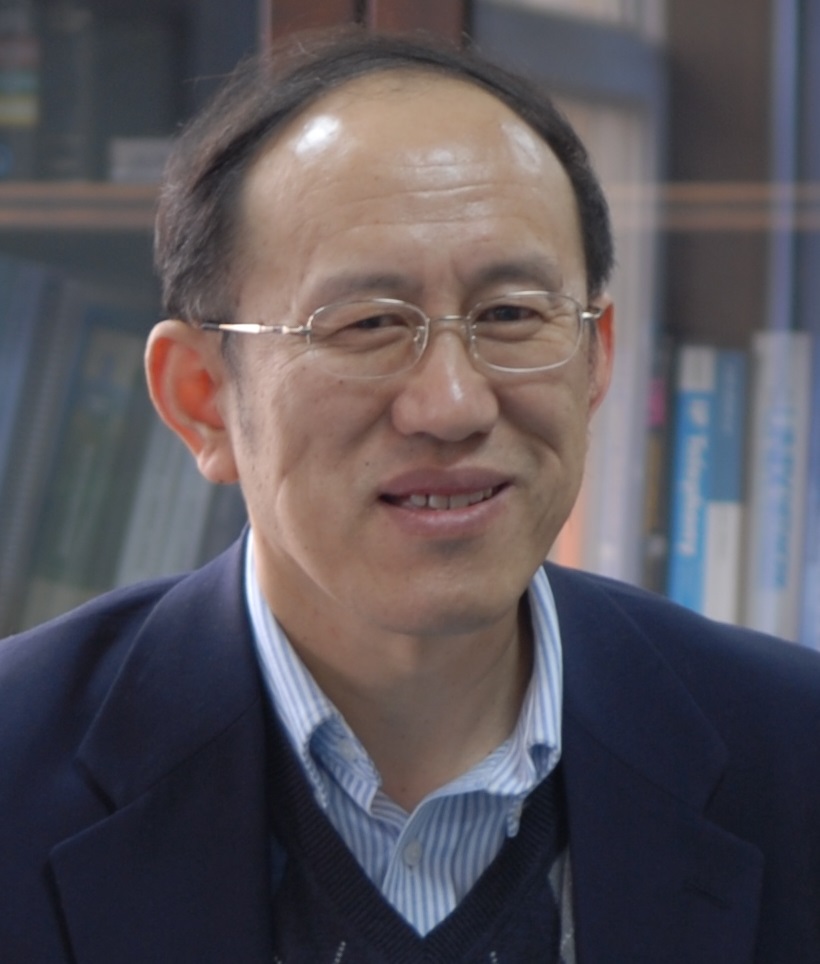}}]{Xiaohong Guan}
(Life Fellow, IEEE) received Ph.D. degree in Electrical and Systems Engineering from the University of Connecticut in 1993. He was a senior consulting engineer with Pacific Gas and Electric from 1993 to 1995. He visited the Division of Engineering and Applied Science, Harvard University from 1999 to 2000. From 1985 to 1988 and since 1995 he has been with Xi'an Jiaotong University, Xi'an, China, and has been as the Cheung Kong Professor of Systems Engineering since 1999, was the director of the State Key Lab for Manufacturing Systems from 1999 to 2009, Dean of School of Electronic and Information Engineering from 2008 to 2018, and Dean of Faculty of Electronic and Information Engineering since 2019. Since 2001 he has also been with the Center for Intelligent and Networked Systems, Tsinghua University, Beijing, China, and served as the Head of Department of Automation, Tsinghua University from 2003 to 2008. Professor Guan is the member of Chinese Academy of Science, and served as an Associate Editor of IEEE Transactions on Smart Grid. His research interests include economics and security of networked systems, optimization based planning and scheduling of power and energy systems.
\end{IEEEbiography}

\begin{IEEEbiography}[{\includegraphics[width=1in,height=1.25in,clip,keepaspectratio]{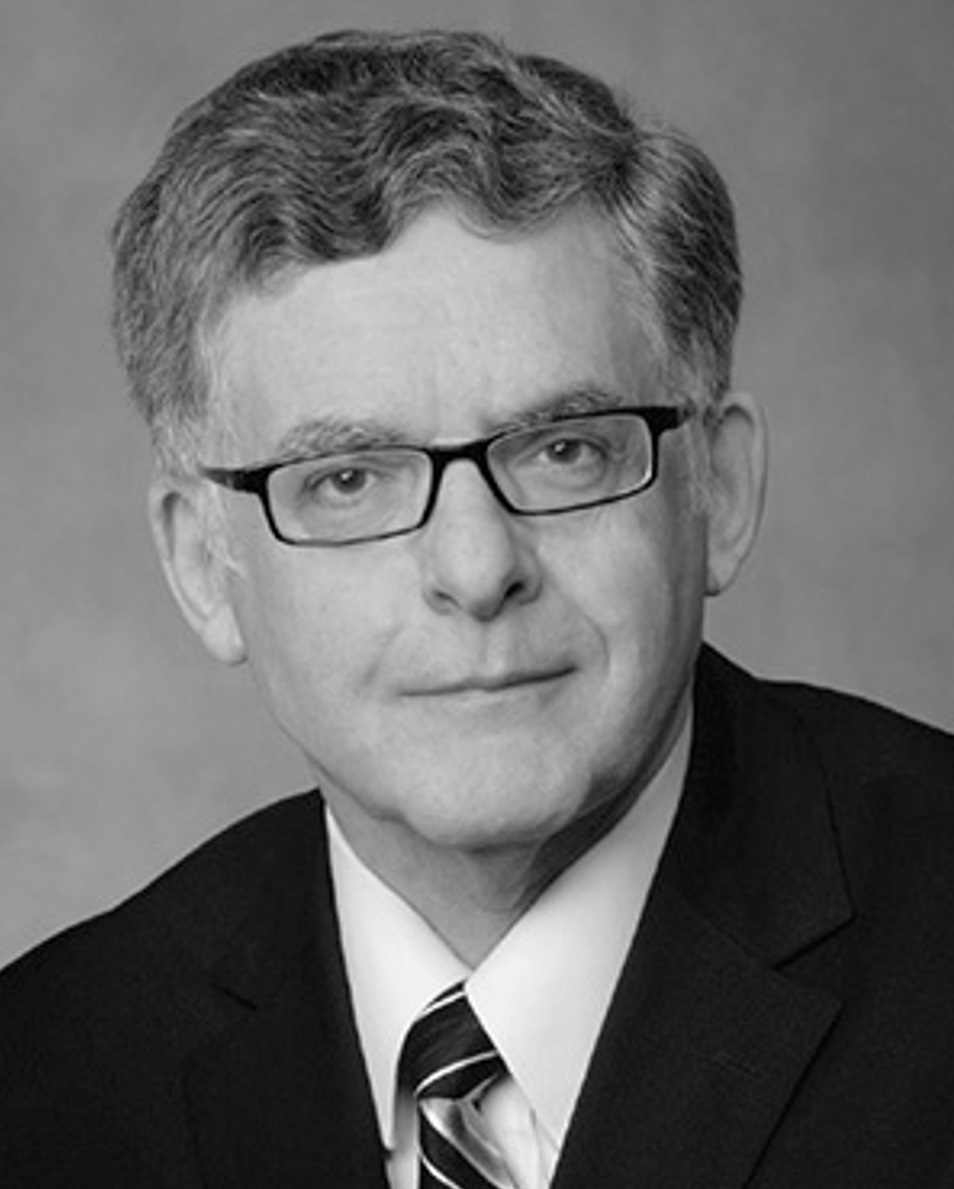}}]{Gerhard P. Hancke} (Life Fellow, IEEE) received the B.Sc., B.Eng. (1970) and M.Eng. (1973) from the University of Stellenbosch, and the D.Eng. (1983) from the University of Pretoria (UP), both in South Africa. He is a Professor with the College of Automation and College of Artificial Intelligence at the Nanjing University of Posts and Telecommunications, China, as well as with the Department of Electrical, Electronic and Computer Engineering at UP, South Africa. He is recognized internationally as a pioneer and leading scholar in Industrial Wireless Sensor Network (IWSN) research. He co-edited a textbook, Industrial Wireless Sensor Networks: Applications, Protocols and Standards (2013), the first on the topic. He initiated and co-edited the first Special Issues on IWSN in the IEEE Transactions on Industrial Electronics (TIE) in 2009 and the IEEE Transactions on Industrial Informatics (TII) in 2013. He received the TII Best Paper Award (For a paper on IWSN) in 2012. He has acted as AE and GE for TII, TIE and IEEE Access, Co-EiC for TII (2019-22), and Senior Editor for IEEE Access (2019-22). Currently, he is the Editor-in-Chief of the IEEE TII (since 2023).
\end{IEEEbiography}


\begin{thebibliography}{1}
\bibitem{YuY2024}
Y. Yu, G. Liu, Y. Huang, \emph{et al.}, ``A blockchain consensus mechanism for real-time regulation of renewable energy power systems," \emph{Nature Communications}, vol. 15, no. 1, pp. 1-15, 2024.


\bibitem{Malley2020}
M.J. O'Malley \emph{et al.}, ``Multicarrier Energy Systems: Shaping Our Energy Future," \emph{Proceedings of the IEEE}, vol. 108, no. 9, pp. 1437-1456, Sept. 2020


\bibitem{Perera2020}
A. Perera, V. Nik, D. Chen, J. Scartezzini, and T. Hong, ``Quantifying the impacts of climate change and extreme climate events on energy systems,Â¡Â± \emph{Nature Energy}, vol. 5, no. 2, pp. 150-159, 2020.

\bibitem{Nik2021}
V. Nik, A. Perera, D. Chen, ``Towards climate resilient urban energy systems: a review," \emph{National Science Review}, vol. 8, no. 3, pp. 1-18, 2021.


\bibitem{Stankovi2023}
A.M. Stankovi$\acute{c}$, ``Methods for analysis and quantification of power system resilience," \emph{IEEE Transactions on Power Systems}, vol. 38, no. 5, pp. 4774-4787, 2023.


\bibitem{Do2023}
V. Do, H. McBrien, N. Flores, \emph{et al.}, ``Spatiotemporal distribution of power outages with climate events and social vulnerability in the USA," \emph{Nature Communications}, vol. 14, pp. 2470-2482, 2023.

%
%

\bibitem{XuLuo2024}
L. Xu, K. Feng, N. Lin, A. Perera, H. Vincent Poor, L. Xie, C. Ji, X. Sun, Q. Guo, and M. Malley, ``Resilience of renewable power systems under climate risks," \emph{Nature Reviews in Electrical Engineering}, vol. 1, pp. 53-66, 2024.


\bibitem{Jasi2021}
J. Jasi$\bar{u}$nas, P. Lund, and J. Mikkola, ``Energy system resilience-A review," \emph{Renewable and Sustainable Energy Reviews}, vol. 150, no. 111476, pp. 1-18, 2021.



\bibitem{Afgan2012}
N. Afgan, A. Veziroglu, ``Sustainable resilience of hydrogen energy system," \emph{International Journal of Hydrogen Energy}, vol. 37, pp. 5461-5467, 2012.


\bibitem{Yang2022}
B. Yang, S. Ge, H. Liu, J. Li, and S. Zhang, ``Resilience assessment methodologies and enhancement strategies of multi-energy cyber-physical systems of the distribution network," \emph{IET Energy Systems Integration}, vol. 4, pp. 171-191, 2022.

\bibitem{Yang2024}
L. Yang, Z. Li, T. Liu, N. An, W. Zhou, and Y. Si, ``Two-stage robust resilience enhancement of distribution system against line failures via hydrogen tube trailers," \emph{Energies}, vol. 17, no. 20, pp. 5028-5041, 2024.

\bibitem{SunC2025}
C. Sun, Z. He, X. Chen, \emph{et al.}, ``Thermodynamic, sensitivity analyses and optimization of a dual-stage diaphragm compressor system: A model-based and experimental study," \emph{Energy}, vol. 330, no. 136848, pp. 1-20, 2025.

\bibitem{MaY2025}
Y. Ma, M. Huang, Y. Zhou, \emph{et al.}, ``Modeling of the multi-period two-stage green hydrogen supply chain with consideration of supply and demand uncertainties," \emph{International Journal of Hydrogen Energy}, vol. 129, pp. 297-314, 2025.

\bibitem{JiaW2025}
W. Jia, T. Ding, Y. Yuan, \emph{et al.}, ``Decentralized distributionally robust chance-constrained operation of integrated electricity and hydrogen transportation networks," \emph{Applied Energy}, vol. 377, no. 124369, pp. 1-16, 2025.

\bibitem{ReddiK2014}
K. Reddi, A. Elgowainy, and E. Sutherland, ``Hydrogen refueling station compression and storage optimization with tube-trailer deliveries," \emph{International Journal of Hydrogen Energy}, vol. 39, no. 33, pp. 19169-19181, 2014.

\bibitem{XiaW2024}
W. Xia, Z. Ren, H. Li, \emph{et al.}, ``A data-driven probabilistic evaluation method of hydrogen fuel cell vehicles hosting capacity for integrated hydrogen-electricity network," \emph{Applied Energy}, vol. 376, no. 123895, pp. 1-16, 2024.

\bibitem{Lv2024}
C. Lv, X. Jiang, X. Yan, \emph{et al.}, ``Optimal economic dispatch of hydrogen-electricity coupled multi-microgrid considering hydrogen tube trailers' transportation," Proc. of China International Conference on Electricity Distribution (CICED), pp. 111-116, 2024.

\bibitem{LiB2023}
B. Li, J. Li, B. Jian, ``Improve multi-energy supply microgrid resilience using mobile hydrogen trucks based on transportation network," \emph{eTransportation}, vol. 18, no. 100265, pp. 1-18, 2023.

\bibitem{QianH2024}
H. Qian, D. Tang, S. Sun, K. Yang and H. Wang, ``Resilience assessment and enhancement for distribution network by hydrogen transportation," \emph{2024 7th International Conference on Power and Energy Applications (ICPEA)}, Taiyuan, China, pp. 559-564, 2024.

\bibitem{SongJ2024}
J. Song, K. Qi, X. Liu, \emph{et al.}, ``Hydrogen leakage and diffusion in the operational cabin of hydrogen tube bundle containers: A CFD study," \emph{International Journal of Hydrogen Energy}, vol. 88, pp. 986-1002, 2024.


\bibitem{ZhangP2024}
P. Zhang, S. A. Mansouri, A. R. Jordehi, et al., ``An ADMM-enabled robust optimization framework for self-healing scheduling of smart grids integrated with smart prosumers," \emph{Applied Energy}, vol. 363, p. 123067, 2024.

\bibitem{Dong2023}
Y. Dong, W. Zheng, X. Cao, X. Sun, and Z. He, ``Co-planning of hydrogen-based microgrids and fuel-cell bus operation centers under low-carbon and resilience considerations," \emph{Applied Energy}, vol. 336, no. 120849, pp. 1-15, 2023.

\bibitem{ChenF2024}
F. Chen, M. Xia, Q. Chen, and L. Yang, ``Resilience enhancement method against persistent extreme weather with low temperatures in self-sustained highway transportation energy system," \emph{IEEE Transactions on Industry Applications}, vol. 60, no. 1, pp. 996-1009, 2024.


\bibitem{ChenS2024}
S. Chen, J. Zhang, Z. Wei, H. Cheng, and S. Lv, ``Towards renewable-dominated energy systems: Role of green hydrogen," \emph{Journal of Modern Power Systems and Clean Energy}, vol. 12, no. 6, pp. 1697-1709, 2024.

\bibitem{YuL2024}
L. Yu, D. Yue, Z. Chen, S. Zhang, Z. Xu and X. Guan, ``Online operation optimization for hydrogen-based building energy systems under uncertainties," \emph{IEEE Transactions on Smart Grid}, vol. 15, no. 5, pp. 4589-4601, 2024.

\bibitem{YueM2021}
M. Yue, H. Lambert, E. Pahon, \emph{et al.} ``Hydrogen energy systems: A critical review of technologies, applications, trends and challenges," \emph{Renewable and Sustainable Energy Reviews}, vol. 146, no. 111180, pp. 1-21, 2021.

\bibitem{GenoveseM2023}
M. Genovese, A. Schl$\ddot{u}$ter, E. Scionti, F. Piraino, O. Corigliano, and P. Fragiacomo, ``Power-to-hydrogen and hydrogen-to-X energy systems for the industry of the future in Europe," \emph{International Journal of Hydrogen Energy}, vol. 48, no. 44, pp. 16545-16568, 2023.

\bibitem{ZhaoH2022}
H. Zhao \emph{et al.}, ``Resilience assessment of hydrogen-integrated energy system for airport electrification," \emph{IEEE Transactions on Industry Applications}, vol. 58, no. 2, pp. 2812-2824, 2022.

\bibitem{Liu2024}
X. Liu, Z. Hua, X. Yuan, and J. Liu, ``Stochastic two-stage resilient scheduling of hydrogen system in power distribution grids with high penetration of solar PV systems: A convex worst case analysis," \emph{Sustainable Cities and Society}, vol. 112, no. 105581, pp. 1-22, 2024.



\bibitem{ZhaoY2024}
Y. Zhao, J. Lin, Y. Song, and Y. Xu, ``A robust microgrid formation strategy for resilience enhancement of hydrogen penetrated active distribution networks," \emph{IEEE Transactions on Power Systems}, vol. 39, no. 2, pp. 2735-2748, 2024.

\bibitem{ZhuR2024}
R. Zhu, H. Liu, W. Yu, W. Gu, and L. Sun, ``Resilience-oriented operation of integrated electricity-natural gas systems using hydrogen enriched compressed natural gas," \emph{IEEE Systems Journal}, vol. 18, no. 2, pp.953-964, 2023.

\bibitem{Wu2022}
Z. Wu, J. Wang, M. Zhou, Q. Xia, C.-W. Tan, and G. Li, ``Incentivizing frequency provision of power-to-hydrogen toward grid resiliency enhancement," \emph{IEEE Transactions on Industrial Informatics}, vol. 19, no. 9, pp. 9370-9381, 2022.

\bibitem{Han2023}
J. Han, J. Wang, Z. He, \emph{et al.} , ``Hydrogen-powered smart grid resilience," \emph{Energy Conversion and Economics}, vol. 4, no. 2, pp. 89-104, 2023.

\bibitem{Bie2017}
Z. Bie, Y. Lin, G. Li, and F. Li, ``Battling the extreme: A study on the power system resilience," \emph{Proceedings of the IEEE}, vol. 105, no. 7, pp. 1253-1266, 2017.



\bibitem{Li2017}
Z. Li, M. Shahidehpour, F. Aminifar, A. Alabdulwahab, and Y. Al-Turki, ``Networked microgrids for enhancing the power system resilience," \emph{Proceedings of the IEEE}, vol. 105, no. 7, pp. 1289-1310, 2017.

\bibitem{Mahzarnia2020}
M. Mahzarnia, M. P. Moghaddam, P. T. Baboli, and P. Siano, ``A review of the measures to enhance power systems resilience," \emph{IEEE Systems Journal}, vol. 14, no. 3, pp. 4059-4070, 2020.

\bibitem{Hossain2021}
E. Hossain, S. Roy, N. Mohammad, N. Nawar, and D. R. Dipta, ``Metrics and enhancement strategies for grid resilience and reliability during natural disasters," \emph{Applied Energy}, vol. 290, no. 116709, pp. 1-24, 2021.

\bibitem{Xu2021}
L. Xu, Q. Guo, Y. Sheng, S. Muyeen, H. Sun, ``On the resilience of modern power systems: A comprehensive review from the cyber-physical perspective," \emph{Renewable and Sustainable Energy Reviews}, vol. 151, no. 111642, pp. 1-23, 2021.

\bibitem{Ma2021}
X. Ma, H. Zhou, and Z. Li, ``On the resilience of modern power systems: A complex network perspective," \emph{Renewable and Sustainable Energy Reviews}, vol. 152, no. 111646, pp. 1-17, 2021.

\bibitem{Shi2022}
Q. Shi, W. Liu, B. Zeng, H. Hui, and F. Li, ``Enhancing distribution system resilience against extreme weather events: Concept review, algorithm summary, and future vision," \emph{International Journal of Electrical Power \& Energy Systems}, vol. 138, no. 107860, pp. 1-23, 2022.


\bibitem{Younesi2022}
A. Younesi, H. Shayeghi, Z. Wang, P. Siano, A. Mehrizi-Sani, and A. Safari, ``Trends in modern power systems resilience: State-of-the-art review," \emph{Renewable and Sustainable Energy Reviews}, vol. 162, no. 112397, pp. 1-15, 2022.

\bibitem{Amini2023}
F. Amini, S. Ghassemzadeh, N. Rostami, and V. S. Tabar, ``Electrical energy systems resilience: A comprehensive review on definitions, challenges, enhancements and future proceedings," \emph{IET Renewable Power Generation}, vol. 17, pp. 1835-1858, 2023.

\bibitem{Modaberi2023}
S. A. Modaberi, S. Tohidi, S. G. Zadeh, and T. G. Bolandi, ``A review of power system resilience assessment and enhancement approaches by focusing on wind farms and wind turbines," \emph{IET Renewable Power Generation}, vol. 17, pp. 2391-2410, 2023.

\bibitem{Huang2024}
H. Huang \emph{et al.}, ``Toward resilient modern power systems: from single-domain to cross-domain resilience enhancement," \emph{Proceedings of the IEEE}, vol. 112, no. 4, pp. 365-398, 2024.


\bibitem{Zidane2025}
T. E. K. Zidane \emph{et al.}, ``Power systems and microgrids resilience enhancement strategies: A review," \emph{Renewable and Sustainable Energy Reviews}, vol. 207, no. 114953, pp. 1-27, 2025.

\bibitem{Gu2024}
Y. Gu, \emph{et al.}, ``Robust resiliency-oriented planning of electricity-hydrogen island energy systems under contingency uncertainty," \emph{Journal of Cleaner Production}, no. 143678, pp. 1-13, 2024.

\bibitem{WenZ2024}
Z. Wen, X. Zhang, G. Wang, Y. Li, J. Qiu, and F. Wen, ``Data-driven stochastic-robust planning for resilient hydrogen-electricity system with progressive hedging decoupling," \emph{IEEE Transactions on Sustainable Energy}, vol. 16, no. 3, pp. 1545-1561, 2025.

\bibitem{Hassan2023}
Q. Hassan and A. Sameen and H. Salman, ``Hydrogen energy future: Advancements in storage technologies and implications for sustainability," \emph{Journal of Energy Storage}, vol. 72, no. 108404, pp. 1-16, 2023.



\bibitem{CaoY2021}
Y. Cao, Y. Yang, X. Zhao, \emph{et al.} ``A review of seasonal hydrogen storage multi-energy systems based on temporal and spatial characteristics," \emph{Journal of Renewable Materials}, vol. 9, no. 11, pp. 1823-1842, 2021.

\bibitem{Haggi2022}
H. Haggi, W. Sun, J. Fenton, and P. Brooker, ``Proactive rolling-horizon-based scheduling of hydrogen systems for resilient power grids," \emph{IEEE Transactions on Industry Applications}, vol. 58, no. 2, pp. 1737-1746, 2022.

\bibitem{CicekA2025}
A. $\c{C}$i$\c{c}$ek, ``Resilience-oriented energy operation of a country house sustained by fishing with renewable sources and hydrogen-based technologies," \emph{International Journal of Hydrogen Energy}, vol. 144, pp. 947-963, 2025.

\bibitem{resiliency}
https://www.ci.calistoga.ca.us/city-hall/departments-services/public-works/calistoga-resiliency-center



\bibitem{WuC2024}
C. Wu, X. Li, R. Jiang, \emph{et al.}, ``Understanding carbon resilience under public health emergencies: a synthetic difference-in-differences approach," \emph{Scientific Reports}, vol. 14, no. 1, p. 20581, 2024.

\bibitem{WangY2025}
Y. Wang, R. Tian, S. Zheng, \emph{et al.}, ``Multi-objective optimal scheduling of islands considering offshore hydrogen production," \emph{Scientific Reports}, vol. 15, no. 1, p. 27371, 2025.



\bibitem{YangT2025}
T. Yang, Y. Geng, and H. Wang, ``Dynamic energy flow optimization of electricity and hydrogen-enriched natural gas integrated energy systems," \emph{International Journal of Hydrogen Energy}, vol. 109, pp. 1274-1285, 2025.

\bibitem{Shahid2025}
S. M. I. Shahid, M. Farhan, A. Rao, \emph{et al.}, ``Hydrogen production by waste heat recovery of hydrogen-enriched compressed natural gas via steam methane reforming process," \emph{International Journal of Hydrogen Energy}, vol. 117, pp. 374-392, 2025.


\bibitem{LuJ2025}
J. Lu, L. Fang, F. Jiang, and X. Li, ``A Black Start Strategy for Hydrogen-Integrated Renewable Grids with Energy Storage Systems,'' \emph{Proc. 2025 IEEE Int. Conf. Energy Technologies for Future Grids (ETFG)}, 2025.

\bibitem{Le2024}
T. Le, P. Sharma, B. Bora, \emph{et al.} ``Fueling the future: A comprehensive review of hydrogen energy systems and their challenges," \emph{International Journal of Hydrogen Energy}, vol. 54, pp. 791-816, 2024.



\bibitem{RiziD2025}
D. Rizi, M. Nazari, and S. Hosseinian, \emph{et al.} ``Strengthening energy hub resilience: KNN-based cyber protection with demand-side data modification for hydrogen, electrical, and thermal storage," \emph{Journal of Energy Storage}, vol. 132, no. 117625, 2025.

\bibitem{HuaD2025}
D. Hua, H. Huang, P. Yan, \emph{et al.}, ``A multi-stage NSGA-III optimization model for false data injection attacks in integrated power-hydrogen cyber-physical systems, \emph{IET Renewable Power Generation}, vol. 19, no. 1, e70022, 2025.

\bibitem{ZhouY2024}
Y. Zhou, Q. Zhai, Z. Xu, L. Wu, and X. Guan, ``Multi-stage adaptive stochastic-robust scheduling method with affine decision policies for hydrogen-based multi-energy microgrid," \emph{IEEE Transactions on Smart Grid}, vol. 15, no. 3, pp. 2738-2750, May 2024.




\bibitem{ChenTSG2025}
Z. Chen \emph{et al.}, ``Reliability and comfort-aware operation optimization for hydrogen-based building energy systems in off-grid mode," \emph{IEEE Transactions on Smart Grid}, vol. 16, no. 4, pp. 2884-2899, 2025.

\bibitem{Yu2023}
L. Yu, Z. Xu, X. Guan, Q. Zhao, C. Dou and D. Yue, ``Joint optimization and learning approach for smart operation of hydrogen-based building energy systems," \emph{IEEE Transactions on Smart Grid}, vol. 14, no. 1, pp. 199-216, 2023.

\bibitem{ZhangY2022}
Y. Zhang, S. Cao, L. Zhao, \emph{et al.}, ``A case application of WRF-UCM models to the simulation of urban wind speed profiles in a typhoon," \emph{Journal of Wind Engineering and Industrial Aerodynamics}, vol. 220, pp. 104874, 2022.

\bibitem{Mancarella2014}
P. Mancarella, ``MES (multi-energy systems): An overview of concepts and evaluation models," \emph{Energy}, vol. 65, pp. 1-17, 2014.

\bibitem{WangX2024}
X. Wang, J. Huang, Z. Xu, C. Zhang and X. Guan, ``Real-world scale deployment of hydrogen-integrated microgrid: design and control," \emph{IEEE Transactions on Sustainable Energy}, vol. 15, no. 4, pp. 2380-2392, 2024.


\bibitem{FanG2024}
G. Fan, B. Yu, B. Sun, \emph{et al.}, ``Multi-time-space scale optimization for a hydrogen-based regional multi-energy system" \emph{Applied Energy}, vol. 371, no. 123430, pp. 1-24, 2024.

\bibitem{Carrington2021}
N. K. Carrington, I. Dobson, and Z. Wang, ``Extracting resilience metrics from distribution utility data using outage and restore process statistics," \emph{IEEE Transactions on Power Systems}, vol. 36, no. 6, pp. 5814-5823, Nov. 2021.

\bibitem{Dobson2024}
I. Dobson and S. Ekisheva, ``How long is a resilience event in a transmission system?: Metrics and models driven by utility data," \emph{IEEE Transactions on Power Systems}, vol. 39, no. 2, pp. 2814-2826, Mar. 2024.

\bibitem{Ahmad2025}
A. Ahmad and I. Dobson, ``Quantifying distribution system resilience from utility data: large event risk and benefits of investments," \emph{IET Conference Proceedings}, vol. 2024, no. 27, pp. 118-122, Jan. 2025.

\bibitem{Lin2024}
C. Lin, P. Zhang, Y. A. Shamash, Z. Lin, and X. Lu, ``Resilience-assuring hydrogen-powered microgrids," \emph{iEnergy}, vol. 3, pp. 77-81, 2024.

\bibitem{Cai2023}
W. Cai, S. A. Mansouri, A. R. Jordehi, \emph{et al.}, ``Resilience of hydrogen fuel station-integrated power systems with high penetration of photovoltaics," \emph{Journal of Energy Storage}, vol. 73, no. 108909, pp. 1-13, 2023.

\bibitem{Jordehi2024}
A. Jordehi, S. Mansouri, M. Tostado-V$\acute{e}$liz, A. Iqbal, M. Marzband, and F. Jurado, ``Industrial energy hubs with electric, thermal and hydrogen demands for resilience enhancement of mobile storage-integrated power systems," \emph{International Journal of Hydrogen Energy}, vol. 50, pp. 77-91, 2024.

\bibitem{Afsari2024}
N. Afsari, S. SeyedShenava, and H. Shayeghi, ``Chance constrained robust hydrogen storage management to service restoration in microgrids considering the methanation process model," \emph{International Journal of Hydrogen Energy}, vol. 50, pp. 1463-1476, 2024.


\bibitem{Xie2024}
P. Xie \emph{et al.}, ``Resilience enhancement strategies for power distribution network based on hydrogen storage and hydrogen vehicle," \emph{IET Energy Systems Integration}, vol. 2024, pp. 1-10, 2024.

\bibitem{MehrjerdiH2022}
H. Mehrjerdi, R. Hemmati, S. Mahdavi, \emph{et al.}, ``Multicarrier microgrid operation model using stochastic mixed integer linear programming," \emph{IEEE Transactions on Industrial Informatics}, vol. 18, no. 7, pp. 4674-4687, 2022.

\bibitem{Cao2023}
X. Cao, T. Cao, Z. Xu, B. Zeng, G. Feng, and X. Guan, ``Resilience constrained scheduling of mobile emergency resources in electricity-hydrogen distribution network," \emph{IEEE Transactions on Sustainable Energy}, vol. 14, no. 2, pp. 1269-1284, 2023.


\bibitem{ZhaoY2023}
Y. Zhao, J. Lin, Y. Song, and Y. Xu, ``A hierarchical strategy for restorative self-healing of hydrogen-penetrated distribution systems considering energy sharing via mobile resources," \emph{IEEE Transactions on Power Systems}, vol. 38, no. 2, pp. 1388-1404, 2023.



\bibitem{Yuan2024}
Z. Yuan and J. Li, ``Photovoltaic-penetrated power distribution networksÂ¡Â¯ resiliency-oriented day-ahead scheduling equipped with power-to-hydrogen systems: A risk-driven decision framework," \emph{Energy}, vol. 299, no. 131115, pp. 1-17, 2024.


\bibitem{ZhuS2024}
S. Zhu, Y. Xu, Y. Wang, and J. He, ``Resilience-motivated service restoration of interdependent power and hydrogen distribution system," \emph{IEEE Access}, vol. 12, no. 149528, pp. 1-15, 2024.


\bibitem{Su2024}
J. Su, R. Zhang, P. Dehghanian, M. H. Kapourchali, S. Choi, and Z. Ding, ``Renewable-dominated mobility-as-a-service framework for resilience delivery in hydrogen-accommodated microgrids," \emph{International Journal of Electrical Power \& Energy Systems}, vol. 159, no. 110047, pp. 1-12, 2024.

\bibitem{ZouX2024}
X. Zou, Y. Wang, and G. Strbac, ``A resilience-oriented pre-positioning approach for electric vehicle routing and scheduling in coupled energy and transport sectors," \emph{Sustainable Energy, Grids and Networks}, vol. 39, no. 101484, pp. 1-16, 2024.




\bibitem{WangYuze2024}
Y. Wang, J. Su, Y. Xue, X. Chang, Z. Li, and H. Sun, ``Toward on rolling optimal dispatch strategy considering alert mechanism for antarctic electricity-hydrogen-heat integrated energy system," \emph{IEEE Transactions on Sustainable Energy}, vol. 15, no. 4, pp. 2754-2471, 2024.


\bibitem{Cicek2024}
A. $\c{C}$i$\c{c}$ek, ``A novel resilience-oriented energy management strategy for hydrogen-based green buildings," \emph{Journal of Cleaner Production}, vol. 470, no. 143297, pp. 1-23, 2024.

\bibitem{Shahbazbegian2023}
V. Shahbazbegian, M. Shafie-khah, H. Laaksonen, G. Strbac, and H. Ameli, ``Resilience-oriented operation of microgrids in the presence of power-to-hydrogen systems," \emph{Applied Energy}, vol. 348, no. 121429, pp. 1-20, 2023.

\bibitem{Sharifpour2023}
M. Sharifpour, M. T. Ameli, H. Ameli, and G. Strbac, ``A resilience-oriented approach for microgrid energy management with hydrogen integration during extreme events," \emph{Energies}, vol. 16, no. 24, pp. 8099-8099, 2023.


\bibitem{Liu2021}
J. Liu, X. Cao, Z. Xu, X. Guan, X. Dong, and C. Wang, ``Resilient operation of multi-energy industrial park based on integrated hydrogen-electricity-heat microgrids," \emph{International Journal of Hydrogen Energy}, vol. 46, no. 57, pp. 28855-28869, 2021.


\bibitem{Tang2022}
W. Tang, Z. Wang, L. Zhang, B. Zhang, J. Liang, K. Liu, and W. Sheng, ``Optimal allocation strategy of electric EPSVs and hydrogen fuel cell EPSVs balancing resilience and economics," \emph{CSEE Journal of Power and Energy Systems}, vol. 11, no. 3, pp. 1270-1283, 2025.

\bibitem{AminHashemifar2022}
S. Hashemifar, M. Joorabian, M. Javadi, \emph{et al.}, ``Two-layer robust optimization framework for resilience enhancement of microgrids considering hydrogen and electrical energy storage systems," \emph{International Journal of Hydrogen Energy}, vol. 47, pp. 33597-33618, 2022.

\bibitem{Huangchunjun2023}
C. Huang, Y. Zong, S. You, \emph{et al.}, ``Economic and resilient operation of hydrogen-based microgrids: An improved MPC-based optimal scheduling scheme considering security constraints of hydrogen facilities," \emph{Applied Energy}, vol. 335, no. 120762, pp. 1-13, 2023.

\bibitem{AlMuhaini2024}
M. AlMuhaini, ``Resilience quantification model for cyber-physical power systems," \emph{IET Cyber-Physical Systems: Theory \& Applications}, vol. 9, no. 4, pp. 454-462, 2024.

\bibitem{Kaloti2023}
S. Kaloti, and B. Chowdhury, ``Toward reaching a consensus on the concept of power system resilience: Definitions, assessment frameworks, and metrics,"  \emph{IEEE Access}, vol. 11, pp. 81401-81418, 2023.

\bibitem{Amiri2024}
M. Amiri, F. Gu¨¦niat, ``Towards a framework for measurements of power systems resiliency: Comprehensive review and development of graph and vector-based resilience metrics," \emph{Sustainable Cities and Society}, no.105517, 2024.

\bibitem{Amini2024}
S. Amini, Omid Safarzadeh, and Babak Mozafari, ``Preventive framework for resilience enhancement in networked microgrids: A focus on hydrogen integration and optimal energy management," \emph{IET generation, transmission \& distribution}, vol. 18, no. 8, pp. 1653-1662, 2024.


\bibitem{Haggi2021}
H. Haggi, \emph{et al.}, ``Proactive scheduling of hydrogen systems for resilience enhancement of distribution networks," \emph{2021 IEEE Kansas Power and Energy Conference (KPEC)}, IEEE, 2021.

\bibitem{MaH2025}
H. Ma and H. Wang, ``Optimal resilient scheduling strategy for electricity-gas-hydrogen multi-energy microgrids considering emergency islanding,'' \emph{Energy}, vol. 324, no. 135732, 2025.


\bibitem{Gao2024}
F. Gao, \emph{et al.}, ``A two-stage resilience planning for integrated electricity-gas energy system considering hydrogen refueling stations during typhoons," \emph{Smart Power and Energy Security}, vol. 1, no. 1, pp. 25-35, 2025.

\bibitem{SunYong2023}
Y. Sun, B. Li, J, Yang, \emph{et al.}, ``Resilience Improvement of active distribution network with electric-hydrogen hybrid energy storage microgrids under ice disaster," \emph{2023 IEEE 7th Conference on Energy Internet and Energy System Integration (EI2)}, 2023.

\bibitem{Bennett2021}
J. Bennett \emph{et al.}, ``Extending energy system modelling to include extreme weather risks and application to hurricane events in Puerto Rico," \emph{Nature Energy}, vol. 6, no. 3, pp. 1-10, 2021.

\bibitem{Shao2023}
Z. Shao, X. Cao, Q. Zhai, and X. Guan, ``Risk-constrained planning of rural-area hydrogen-based microgrid considering multiscale and multi-energy storage systems," \emph{Applied Energy}, vol. 334, no. 120682, pp. 1-16, 2023.

\bibitem{LiuKun2023}
K. Liu, Z. Xu, F. Gao, J. Wu, and X. Guan, ``Coordination optimization of hydrogen-based multi-energy system with multiple storage for industrial park," \emph{IET Generation, Transmission \& Distribution}, vol. 17, no. 6, pp. 1190-1203, 2023.

\bibitem{ChenZhe2024}
Z. Chen, Z. Sun, D. Lin, Z. Li, and J. Chen, ``Optimal configuration of multi-energy storage in an electric–thermal–hydrogen integrated energy system considering extreme disaster scenarios," \emph{Sustainability}, vol. 16, no. 2276, pp. 1-24, 2024.


\bibitem{Artime2024}
O. Artime, M. Grassia, M. Domenico, \emph{et al.}, ``Robustness and resilience of complex networks," \emph{Nature Reviews Physics}, vol. 6, pp. 114-131, 2024.

\bibitem{Yan2025}
J. Ye, J. Song, S. Yan, H. Ma, Q. Yang, D. Li, and W. Yin, ``Resilient planning for battery and hydrogen energy storage in power systems against extreme heatwave events,'' \emph{Journal of Energy Storage}, vol. 139, part B, no. 118850, 2025.


\bibitem{Qu2024}
J. Qu, K. Hou, Z. Liu, H. Jia, and L. Zhu, ``Enhancing distribution network resilience: a multi-level cooperative planning approach for hydrogen-electrical microgrids and sop," \emph{2024 IEEE Power \& Energy Society General Meeting (PESGM)}, pp. 1-5, Jul. 2024.

\bibitem{AlizadE2024}
E. Alizad, F. Hasanzad, and H. Rastegar, ``A tri-level hybrid stochastic-IGDT dynamic planning model for resilience enhancement of community-integrated energy systems," \emph{Sustainable Cities and Society}, vol. 117, p. 105948, 2024.



\bibitem{BerneckerM2026}
M. Bernecker, S. Sgarciu, X. Kan, \emph{et al.}, ``Adaptive robust optimization for european electricity system planning considering regional dunkelflaute events," \emph{Applied Energy}, vol. 412, article no. 127671, 2026.


\bibitem{Oh2024}
B. Oh, Y. Son, D. Zhao, et al. ``A bi-level approach for networked microgrid planning considering multiple contingencies and resilience," \emph{IEEE Transactions on Power Systems}, vol. 39, no. 4, pp. 5620-5630, 2024.


\bibitem{Ameli2024}
H. Ameli, D. Pudjianto, and G. Strbac, ``The impact of hydrogen on decarbonisation and resilience in integrated energy systems," \emph{Advances in Applied Energy}, vol. 17, no. 100200, 2025.

\bibitem{XieC2024}
C. Xie, P. Dehghanian, A. Estebsari, ``Fueling the seaport of the future: Investments in low-carbon energy technologies for operational resilience in seaport multi-energy systems," \emph{IET Generation, Transmission \& Distribution}, vol. 18, no. 2, pp. 248-265, 2024.



\bibitem{WangT2025}
T. Wang, Y. Dong, X. Cao, Z. Xu, F. Gao, and X. Guan, ``Economic and resilient planning of hydrogen-enriched power distribution network with mobile hydrogen energy resources," \emph{International Journal of Hydrogen Energy}, vol. 144, pp. 1001-1008, 2025.

\bibitem{LiuS2026}
S. Liu, B. Yang, X. Li, X. Yang, Z. Wang, D. Zhu, and X. Guan, ``Defense hardening of electricity-hydrogen networks for extreme weather: A risk limitation approach with endogenous uncertainties,'' \emph{Cyber-Physical Energy Systems}, vol. 2, no. 1, pp. 42-52, 2026.


\bibitem{DongY2026}
Y. Dong, Z. Lu, X. Cao, \emph{et al.}, ``Distributionally robust planning of hydrogen-electrical microgrids for sea islands,'' \emph{Applied Energy}, vol. 411, no. 127630, 2026.



\bibitem{WangHong2025}
H. Wang, B. Qin, S. Hong, \emph{et al.}, ``Optimal planning of hybrid hydrogen and battery energy storage for resilience enhancement using bi-layer decomposition algorithm," \emph{Journal of Energy Storage}, vol. 110, no. 115367, pp. 1-12, 2025.



\bibitem{Sturmer2024}
J. St$\ddot{Â¨Â¹}$rmer \emph{et al.}, ``Increasing the resilience of the Texas power grid against extreme storms by hardening critical lines," \emph{Nature Energy}, vol. 9, no. 5, pp. 1-10, Mar. 2024.

\bibitem{Arjomandi2020}
A. Arjomandi-Nezhad, \emph{et al.}, ``Modeling and optimizing recovery strategies for power distribution system resilience," \emph{IEEE Systems Journal}, vol. 15, no. 4, pp. 4725-4734, 2020.

\bibitem{LongoE2026}
E. Longo, A. Ficchì, M. Verlaan, \emph{et al.}, ``A deep learning framework for extreme storm surge modeling under future climate scenarios," \emph{Earth's Future}, vol. 14, no. 3, pp. 1-19, 2026.

\bibitem{HuJ2026}
J. Hu, B. Xu, H. Zhou, \emph{et al.}, ``A spatial-clustering conditional variational auto-encoder framework for high-dimensional scenario generation of large-scale multi-site hybrid energy systems," \emph{Energies}, vol. 19, no. 6, article no. 1520, 2026.

\bibitem{ZhangJ2025}
J. Zhang, Y. Zhang, J. Teng, \emph{et al.}, ``Optimal energy storage allocation for power systems with high-wind-power penetration against extreme-weather events," \emph{Energies}, vol. 19, no. 1, article no. 146, 2025.

\bibitem{WuX2026}
X. Wu, Y. Hao, J. Zhou, X. Zhang, and Y. Zhang, ``A scenario generation method for wind/pv power outputs and load sequences preserving extreme scenario characteristics," \emph{IET Renewable Power Generation}, vol. 20, no. 1, e70185.

\bibitem{HuaD2026}
D. Hua, \emph{et al.}, ``An expectile-based framework for risk-calibrated credible capacity evaluation of virtual power plants under wind and PV forecast uncertainties," \emph{Scientific Reports}, https://www.nature.com/articles/s41598-026-44559-5, 2026.

\bibitem{OrtegaJ2009}
J. Ortega, R. Pullirsch, J. Teichmann, and J. Wergieluk, ``A new approach for scenario generation in risk management," https://arxiv.org/abs/0904.0624, 2009.

\bibitem{ZhaoZ2026}
Z. Zhao, J. Yang, X. Su, \emph{et al.}, ``Applications of AI for the optimal operations of power systems under extreme weather events: a task-driven and methodological review," \emph{Energies}, vol. 19, no. 2, article no. 506, 2026.

\bibitem{SerreC2026}
C. Serre-Combe, N. Meyer, T. Opitz, and G. Toulemonde, ``Spatio-temporal modeling of urban extreme rainfall events at high resolution," https://arxiv.org/abs/2602.19774, 2026.

\bibitem{GuJ2025}
J. Gu, X. Zhang, and G. Wang, ``Beyond the norm: A survey of synthetic data generation for rare events," https://arxiv.org/abs/2506.06380, 2025.

\bibitem{IizumiT2012}
T. Iizumi, \emph{et al.}, ``ELPIS-JP: a dataset of local-scale daily climate change scenarios for Japan," \emph{Philosophical Transactions of the Royal Society A: Mathematical, Physical and Engineering Sciences}, vol. 370, no. 1962, pp. 1121-1139, 2012.

\bibitem{NajibiN2024}
N. Najibi, \emph{et al.}, ``A statewide, weather-regime based stochastic weather generator for process-based bottom-up climate risk assessments in California-Part I: Model evaluation," \emph{Climate Services}, vol. 34, article no. 100489, 2024.



\bibitem{WangH2025}
H. Wang et al., ``Enhanced GAN-based joint wind-solar-load scenario generation with extreme weather labelling," \emph{IEEE Transactions on Smart Grid}, vol. 16, no. 5, pp. 4213-4224, 2025.

\bibitem{YangR2025}
R. Yang, Y. Li, C. Hu and F. Y. Hou, ``ExDiffusion: Classifier-guidance diffusion model for extreme load scenario generation with extreme value theory," \emph{IEEE Transactions on Smart Grid}, vol. 16, no. 5, pp. 3887-3903, 2025.


\bibitem{FuX2026}
X. Fu, \emph{et al.}, ``Redesigning the decoder and loss function of diffusion transformer for pv temporal simulation," \emph{IEEE Transactions on Smart Grid}, vol. 17, no. 2, pp. 1629-1638, 2026.


\bibitem{LiY2025EI}
Y. Li, L. Li, J. Yan, W. Mao, G. Gao, and K. Zhu, ``A review and validation of power systems extreme scenario generation methods oriented towards dispatch decision," \emph{Proc. of IEEE 9th Conference on Energy Internet and Energy System Integration (EI2)}, pp. 2945-2948), 2025.

\bibitem{DeF2025}
F. De Marco, J. Mannhardt, A. Oneto, and G. Sansavini, ``Climate-resilient energy systems planning via system-informed identification of stressful events," \emph{Advances in Applied Energy}, vol. 19, article no. 100235, 2025.



\bibitem{JiaF2017}
F. Jia, L. Guo, and H. Liu, ``Mitigation strategies for hydrogen starvation under dynamic loading in proton exchange membrane fuel cells," \emph{Energy Conversion and Management}, vol. 139, pp. 175-181, 2017.

\bibitem{ChangS2025}
S. Chang, G. Li, T. Zhang, \emph{et al.}, ``Characterising the resilience of electro-hydrogen coupled system via convex hull estimation," \emph{Energy Conversion and Economics}, vol. 6, no. 1, pp. 13-25, 2025.

\bibitem{Wangz2021}
Z. Wang, T. Ding, W. Jia, \emph{et al.}, ``Multi-period restoration model for integrated power-hydrogen systems considering transportation states," \emph{IEEE Transactions on Industry Applications}, vol. 58, no. 2, pp. 2694-2706, Oct. 2021.

\bibitem{GuZhong2025}
Z. Gu \emph{et al.}, ``Optimal scheduling and resilience enhancement of electricity-hydrogen-heat coupled systems with flexible reversible solid oxide cells operation," \emph{International Journal of Hydrogen Energy}, vol. 99, pp. 1065-1078, 2025.



\bibitem{TangD2025}
D. Tang \emph{et al.}, ``Economic and resilience-oriented operation of coupled hydrogen-electricity energy systems at ports," \emph{Applied Energy}, vol. 390, pp. 125825-125825, 2025.

\bibitem{LiL2025}
L. Li, C. Li, Y. Z. Alharthi, Y. Wang, and Murodbek Safaraliev, ``A two-layer economic resilience model for distribution network restoration after natural disasters," \emph{Applied Energy}, vol. 377, pp. 124605-124605, 2025.


\bibitem{ChenJ2025}
J. Chen, W. Gu, Y. Z. Alharthi, S. Huang, and S. A. Mansouri, ``A decentralized framework for self-healing in hydrogen-integrated energy systems," \emph{Energy}, vol. 331, pp. 137033-137033, 2025.





\bibitem{FanG2025}
G. Fan, W. Zhang, H. Bie, \emph{et al.}, ``Collaborative planning for power and hydrogen networks considering hydrogen pipeline slow dynamic and pipe storage characteristics, \emph{International Journal of Hydrogen Energy}, vol. 125, pp. 214-232, 2025.

\bibitem{Bah2026}
M. M. Bah, S. Wang, M. Kia, \emph{et al.}, ``Modeling hydrogen integration in energy system models: Best practices for policy insights,'' \emph{International Journal of Hydrogen Energy}, vol. 217, no. 153777, 2026.

\bibitem{YaoP2025}
P. Yao, B. Yan, and Q. Yang, ``Game theoretical decision-making of dynamic defense in cyber-physical power systems under cyber-attacks," \emph{ACM Transactions on Cyber-Physical Systems}, vol. 9, no. 2, pp. 1-21, 2025.

\bibitem{LiJ2025}
J. Li, Y. Li, and Q. Su, ``Sequential recovery of cyber-physical power systems considering cyber-attacks," \emph{Information Sciences}, vol. 717, pp. 122310, 2025.

\bibitem{FengY2025}
Y. Feng, R. Huang, W. Zhao, \emph{et al.}, ``A survey on coordinated attacks against cyber¨Cphysical power systems: Attack, detection, and defense methods," \emph{Electric Power Systems Research}, vol. 241, pp. 111286, 2025.

\bibitem{LiuF2025}
F. Liu, L. Wu, Q. Liu, and D. Sidorov, ``Dynamic-memory event-triggered secure control for cyber-physical power systems under hybrid attacks," \emph{IEEE Transactions on Network Science and Engineering}, vol. 12, no. 5, pp. 3850-3863, 2025.



\bibitem{AghajanEshkevariS2024}
S. Aghajan-Eshkevari, M. T. Ameli, S. Azad, and N. T. Pashiri, ``Resilience enhancement of distribution networks by optimal scheduling of hydrogen systems," \emph{Future Modern Distribution Networks Resilience}, Elsevier, pp. 353-374, 2024.



\bibitem{LiuJin2021}
J. Liu, Z. Xu, J. Wu, K. Liu, and X. Guan, ``Optimal planning of distributed hydrogen-based multi-energy systems," \emph{Applied Energy}, vol. 281, no. 116107, pp. 1-12, 2021.

\bibitem{Mehrjerdi2021}
H. Mehrjerdi, R. Hemmati, S. Mahdavi, \emph{et al.} ``Multicarrier microgrid operation model using stochastic mixed integer linear programming," \emph{IEEE Transactions on Industrial Informatics}, vol. 18, no. 7, pp. 4674-4687, 2021.

\bibitem{WangH2022}
H. Wang, K. Hou, J. Zhao, X. Yu, H. Jia, Y. Mu, ``Planning-Oriented resilience assessment and enhancement of integrated electricity-gas system considering multi-type natural disasters, \emph{Applied Energy}, vol. 315, no. 118824, pp. 1-18, 2022.

\bibitem{Yodo2017}
N. Yodo, P. Wang and Z. Zhou, ``Predictive resilience analysis of complex systems using dynamic bayesian networks," \emph{IEEE Transactions on Reliability}, vol. 66, no. 3, pp. 761-770, 2017.

\bibitem{HuY2024}
Y. Hu, J. Xue, J. Zhao, \emph{et al.} ``Dynamic Bayesian networks for spatiotemporal modeling and its uncertainty in tradeoffs and synergies of ecosystem services: a case study in the Tarim River Basin, China" \emph{Stochastic Environmental Research and Risk Assessment}, vol. 38, no. 11, pp. 4311-4329, 2024.

\bibitem{BaiK2025}
K. Bai, W. Zhang, S. Wen, D. Jia and W. Meng, ``An Interpretable Data-Driven Fuzzy Petri Net Method for Industrial Domain Knowledge Modeling of Energy Efficiency Management," \emph{IEEE Transactions on Automation Science and Engineering}, vol. 22, pp. 16603-16615, 2025.

\bibitem{Panteli2017}
M. Panteli, D. N. Trakas, P. Mancarella and N. D. Hatziargyriou, ``Power systems resilience assessment: hardening and smart operational enhancement strategies," \emph{Proceedings of the IEEE}, vol. 105, no. 7, pp. 1202-1213, 2017.

\bibitem{JiaQ2025}
Q. Jia, T. Zhang, Z. Zhu, \emph{et al.}, ``Harnessing hydrogen energy storage for renewable energy stability in China: A path to carbon neutrality", \emph{International Journal of Hydrogen Energy}, vol. 118, pp. 93-101, 2025.

\bibitem{Evro2024}
S. Evro, A. Oni, O. Tomomewo, ``Carbon neutrality and hydrogen energy systems," \emph{International Journal of Hydrogen Energy}, vol. 78, pp. 1449-1467, 2024.

\bibitem{IEA2024}
International Energy Agency (IEA). World Energy Outlook 2024. [Online]. Available:https://www.iea.org/reports/world-energy-outlook-2024.

\bibitem{ZhangS2024}
S. Zhang, W. Chen, Q. Zhang, \emph{et al.} ``Targeting net-zero emissions while advancing other sustainable development goals in China," \emph{Nature Sustainability}, vol. 7, no. 9, pp. 1107-1119, 2024.



\bibitem{Nguyen2020}
T. T. Nguyen, N. D. Nguyen, P. Vamplew, S. Nahavandi, R. Dazeley, and C. P. Lim, ``A multi-objective deep reinforcement learning framework," \emph{Engineering Applications of Artificial Intelligence}, vol. 96, p. 103915, 2020.

\bibitem{WangZ2023}
Z. Wang, T. Zeng, X. Chu, \emph{et al.}, ``Multi-objective deep reinforcement learning for optimal design of wind turbine blade," \emph{Renewable Energy}, vol. 203, pp. 854-869, 2023.

\bibitem{Coskun2024}
S. Coskun, O. Yazar, F. Zhang, L. Li, C. Huang, H. Karimi, ``A multi-objective hierarchical deep reinforcement learning algorithm for connected and automated HEVs energy management," \emph{Control Engineering Practice}, vol. 153, no. 106104, pp. 1-14, 2024.

\bibitem{HeQing2023}
Q. He, T. Gao, Y. Gao, H. Li, P. Schonfeld, Y. Zhu, Q. Li, P. Wang, ``A bi-objective deep reinforcement learning approach for low-carbon-emission high-speed railway alignment design," \emph{Transportation Research Part C: Emerging Technologies}, vol. 147, no. 104006, pp. 1-19, 2023.








\bibitem{Wen2025}
X. Wen, X. Zhang, H. Li, \emph{et al.}, ``An improved NSGA-II algorithm based on reinforcement learning for aircraft moving assembly line integration optimization problem," \emph{Swarm and Evolutionary Computation}, vol. 94, no. 101911, 2025.

\bibitem{WuR2025}
R. Wu, R. Wang, J. Hao, \emph{et al.}, ``Multiobjective vehicle routing optimization with time windows: A hybrid approach using deep reinforcement learning and NSGA-II," \emph{IEEE Trans. Intelligent Transportation Systems}, vol. 26, no. 3, pp. 4032-4047, 2025.

\bibitem{Bi2023}
K. Bi, L. Xie, H. Zhang, X. Chen, X. Gu, and Q. Tian, ``Accurate medium-range global weather forecasting with 3D neural networks," \emph{Nature}, vol. 619, no. 7970, pp. 533-538, 2023.

\bibitem{Ding2024}
D. J. Ding, \emph{et al.}, ``Forecasting of tropospheric delay using AI foundation models in support of microwave remote sensing," \emph{IEEE Transactions on Geoscience and Remote Sensing}, vol. 62, no. 5802019, 2024.


%
%
%
%
%
%
%
%
%
%
%
%
%
%
%
%
%
%
%
%
%
%
%

\end{thebibliography}
\end{document}